\documentclass[
footinbib,
a4paper,
aps,
prx,
superscriptaddress,
reprint,
twocolumn,
preprintnumbers,
nobalancelastpage,
floatfix,
10pt]{revtex4-2}

\usepackage{xcolor}
\usepackage{graphicx}
 	\graphicspath{ {figuras/} }
\usepackage{dcolumn}
\usepackage{tikz}
\usepackage{float}
\definecolor{cmblue}{rgb}{0.12156862745098039, 0.4666666666666667, 0.7058823529411765}
\definecolor{mygrey}{gray}{0.35}
\definecolor{myblue}{rgb}{0.2,0.2,0.8}
\definecolor{mygreen}{rgb}{0.2,0.8,0.5}
\definecolor{myzard}{cmyk}{0,0,0.05,0}
\definecolor{mywhite}{rgb}{1,1,1}
\definecolor{myred}{rgb}{1,0.,0.3}

\usepackage[utf8]{inputenc}
\usepackage{dsfont}
\usepackage[normalem]{ulem}
\usepackage{bbold}
\usepackage{soul}
\usepackage{slashed}

\usepackage{amsmath, amsthm, amsfonts, amssymb,amstext}
\usepackage{mathptmx}
\usepackage{bm}
\usepackage{xfrac}
\usepackage{mathrsfs}
\usepackage{latexsym}
\usepackage{array}

\def\dd{\mathord{\rm d}} 
 \def\ee{\mathord{\rm e}}
 
 \def\ii{\mathord{\rm i}}

\def\half{\textstyle\frac{1}{2}}

\def\fourth{\textstyle\frac{1}{4}}

\newcommand{\sinc}{{\rm{sinc}}\,}

\def\beq{\begin{equation}}
\def\eeq{\end{equation}}
\def\barray{\begin{eqnarray}}
\def\earray{\end{eqnarray}}

\usepackage{physics}
\usepackage{braket}
\usepackage{graphicx}
\usepackage{dcolumn}
\usepackage{bm}
\usepackage{hyperref}
\hypersetup{
  colorlinks   = true, 
  urlcolor     = blue, 
  linkcolor    = blue, 
  citecolor   = blue
}
\usepackage{physics}
\usepackage{stackengine}
\usepackage[caption=false]{subfig}

\begin{document}


\title{  Lindblad-like  quantum  tomography for non-Markovian  quantum  dynamical  maps}

\author{S. Varona}
\email{svarona@physik.rwth-aachen.de}
\affiliation{Institute for Quantum Information, RWTH Aachen University, 52056 Aachen, Germany}
\affiliation{Instituto de Física Teórica, UAM-CSIC, Universidad Autónoma de Madrid, Cantoblanco, 28049 Madrid, Spain}
\author{M. M\"uller}
\affiliation{Institute for Quantum Information, RWTH Aachen University, 52056 Aachen, Germany}
\affiliation{Peter Grünberg Institute, Theoretical Nanoelectronics, Forschungszentrum Jülich, 52425 Jülich, Germany} 
\author{A. Bermudez}

\affiliation{Instituto de Física Teórica, UAM-CSIC, Universidad Autónoma de Madrid, Cantoblanco, 28049 Madrid, Spain}


\begin{abstract}
We 
introduce Lindblad-like quantum tomography (L$\ell$QT) as a quantum characterization technique of time-correlated noise in quantum information processors.  This approach enables the estimation of time-local master equations, including their possible negative decay rates, by maximizing a likelihood function subject to dynamical constraints.  We  discuss  L$\ell$QT  for the dephasing dynamics of single qubits in detail, which allows for a neat understanding of   the importance of including multiple snapshots of the quantum evolution in the likelihood function, and how these need to be  distributed in time depending on the noise characteristics. By a detailed comparative study employing both frequentist and Bayesian approaches, we assess the accuracy and precision of L$\ell$QT of a dephasing quantum dynamical map that goes beyond the Lindblad limit,    focusing on  two different microscopic  noise models that can be realised in either trapped-ion or superconducting-circuit architectures.  We explore the optimization of the distribution of measurement times to minimize the estimation errors,   assessing the  superiority of each  learning scheme conditioned on the  degree of non-Markovinity of  the  noise, and setting the stage for future experimental designs of non-Markovian quantum tomography.  

\end{abstract}

\maketitle

\setcounter{tocdepth}{2}
\begingroup
\hypersetup{linkcolor=black}
\tableofcontents
\endgroup

\section*{ \bf Introduction}
\label{sec:intro}

The field of quantum information processing  has witnessed a remarkable progress in the last years~\cite{nielsen_chuang_2010,doi:10.1063/1.5088164,10.1116/5.0036562,Wendin_2017,Arute2019}, as evidenced by the key advances reported in~\cite{doi:10.1126/science.abe8770,Pino2021,Egan2021,Postler2022,Krinner2022,PhysRevLett.129.030501,Acharya2023,Kim2023,Bluvstein2024,Gupta2024, wang2023faulttolerant, Yamamoto_2024, https://doi.org/10.48550/arxiv.2208.01863}. This progress lays the groundwork for the eventual demonstration of practical quantum advantage in real-world applications \cite{dalzell2023quantum}. Central to these advancements and  future breakthroughs  is the exceptional level of isolation and control achieved over quantum information processors (QIPs), enabling the application and integration of various strategies to fight against the accumulation of errors during quantum computations. These strategies can be implemented either during the processing of quantum information or, alternatively, post-measurement,  falling into three distinct categories: quantum error suppression (QES) \cite{PhysRevLett.82.2417,PhysRevLett.95.180501,PhysRevLett.102.080501,review_dd}, quantum error mitigation (QEM) \cite{PhysRevLett.119.180509,PRXQuantum.2.040326,vandenBerg2023,RevModPhys.95.045005}, and quantum error correction (QEC) \cite{PhysRevA.54.1098,PhysRevLett.77.793,RevModPhys.87.307}.

The development and optimization of these techniques for specific  architectures greatly benefits from a comprehensive understanding  of the underlying sources of noise,  including a thorough  quantum characterization, verification and validation (QCVV) of the noise models~\cite{Eisert2020,PRXQuantum.2.010201,Gebhart2023}. By addressing the noise characteristics, researchers can tailor their strategies to suit the specific requirements of different platforms, thereby enhancing the reliability and performance of QIPs. For instance, the presence of spatial and temporal noise correlations  is a critical consideration in some techniques of QES, such as decoherence-free subspaces~\cite{PhysRevLett.79.3306,PhysRevLett.81.2594,PhysRevLett.84.2525,PhysRevA.63.042307,Lidar_2003} and dynamical decoupling  \cite{PhysRevA.58.2733,PhysRevLett.93.130406,Gordon_2007,PhysRevLett.98.100504,PhysRevB.77.174509,Biercuk_2011}, respectively. Likewise, in the context of QEC, the presence of spatial~\cite{PhysRevA.69.062313,PhysRevLett.95.230503,PhysRevLett.96.050504} and temporal~\cite{PhysRevA.71.012336,10.5555/2011665.2011666,PhysRevA.79.032318} noise correlations  must be carefully accounted for when considering fault-tolerant quantum computation beyond the idealized regime of independent and identically distributed errors. In this work, we will focus on the characterization of qubit dynamics under temporally correlated noise. This can actually  lead to a non-Markovian quantum evolution, which will require reconsidering some of the established characterization tools for the dynamics of Markovian open quantum systems.
Before delving into more specific details about the characterization of non-Markovian quantum evolutions, we  note that the degree of non-Markovianity~\cite{Rivas_2014,RevModPhys.88.021002,Li_2018}
can play a role in the effectiveness of QES~\cite{Addis_2015,Berk2023}
and QEM~\cite{PhysRevA.103.012611,donvil2023quantum} techniques.

Within the set of QCVV techniques~\cite{Eisert2020,PRXQuantum.2.010201,Gebhart2023}, quantum process tomography (QPT) aims at characterizing   
 the most generic type of process that can account for the evolution of a quantum system, solely constrained by the laws of quantum mechanics~\cite{doi:10.1080/09500349708231894,PhysRevLett.78.390,PhysRevA.63.020101,PhysRevA.63.054104,PhysRevA.68.012305}. For a specific evolution time, after preparing and measuring the system in an informationally complete setting, one can make an estimate of the completely-positive trace-preserving (CPTP) quantum channel~\cite{nielsen00,watrous_2018} that determines a snapshot of the quantum evolution. Therefore, QPT has been applied for the characterization of quantum gates in various experimental QIPs~\cite{Childs2001,10.1063/1.1785151,PhysRevLett.91.120402,PhysRevLett.93.080502,PhysRevLett.97.220407,doi:10.1126/science.1177077,PhysRevLett.102.040501,Bialczak2010,PhysRevLett.109.240505}, typically restricted to  small number of qubits. In addition to the inherent complexity of QPT as the number of qubits increases, this characterization must be repeated for each instant of time of interest, in order to obtain a coarse-grained reconstruction of the full, i.e.~a one-parameter family of CPTP channels~\cite{Breuer2002,RevModPhys.89.015001,chruscinski2022dynamical} that governs the time evolution of the quantum system. Although repeating QPT can allow to characterise non-Markovian  evolutions, the associated overhead  can limit the precision  in architectures where the number of measurements shots cannot be sufficiently large~\cite{velazquez2024dynamical}.  

 A strategy to overcome this limitation is to focus on the estimation of the generators of the noisy dynamics, rather than  on the various coarse-grained snapshots. For time-homogeneous quantum dynamical maps, which form a semigroup, the time evolution can be described by the exponential of a Lindblad super-operator~\cite{Lindblad1976,Gorini1976,Breuer2002}. Although one may  expect that the generators of this type of   Markovian noise can be obtained by simple algebraic manipulations of a single QPT snapshot at any arbitrary time, this approach can  lead to inconsistencies~\cite{wolf2008,PhysRevA.67.042322,
Howard_2006,onorati2021fitting} due to the branches of the complex logarithm. Therefore,   alternative QCVV techniques are  required. One possibility is use the Lindbladian generators for a parametrization of the quantum evolution, which can then be inferred using different  learning strategies~\cite{Childs2001,PhysRevA.67.042322,Howard_2006,Bairey_2020,PRXQuantum.3.030324,franca2022efficient,PhysRevApplied.18.064056,PhysRevApplied.17.054018,PhysRevA.101.062305,dobrynin2024compressedsensing}. Lindblad learning aims to estimate the Hamiltonian under which the system evolves and, additionally, the jump operators and dissipation rates  that govern the non-unitary part of the dynamics of the noisy QIP. However, it is important to note that Lindblad learning is based on the Lindblad master equation, valid only for Markovian system-environment interactions, i.e., memoryless interactions in which information flows from the system to the environment but never flows back.

However, noise in real QIPs does not  always fall in this category, and temporal correlations and even non-Markovianity  can play an important role, as alluded above in connection to QES, QEM and QEC. Hence, it would be desirable to extend  the Lindblad learning to encompass non-Markovian noise scenarios, such that the quantum dynamical maps are no longer a semigroup, nor can they be divided  into the composition of sequential CP channels at any intermediate time~\cite{RevModPhys.89.015001,chruscinski2022dynamical}. 
Several  characterization techniques for related problems have been developed in recent years. Quantum noise spectroscopy protocols \cite{alvarezMeasuringSpectrumColored2011, yugeMeasurementNoiseSpectrum2011} have evolved into characterization tools of non-Markovian systems, as in Ref.\ \cite{PRXQuantum.2.030315}, where the authors develop a filter-function formalism based on frames. A general non-Markovian quantum process framework was proposed by Ref.\ \cite{PhysRevA.97.012127} allowing to characterize non-Markovian quantum processes using process tensor tomography (PTT). This later led to proposals of quantum non-Markovian process tomography \cite{PRXQuantum.3.020344}. Reference \cite{white2023unifying} introduced into this PTT frameworks the possibility of memory effects within the system control itself and from undesirable interactions between the control and the environment. This  contrasts Ref.\ \cite{PhysRevA.97.012127} where perfect gates were assumed. Reference \cite{PhysRevApplied.21.024018} characterizes certain type of non-Markovian errors in quantum gates and \cite{PRXQuantum.5.010306} proposes also characterization of quantum gates but with a compressed approach that reduces the resources needed in comparison to, for instance, Ref.\ \cite{white2023unifying}.

In general, any  quantum evolution that results from the coupling of a quantum system to a larger environment, or to a set of noisy controls modeled by stochastic processes, can be expressed in terms of a time-local master equation by using a time-convolutionless formulation~\cite{Chaturvedi1979} of the Nakajima-Zwanzig integro-differential equation~\cite{10.1143/PTP.20.948,10.1063/1.1731409,Breuer2002}. These time-local master equations generalize the aforementioned Lindblad master equation~\cite{Lindblad1976,Gorini1976}, and can be expressed in a canonical form that connects directly with the degree of non-Markovianity~\cite{PhysRevA.89.042120}. In essence, the characterization of these time-local master equations would require a time-dependent parametrization of the Hamiltonian, jump operators and dissipation rates, which can then be incorporated into a maximum-likelihood estimation that parallels the Markovian Lindblad limit~\cite{PhysRevApplied.18.064056,PhysRevApplied.17.054018,PhysRevA.101.062305,dobrynin2024compressedsensing}. 
In this work, we call this QCVV technique  {\it Lindblad-like quantum tomography} (L$\ell$QT), and develop it in the simplest possible scenario: the  dephasing dynamics of a single qubit. We present a detailed comparative study of this QCVV technique, considering both a frequentist and a Bayesian approach for the statistical inference. We consider minimal dephasing models, both semi-classical and fully quantum-mechanical, in which the temporal correlations and degree of non-Markovianity can be independently controlled. By making a careful connection to the the theory of asymptotic inference and Bayesian estimation, we quantify both the accuracy and precision of L$\ell$QT. We discuss how the amount of temporal correlations and the degree of non-Markovianity can play a key role in deciding which of the two approaches is preferable when  learning the non-Lindblad qubit dephasing.

This article is organized as follows. In subsection ``Learning time-local master equations by maximum-likelihood estimation'' we review the techniques of Lindblad quantum tomography and present L$\ell$QT, a generalization of Lindblad learning that allows us to characterize non-Markovian noise. In subsections ``Frequentist approach to non-Markovian inference'' and ``Bayesian approach to non-Markovian inference'' two approaches to L$\ell$QT are presented. Subsection ``Lindblad-like quantum tomography for non-Markovian dephasing'' presents L$\ell$QT applied to dephasing noise. The frequentist and Bayesian approaches are later compared in a performance analysis in subsections ``Markovian semi-classical dephasing'' and ``Non-Markovian quantum dephasing'',  where we also study how measurement times should be selected to reduce the number of necessary measurements and the error in the estimation of noise parameters.

\section*{\bf Results}
\subsection*{Learning  time-local master equations by maximum-likelihood estimation }
\label{sec:LlQT}

The Lindblad master equation generalizes the Schrödinger equation to open and noisy quantum systems~\cite{Lindblad1976,Gorini1976,Breuer2002}, and describes the non-unitary time evolution of the  density matrix  of the system, defined as a positive-definite unit-trace linear operator $\rho\in\mathsf{D}(\mathcal{H}_{\rm S})\subset\mathsf{L}(\mathcal{H}_{\rm S})$ in a Hilbert space of dimension $d={\rm dim }{\mathcal{H}_{\rm S}}$~\cite{watrous_2018}. This master equation can be written in terms of an infinitesimal generator  ${\rm d}{\rho}/{{\rm d}t} = \mathcal{L}_{H,G}(\rho)$, namely 
\begin{equation}\label{eq:linblad_master}
   \mathcal{L}_{H,G}(\rho) = - \mathrm{i} \left[ H, \rho \right] + \!\!\sum_{\alpha,\beta=1}^{d^2-1} G_{\alpha\beta} \!\!\left( E^{\phantom{\dagger}}_\alpha \rho E_\beta^\dagger - \half \big\{ E_\beta^\dagger E^{\phantom{\dagger}}_\alpha, \rho \big\} \!\right)\!,
\end{equation}
where the Hamiltonian $H\in\mathsf{Herm}(\mathcal{H}_{\rm S})$ is a Hermitian operator, and we have introduced   the so-called dissipation Lindblad matrix, a positive semidefinite matrix $G\in\mathsf{Pos}(\mathbb{C}^{d^2-1})$. Here,  $\mathcal{B}=\{E_0=\mathbb{1}_d,E_\alpha: \alpha\in\{1,\cdots, d^2-1\}\}$ forms an operator basis $\mathsf{L}(\mathcal{H}_{\rm S})={\rm span}\{\mathcal{B}\}$ and, together with the Lindblad matrix, determines the dissipative non-unitary dynamics of the system. Diagonalizing the Lindblad matrix, we obtain
\begin{equation}\label{eq:linblad_master_diag}
 \mathcal{L}_{H,G}(\rho) = - \mathrm{i} \left[ H, \rho \right] + \sum_{n=1}^{d^2-1} \gamma_{n}\! \left( L_n \rho L_n^\dagger - \half \left\{ L_n^\dagger L_n, \rho \right\} \right),
\end{equation}
 where $\{L_n:\, n\in\{1,\cdots,d^2-1\}\}\in\mathsf{L}(\mathcal{H}_{\rm S})$ are the jump operators responsible of generating the different noise processes with dissipative decay rates $\gamma_n\in\mathbb{R}^+$\!. 
The goal of Lindblad learning  is to estimate $\mathcal{L}_{H,G}$ or, equivalently the system Hamiltonian and the dissipation rates and jump operators $\{H,\gamma_n,L_n\}$, using a finite number of measurements~\cite{PhysRevA.58.1723,Childs2001,PhysRevA.67.042322,Howard_2006,Bairey_2020,PhysRevApplied.17.054018,PhysRevApplied.18.064056,PhysRevA.101.062305,PRXQuantum.3.030324,franca2022efficient}. In particular, our work starts from a maximum-likelihood  approach~\cite{alma9924122381802466} to Lindbladian  quantum tomography (LQT)~\cite{PhysRevApplied.18.064056,PhysRevA.101.062305, dobrynin2024compressedsensing}.

In the general case, LQT involves preparing an informationally complete set of initial states $s\in\mathbb{S}_0$, allowing the system to evolve over a set of times $i\in\mathbb{I}_t$, and performing measurements in different basis $b\in\mathbb{M}_b$, with corresponding outcomes $m_b\in\mathbb{M}_{m_b}$ (or more generally using a POVM). These independent configurations, consisting of initial states, evolution times, and measurement outcomes, provide the necessary data to estimate the Hamiltonian, dissipation rates, and jump operators, ensuring a complete reconstruction of the Lindbladian dynamics. LQT makes use of a total of $N_{\rm shot}$ measurement shots, also known as trials in the context of statistics, which will be distributed among the different initial states, instants of time and measurement basis $N_{\rm shot}=\sum_{s,i,b}N_{s,i,b}$. Therefore,  the total number of measurement shots is a function of number of initial states, measurement basis, measurement outcomes and measurement times, expressed as $N_{\rm shot}(|\mathbb{S}_0|, |\mathbb{M}_b|, |\mathbb{M}_{m_b}|, |\mathbb{I}_t|)$, where $|\mathbb{A}|$ denotes the cardinality of the set $\mathbb{A}$, i.e., the number of elements in the set. In the experiment, one would count the number  of times $N_{s,i,b,m_b}$  that the $m_b$ outcome is obtained for each of the configurations, such that  $N_{s,i,b}=\sum_{m_b}N_{s,i,b,m_b}$. This provides a data set $\mathbb{D}=\{N_{s,i,b,m_b}\}$ that can be understood as a random sample of the corresponding random variable $\tilde{N}_{s,i,b,m_b}$ obtained from $N_{\rm shot}$ experimental measurements. We use tildes to refer to stochastic variables. In this case, the probability of obtaining the outcome $m_b$ when measuring in basis $b$ is given by $p_{s,i,b}(m_b) = \mathrm{Tr}\left\{ M_\mu \mathcal{E}_{t_i,t_0} (\rho_{0,s})\right\}$, where $M_\mu$ represents the corresponding projector of the combination $(b,m_b)$. $\mathcal{E}_{t,t_0}\in\mathsf{C}(\mathcal{H}_{\rm S})$ is a one-parameter family of completely-positive trace-preserving (CPTP) channels~\cite{watrous_2018} describing the actual time evolution of the noisy quantum system. Each measurement configuration gives rise to an independent binomial distribution with $N_{s,i,b}$ trials.

The data set $\mathbb{D}$  can  be used to estimate the Hamiltonian $H$ and Lindblad matrix $G$  by maximizing the  likelihood function, which is defined as the probability distribution of the combination of all the independent binomials $p_{s,i,b}(m_b)$. These binomials can be  approximated by the Lindbladian of Eq.~\eqref{eq:linblad_master}, namely $p^{\phantom{L}}_{s,i,b}(m_b)\mapsto p^{\rm L}_{s,i,b}(m_b)=\mathrm{Tr}\{ M_\mu \ee^{(t_i-t_0)\mathcal{L}_{ H,  G} } (\rho_{0,s})  \}$.
The larger this likelihood function is for a given pair $H,G$, the better the Lindbladian description  approximates  the observed data~\cite{alma9924122381802466}. Taking the negative logarithm of the likelihood function we obtain a cost function
\begin{equation}
\label{eq:LQT}
   \mathsf{C}_{\rm L}(H,G) = -\sum_{s,i,b.m_b} N_{s,i,b,m_b} \log p^{\rm L}_{s,i,b}(m_b),
\end{equation}
with the minimum located in the same place as the maximum of the original  likelihood. The optimization problem is thus converted into a non-linear minimization of this  Lindbladian  cost  function. By minimizing this non-convex estimator or, instead,  a convex approximation  based on linearization and/or compressed sensing~\cite{dobrynin2024compressedsensing}, LQT provides an estimate of the generators $\hat{H},\hat{G}$ that yield the best match with the observed data, where we will use hats to refer to estimated quantities. We note that  this minimization is subject to constraints on Hamiltonian hermicity  and Lindblad matrix semidefinite positiveness.

In this work, we describe our first steps in the  development of a learning procedure for non-Markovian quantum dynamical maps that supersedes the above LQT. In particular, we consider the statistical inference of the generators of quantum dynamical maps that need not fulfill CP-divisibility,  the property of a quantum dynamical map where the evolution between any two intermediate times can be described by a completely positive map \cite{PhysRevLett.105.050403,Rivas_2014}. These maps $\rho(t)=\mathcal{E}^{\rm TL}_{t,t_0}(\rho_0)$ no longer have the Lindbladian generator of Eq.~\eqref{eq:linblad_master}, but are instead governed by a {\it time-local master equation} that,  when expressed in a canonical  form~\cite{PhysRevA.89.042120}, reads $\dot{\rho}=-\ii[H(t),\rho]+\mathcal{D}_{\rm TL}(\rho)$ with  
\beq
\label{eq:TLC_meq}
\mathcal{D}_{\rm TL}(\rho)= \sum_{n}{\gamma}_n(t)\big(L_n(t)\rho L^{\dagger}_n(t)-\half\{L^{\dagger}_n(t)L_n(t),\rho\}\! \big)\!.
\eeq
Here, the Hamiltonian $H(t)$,  as well as the  dissipative rates and jump operators ${\gamma}_n(t), L_n(t)$, can be time dependent. It is important to note that the `rates' are no longer required to be positive semidefinite. The possibility of encountering negative rates is directly linked with the non-Markovianity of the quantum evolution~\cite{Rivas_2014,RevModPhys.88.021002,RevModPhys.89.015001}. This time-local master equation can always be written in the form of Eq.~\eqref{eq:linblad_master} by letting $H\mapsto H(t)$ and $G\mapsto G(t)$,
such that the corresponding  quantum dynamical map will depend on the history of the time-dependent generators $\mathcal{E}^{\rm TL}_{t,t_0}=\mathcal{E}^{\rm TL}_{t,t_0}(\{H(t'),G(t')\})$ for all $ t'\in[t,t_0]$. Here, the $t'$ argument reflects the fact that now the dynamical map between initial time $t_0$ and time $t$ depends on all intermediate times. 
We  thus formulate a {\it Lindblad-like quantum} {\it  tomography}~{(L$\ell$QT)}  by upgrading the LQT cost function in Eq.~\eqref{eq:LQT} to a  time-local one that can encompass non-Markovian effects
\beq
\label{eq:NM_cost}
    \mathsf{C}_{\rm TL}\big(\{H(t'),G(t')\}\big) = -\sum_{s,i,b,m_b} N_{s,i,b,m_b} {\rm log}p^{\rm TL}_{s,i,b}(m_b),
\eeq
where the theoretical probabilities are calculated following
\beq\label{eq:p_tl_sib}
 p^{\rm TL}_{s,i,b}(m_b)=\mathrm{Tr}\left\{ M_{b,m_b} \mathcal{E}^{\rm TL}_{t_i,t_0} (\rho_{0,s})\right\}.
\eeq
The above cost function must be  minimized subject to   dynamical constraints 
\beq
\label{eq:LQT_ML_prob}
\big(\hat{{H}}(t),\hat{G}(t)\big)= \texttt{\,argmin}\big\{\mathsf{C}_{\rm TL}\big(\{H(t'),G(t')\}\big)\big\}_{\phantom{l_j}},
\eeq
subject to $ {H}(t')\in\mathsf{Herm}(\mathcal{H}_{\rm S}), G(t')\in\mathsf{Herm}(\mathbb{C}^{d^2-1})$. Therefore, we see that in addition to the time dependences, the dissipation matrix  is no longer required to be semi-positive definite, but only Hermitian, and can thus support negative decay rates and  incorporate non-Markovian effects.

The crucial property that differentiates LQT from other learning approaches such as  quantum process tomography~\cite{doi:10.1080/09500349708231894,PhysRevLett.78.390,PhysRevA.63.020101,PhysRevA.63.054104,PhysRevA.68.012305} is that the estimator includes $|\mathbb{I}_t|$ different instants of time $\{t_i, i\in\mathbb{I}_t\}\subset T$, instead of focusing on a single snapshot of the quantum dynamical map. Although we have shown in Ref.~\cite{dobrynin2024compressedsensing} that, in certain regimes, an accurate LQT  can be obtained by focusing on a single snapshot  for  Lindbladian evolution,  this will not be the case for time-correlated and non-Markovian quantum evolutions. In this case,  it will be crucial to include the information of various snapshots into the cost function. In fact, we address in this work how many snapshots would be required, and which particular instants of time   would be optimal in order to learn about the memory effects of a time-correlated or a  non-Markovian noisy quantum evolution. We note that a black-box approach to L$\ell$QT is a very complicated problem, as the parameters of the Hamiltonian and dissipation matrix can have any arbitrary time dependence. In order to progress further, we instead look into physically-motivated models for L$\ell$QT, allowing us to restrict the search space, and start by focusing on a simple and, yet, very relevant setting: a single qubit subject to time-correlated dephasing noise, which can result in non-Markovian quantum dynamics. Our techniques and conclusions may be useful when generalising to more complicated non-Markovian dynamics, aiming at the characterization of non-Markovian noise in  gate sets of QIPs to optimise a tomographic analysis~\cite{velazquez2024dynamical}. 

As stated in Eq.~\eqref{eq:LQT_ML_prob}  we can follow a frequentist approach similarly to the one in Markovian Lindblad quantum tomography~\cite{PhysRevApplied.18.064056,PhysRevA.101.062305, dobrynin2024compressedsensing}, but now  taking into account  the time-dependence of the decay rates, which will require an a priori selection of the evolution times at which the system is probed. In light of the L$\ell$QT estimation problem of Eq.~\eqref{eq:LQT_ML_prob}, we do not need to consider arbitrary time-dependent functions for  the Hamiltonian and dissipation matrix, but can actually find an effective parametrization that reduces drastically the search space. This means that the minimization in Eq.~\eqref{eq:LQT_ML_prob} can be over the parameters $\boldsymbol{\theta}$ of the Hamiltonian and dissipation matrix. 
Although several intermediate times can be taken, a single time instant actually suffices for an accurate  learning in  LQT~\cite{dobrynin2024compressedsensing}, which makes it simpler and  computationally lighter. As the noise becomes time correlated,  reconstructing the  more complex time-dependent dynamics via L$\ell$QT,  demands multiple time steps to achieve high accuracies. LQT has in principle the same number of parameters $d^2(d^2-1)$ to be learnt as quantum process tomography, and to do so requires at least $d^2(d^2-1)$ measurement configurations. For the general case, where arbitrary time-dependent rates need to be estimated, a naive blind-search approach would require repeated LQT at many time points, leading to high sample complexity. In contrast, L$\ell$QT’s use of a microscopic parameterization reduces the need for extensive measurements by imposing physical constraints, which partially addresses the sample complexity challenge and makes it a more scalable method. Thus in parameterized L$\ell$QT we will need at least as many measurement configurations as the number of parameters to estimate.

Alternatively to this frequentist approach, we can also consider a Bayesian approach that exploits a physically-motivated prior knowledge about these noise parameters, which can be represented by a certain probability distribution. After performing   measurements on the system at certain  instants of time and certain measurement basis, we update this probability distribution with the new acquired information, a  process that is repeated until reaching a target  accuracy for the parameter estimation.  The Bayesian approach has the  advantage of choosing, at each step,  the most convenient subsequent time and measurement basis at which to measure by maximizing  the information one  would gain. In principle, this can lead to a reduction  in the number of measurements  required to reach a certain accuracy with respect to  those required by the frequentist approach. In practice, however, the   non-Markovianity of the quantum evolution can modify this argument, as our probabilistic account of the model parameters can be affected by the actual time correlations of the underlying random process.

\subsection*{\!Frequentist approach to non-Markovian inference}\label{sec:frequentist}
In this section, we focus on the  frequentist approach, which builds on the relative frequencies of observed outcomes. The L$\ell$QT cost function of Eq.~\eqref{eq:NM_cost} can be rewritten  as follows
\begin{align}
\label{eq:ML_cost_parameter}
    \mathsf{C}_{\mathrm{TL}}(\boldsymbol{\theta}) = -\sum_{s,i,b}  N_{s,i,b}\,\sum_{m_b}\, \tilde{f}_{s,i,b}(m_b\,|\boldsymbol{\theta}_\star) \log p^{\rm TL}_{s,i,b}(m_b | \boldsymbol{\theta}), 
\end{align}
where $\tilde{f}_{s,i,b}(m_b\,|\boldsymbol{\theta}_\star)=N_{s,i,b,m_b}/N_{s,i,b}$ is the ratio of the number of $m_b$ measurement outcomes observed $N_{s,i,b,m_b}$ to the total of $N_{s,i,b}$  shots  collected at the instant of time $i$ and initial state $s$ when measuring the system with the POVM element $b$.  Our notation remarks that these relative frequencies carry information about the real noise parameters $\boldsymbol{\theta}_{\!\star}$ we aim at estimating. In addition, the estimator depends on  $p_{s,i,b}(m_b | \boldsymbol{\theta})$ shown in Eq.~\eqref{eq:p_tl_sib}, where  we make  explicit the dependence on the parametrized  noise. 
 
The  minimization in Eq.~\eqref{eq:LQT_ML_prob} can explicitely written as a minimization over the noise parameters. As a consequence,  the frequentist approach can be recast as a statistical problem of parameter point estimation~\cite{alma9924122381802466}, namely
 \beq
 \label{eq:min_freq}
\boldsymbol{\hat{\theta}}_{\rm F}=\texttt{argmin}_{\boldsymbol{\theta}}\{ \mathsf{C}_{\mathrm{TL}}(\boldsymbol{\theta}): \, \boldsymbol{\theta}\in\Theta=\mathbb{R}^{n
}\}.
 \eeq
Instead of the general L$\ell$QT learning  over $d^2(d^2-1)$  parameters, which   increase exponentially $d=2^{n_q}$ with the number of qubits $n_q$ and can arbitrarily change  in time, our  procedure   revolves around the estimation of $n$ noise  parameters, which are independent of the system size and the evolution times. On the other hand, the imprecision of our estimates will indeed depend on our choice of the evolution times, forcing us to go beyond the LQT single-time estimator~\cite{dobrynin2024compressedsensing}, and actually measure at optimal times, initial states and measurement elements for which the estimation imprecision can be minimized.    
Let us  note that the conditions to minimise this  cost function are the same as those that minimize the Kullback-Leibler divergence~\cite{Kullback1951, Mogilevtsev2012}, which is the following   relative entropy 
 \beq 
 \label{eq:KL_div}
 \mathsf{D}_{\rm KL}(\boldsymbol{p}_1||\,\boldsymbol{p}_2)=-\sum_k p_{1,k}\log\left(\frac{p_{2,k}}{p_{1,k}}\right),
 \eeq
  between the experimental $p_{1,k}\in\{ \tilde{f}_{s,i,b}(m_b\,|\boldsymbol{\theta}_\star)\}$ and parameterized theoretical $p_{2,k}\in\{ p^{\rm TL}_{s,i,b}(m_b\,|\boldsymbol{\theta})\}$ probability distributions, provided   one considers variations with respect to the estimation parameters $\boldsymbol{\theta}$. 

The above estimator depends on the data set $\mathbb{D}$, and  is thus also a stochastic variable, which will be  characterized by its mean $\mathbb{E}[\boldsymbol{\hat{\theta}}_{\rm F}]$ and its  moments, such as the covariance matrix
\beq 
\label{eq:cov}
[\mathrm{Cov}(\boldsymbol{\hat{\theta}}_{\rm F})]_{n_1,n_2}=\mathbb{E}\big[\big(\hat{\theta}_{n_1}-\mathbb{E}[\hat{\theta}_{n_1}]\big)\big(\hat{\theta}_{n_2}-\mathbb{E}[\hat{\theta}_{n_2}]\big)\big].
\eeq
We note that the expectation values are taken with respect to the probability distributions for the measurements of $p_{s,i,b}(m_b)$ which,  implicitly, also have the stochastic average over the random dephasing noise in the semi-classical  model, or a partial trace over the environment in the quantum-mechanical one.
The nice property of the maximum-likelihood estimator is that it is asymptotically unbiased, such that $\boldsymbol{\mathsf{B}}_{\boldsymbol{\theta}_\star}\!\!(\boldsymbol{\theta}_{\rm F})=\mathbb{E}[\boldsymbol{\hat{\theta}}_{\rm F}]-\boldsymbol{\theta}_{\star}\to\boldsymbol{0}$ for a sufficiently large $N_{\rm shot}$. Moreover, its asymptotic covariance matrix  saturates the Cram\'er-Rao bound \cite{doi:10.1142/6096} relating the estimation precision   
to the Fisher information matrix, which quantifies the amount of  information in $\mathbb{D}$ about the unknown parameters. If we  momentarily assume that the measurements occur at a single instant of time $t=t_i$, a single initial state and a single measurement basis with outcomes that are independent and identically distributed, the covariance matrix becomes $\mathrm{Cov}(\boldsymbol{\hat{\theta}}_{\rm F}) \approx (N_{s,i,b}{I}_{F,s,i,b}(\boldsymbol{\theta}_{\!\star}))^{-1}$, where
\beq
\label{eq:fisher}
[{I}_{F,s,i,b}(\boldsymbol{\theta}_{\!\star})]_{n_1,n_2} \!= \mathbb{E}\!\!\left[\left. \!\frac{\partial{\rm log} \big(p_{s,i,b}^{\rm TL}(m_b|\boldsymbol{\theta})\big)}{\partial{\theta_{n_1}}} \!\frac{\partial{\rm log} \big(p_{s,i,b}^{\rm TL}(m_b|\boldsymbol{\theta})\big)}{\!\partial\theta_{n_2}}\!\right|_{\boldsymbol{\theta}_{\!\star}}\right]\!.
\eeq 
In this work, we  deal with the more general case in which we measure at several times, therefore the random variables are not identically distributed, and the total number of shots need not be the same for different measurement configurations, i.e., instants of time, initial states and measurement basis. In this case, 
we must take a linear combination of the Fisher information matrices of each measurement configuration weighted by the proportion of measurements taken at each one of these configurations ~\cite{10.1214/aoms/1177693066}. We thus obtain the asymptotic covariance matrix
\begin{equation}\label{eq:asymptotic_covariance}
    \mathrm{Cov}(\boldsymbol{\hat{\theta}}_{\rm F}) \approx \Sigma_{\boldsymbol{\hat{\theta}}} \equiv \left[N_{\rm shot} \sum_{s,i,b} \frac{N_{s,i,b}}{N_{\rm shot}} {I}_{F,s,i,b}(\boldsymbol{\theta}_{\!\star}) \right]^{-1},
\end{equation}
such that, the more the Ramsey estimator varies under changes of the noise parameters, the bigger the amplification of the noise parameter is and, thus, the smaller the imprecision one can achieve. As we can see, the imprecision of the estimate will scale with $1/\sqrt{N_{\rm shot}}$, such that the Ramsey estimator is asymptotically consistent in the $N_{\rm shot}\rightarrow \infty$ asymptotic limit~\cite{alma9924122381802466}. The asymptotic statistics of the maximum-likelihood estimator is further explained in the Methods section.

We also note that, in this limit, the observed relative frequencies will be normally distributed, such that one can consider minimizing a weighted least-squares cost function. For instance, assuming a binary case where $m_b \in \{0,1\}$, corresponding to a qubit measurement, we have
\begin{equation}\label{eq:least_squares_fit}
    \mathsf{C}_{\mathrm{TL}}(\boldsymbol{\theta}) \approx \sum_{s,i,b} \frac{\big(\tilde{f}_{s,i,b}(0\,|\boldsymbol{\theta}_\star) - p_{s,i,b}^{\rm TL}(0 | \boldsymbol{\theta})\big)^2}{\tilde{\sigma}^2_{f_{s,i,b}}},
\end{equation}
where $\tilde{f}_{s,i,b}(0\,|\boldsymbol{\theta}_\star)=N_{s,i,b,0}/N_{s,i,b}$ is the ratio of the number of outcomes observed $N_{s,i,b,0}$ to the total of  shots $N_{s,i,b}$  collected at initial state $s$,  the instant of time $t_i$ and measurement element $b$. Here, $\tilde{\sigma}^2_{f_{s,i,b}}$ is the variance of these measured samples. These $\tilde{f}_{s,i,b}(0\,|\boldsymbol{\theta}_\star)$ follow a binomial proportion distribution with mean given by $p_{s,i,b}^{\rm TL}(0 | \boldsymbol{\theta_{\star}})$ and, in the limit of a large $N_{s,i,b}$,  they can be approximated by a normal distribution with mean $p_{s,i,b}^{\rm TL}(0 | \boldsymbol{\theta_{\star}})$ and variance $\sigma^2_{f_{s,i,b}}$ according to the central limit theorem. The expected variance of the measurement configuration $(s,i,b)$ is
\beq
\label{eq:bin_variance}
\sigma^2_{f_{s,i,b}}=\frac{p_{s,i,b}^{\rm TL}(0\,|\boldsymbol{\theta})\big(1-p_{s,i,b}^{\rm TL}(0\,|\boldsymbol{\theta})\big)}{N_{s,i,b}},
\eeq
since we are sampling from a binomial distribution. This approximation allows us to use a simpler weighted least-squares algorithm, such as the trust-region reflective algorithm implemented in SciPy, where we can optionally set some bounds for the parameters to be estimated. The computational complexity of the trust-region reflective algorithm is dominated in each iteration by solving the trust-region subproblem, which is typically $O(n^3)$, with $n$ the number of parameters. The number of iterations depends on the complexity of the cost function, the accuracy of the initial guess, and the desired level of convergence. However, once 
the algorithm gets close to the solution, the remaining convergence is rapid, showing a  quadratic rate, meaning the distance to the solution decreases at a rate proportional to the square of the distance at the previous step.

Once the properties of the Ramsey  frequentist estimate $\boldsymbol{\hat{\theta}}_{\rm F}$ have been discussed, we can search for the optimal measurement times, initial states and measurement elements that would lead to a Ramsey estimator with the lowest possible imprecision for a given finite $N_{\rm shot}$. Depending on which parameter we are interested in, we may be interested in minimizing a particular component of the covariance matrix $[\Sigma_{\boldsymbol{\hat{\theta}}}]_{n_1,n_1}$ in Eq.~\eqref{eq:asymptotic_covariance} or, alternatively, minimize its determinant as a whole. In the asymptotic limit in which $\boldsymbol{\hat{\theta}}_{\rm F}$ follows a multivariate normal distribution, $\det \Sigma_{\boldsymbol{\hat{\theta}}}$ is proportional to the volume enclosed by the covariance elliptical region, so it is a good measure of the dispersion of the distribution, and a good way to quantify the accuracy of the estimation. Sometimes, we will also use $\det \Sigma_{\boldsymbol{\hat{\theta}}}^{1/2n}$ instead, which can be more easily compared to the individual standard deviations $[\Sigma_{\boldsymbol{\hat{\theta}}}]_{n_1,n_1}^{1/2}$, as both quantities scale with $1/\sqrt{N_{\rm shot}}$. Minimizing $\det \Sigma_{\boldsymbol{\hat{\theta}}}$ has also the advantage that the optimal measurement configurations obtained are independent of the parameters we want to determine, assuming the different parametrizations have the same number of parameters and that a coordinate transformation exists between parametrizations. In this case, the determinants of the covariance matrices are related by $\det \Sigma_{\boldsymbol{\hat{\theta}}} = \det \Sigma_{\boldsymbol{\hat{\theta}}'} \det J^2$, with $J$ the Jacobian of the coordinate transformation between both parametrizations, which does not depend on the measurement configurations and therefore will have no influence in the minimization. Before presenting these results, we discuss an  approach  based on Bayesian inference~\cite{alma9924122381802466}.

\subsection*{\!Bayesian approach to non-Markovian inference}\label{sec:bayesian}

Rather than considering the relative frequencies as approximations of the underlying probability distribution with a certain fixed value of $\boldsymbol{\theta}_\star$, the idea of Bayesian inference is to quantify statistically our knowledge about the noise parameters, and how this knowledge gets updated as we collect more information via measurements.  Hence, the noise parameters become continuous stochastic variables themselves $\boldsymbol{\theta}_\star\mapsto \boldsymbol{\tilde{\theta}}$ that take values $\boldsymbol{\theta}\in\Theta$ according to a prior probability density function (PDF)  $\pi_0(\boldsymbol{\theta})$. This probability distribution  quantifies our  uncertainty about the noise parameters before making any measurement $\mathbb{D}_0=\emptyset$. At each  $\ell>0$ Bayesian  step, we measure the system  enlarging the data set sequentially $\mathbb{D}_{\ell-1}\mapsto\mathbb{D}_\ell=\mathbb{D}_{\ell-1}\cup\delta\mathbb{D}_{\ell}$, where $\delta\mathbb{D}_{\ell}\subset\mathbb{D}=\{N_{i,s,b,m_{b}}\}$ contains a number of measurement outcomes  $|\delta N_{\ell}|$ that is a fraction of the total $N_{\rm shot}$. These outcomes will be labeled as $\delta \mathbb{D}_\ell=\{N_{i_\ell,s_\ell,b_\ell,m_{b_\ell}}\}$. The measurements in this data set are again binary Bernoulli trials, and can be described by a joint distribution of independent binomials $p_{\tilde{N}}(\delta \mathbb{D}_\ell)$, defined in analogy to the likelihood function, but only extended to the configurations measured in the particular Bayesian step. The prior $(\ell-1)$-th probability distribution is then updated by using  Bayes' rule based on the parametrized probability distributions $p^{\rm TL}_\ell(\delta{\mathbb{D}}_\ell|\boldsymbol{\theta})$ being understood as  probabilities conditioned on our statistical knowledge of the noise parameters 
\begin{equation}
\label{eq:posterior}
    \pi_{\ell}(\boldsymbol{\theta}|\mathbb{D}_\ell) = \frac{1}{p^{\rm TL}_{\ell-1}(\delta\mathbb{D}_\ell)} p^{\rm TL}_\ell(\delta\mathbb{D}_\ell| \boldsymbol{\theta}) \pi_{\ell-1}(\boldsymbol{\theta}).
\end{equation}
Here, $p^{\rm TL}_{\ell-1}(\delta\mathbb{D}_\ell) = \int_{\Theta} \dd^n {\theta} p^{\rm TL}_\ell(\delta\mathbb{D}_\ell| \boldsymbol{\theta}) \pi_{\ell-1}(\boldsymbol{\theta})$ is a normalization constant required to interpret  $\pi_{\ell}(\boldsymbol{\theta}|\mathbb{D}_\ell)\mapsto \pi_{\ell}(\boldsymbol{\theta})$ as a probability distribution  describing our updated knowledge about the noise, which will be used as the subsequent prior. 

One of the main differences with respect to the frequentist approach is that we have, at each step, a probability distribution from which one can obtain a Bayesian estimate 
\beq
\label{eq:exp_value}
\boldsymbol{\hat{\theta}}_{{\rm B}}=\mathbb{E}_{\boldsymbol{\theta}}[\boldsymbol{\theta}]=\int_{\Theta}\!{\rm d}^n\theta\,\pi_\ell(\boldsymbol{\theta})\boldsymbol{\theta}.
\eeq
We note that this Bayesian estimator $\hat{\boldsymbol{\theta}}_{\rm B}$ minimizes the Bayesian risk associated to a squared error loss function over all possible estimators $\hat{\tau}$, $\boldsymbol{\hat{\theta}}_{{\rm B}}=\texttt{argmin}\{\mathsf{R}_{\rm TL}(\hat{\tau})\}$~\cite{alma9924122381802466} with
\beq
\mathsf{R}_{\rm TL}(\hat{\tau})=\!\sum_{i_\ell, s_\ell, b_\ell,m_{b_\ell}\in\delta\mathbb{D}_\ell} \int_{\Theta}\!{\rm d}^n\theta\,\pi_\ell(\boldsymbol{\theta})p^{\rm TL}_\ell\big(\delta\mathbb{D}_\ell| \boldsymbol{\theta}\big)(\hat{\tau}-\boldsymbol{\theta})^2.
\eeq
In addition to the expectation value in Eq.~\eqref{eq:exp_value}, since we have the updated probability distributions,  we can quantify how our uncertainty about the noise parameters changes via the associated covariance matrix $[\mathrm{Cov}(\boldsymbol{\hat{\theta}}_{{\rm B}})]_{n_1,n_2}$ or any other statistics, regardless of the size of  the Bayesian data set $\mathbb{D}_\ell$. This differs from  our previous arguments for covariance of the maximum likelihood estimate $\boldsymbol{\hat{\theta}}_{\rm F}$ in Eq.~\eqref{eq:asymptotic_covariance}, which require working in the asymptotic regime $N_{\rm shot}\rightarrow \infty$. In experimental situations in which this regime cannot be reached, we note that one could use  Monte Carlo sampling techniques to estimate precision of $\boldsymbol{\hat{\theta}}_{\rm F}$~\cite{blumekohout2012robust, PhysRevA.99.052311,velazquez2024dynamical}, although these  deal with the estimator based on the full likelihood function.

Another crucial difference of the Bayesian approach  is that, instead of choosing a predefined set of evolution times, initial states and measurement basis,   we can find  the optimal time at which we should measure  to maximize the information gain at each Bayesian step. For each update, we thus solve for 
\begin{equation}\label{eq:optimal_time_information_gain}
    (i_{\ell}, s_\ell, b_\ell) = \texttt{argmax}_{i, s, b}\big\{ \mathbb{E}\left[\mathsf{D}_{\mathrm{KL}}\big(\pi_{\ell}(\boldsymbol{\theta}| \mathbb{D}_{\ell})\,\big|\!\big|\, \pi_\ell(\boldsymbol{\theta})\big)\right]\big\},
\end{equation}
where $\pi_{\ell}(\boldsymbol{\theta}| \mathbb{D}_{\ell})$ is the posterior probability in Eq.~\eqref{eq:posterior}  corresponding to the   measurement results   one would obtain by measuring at a time $t_i$, initial state $s$ and measurement basis $b$, and enlarging the data set as $\mathbb{D}_{\ell'}=\mathbb{D}_{\ell}\cup\delta \mathbb{D}_{\ell'}$. In the expression above, we  are making use of the Kullback-Leibler divergence of Eq.~\eqref{eq:KL_div} between the posterior and the prior, searching for a measurement configuration that maximizes the relative entropy between the prior and any of the possible posteriors, such that one  gains the maximum amount of information at each Bayesian step. Therefore, not only the data set is  enlarged sequentially $\mathbb{D}_\ell\supset\mathbb{D}_{\ell-1}$, but also the specific times are chosen adaptively. In light of the fixed set of measurement times in the frequentist approach $\{t_i:\, i\in\mathbb{I}_t\}$, we note that the  total set of updated times after $N_{\rm B}$ Bayesian steps $\{t_\ell, \ell\in\{0,\cdots,N_{\rm B}\}\}$ can be very different, and that is the reason why we use a different notation. 
Computing Eq.~\eqref{eq:optimal_time_information_gain} can be quite time-consuming, especially when dealing with a large number of parameters. In practice, long computation times may lead to a reduction in the frequency of experimental shots, which is undesirable. To avoid this, we can take tens or hundreds of shots at each step before computing again the optimal measurement configuration. This will not change the results significantly, since a Bayesian update of a single shot does not change the prior much and the optimal measurement time of next step remains very similar to the previous one.

We note that 
the Bayesian approach is ultimately related to the maximum-likelihood estimation in the asymptotic limit $N_{\rm shot}\rightarrow \infty$. When the variance of the prior is small, the maxima of the Kullback-Leibler divergence in Eq.~\eqref{eq:KL_div} between   prior and posterior are localized at the optimal measurement configurations obtained by minimizing the asymptotic covariance matrix of the maximum likelihood-estimation in Eq.~\eqref{eq:asymptotic_covariance}. Additionally, if we make a single Bayesian update in which we take a very big number of shots at several measurement configurations, the likelihood function relating the posterior and the prior will be a joint distribution of independent binomials. Given the big number of shots taken in this single step, the posterior distribution will be mainly shaped by the likelihood function, which contains most of the information about the parameters. The position of the maximum of this PDF will be located at the most likely value of the parameters, and therefore it is in agreement with the maximum-likelihood estimator that takes measurements at those measurement configurations. The Bayesian approach has the advantage that, at each step, we measure at the most convenient configuration that maximizes the expected information gain, and this can lead to an overall reduction of the number of measurements needed. Also, as noted above, the final estimate 
relies on a probability distribution, so that we can immediately derive confidence intervals without requiring any asymptotic limit. 

For the Bayesian approach, where we use sequential Monte Carlo \cite{doucet_book} to implement it numerically, the computational complexity is dominated by the computation of the best measurement configuration. At each step in the Bayesian approach, we need to compute the expected posterior of the measurement configurations, and then select the measurement configuration maximizing the Kullback-Leibler divergence between posterior and prior. The complexity of this calculation depends on the number of particles in the particle filter, $N_{\rm filter}$, the number of initial states, the number of measurement bases, the number of measurement outcomes and the number of measurement times considered. Therefore, we have a complexity $O(N_{\rm filter} \times |\mathbb{S}_0| \times|\mathbb{M}_b| \times |\mathbb{M}_{m_b}| \times |\mathbb{I}_t|)$. At each one of these $N_{\rm filter} \times |\mathbb{S}_0| \times |\mathbb{M}_b| \times |\mathbb{M}_{m_b}| \times |\mathbb{I}_t|$ operations we need to evaluate the likelihood function $p^{\rm TL}_\ell(\delta\mathbb{D}_\ell| \boldsymbol{\theta})$ to compute the posterior weight. Therefore if the posterior function is costly to compute we could think of precomputing it on a grid and then make linear interpolation to reduce the computational cost, as the authors of Ref.\ \cite{PRXQuantum.3.020350} do. $N_{\rm filter}$ may need to increase exponentially with the number of parameters, which makes the frequentist approach more suitable in the case of large number of parameters. However, this can be moderated by the specific structure of the model and how concentrated the posterior is.

\subsection*{Lindblad-like quantum tomography for non-Markovian dephasing } \label{sec:llqt_non_markovian}

\begin{figure*}[!t]
  \centering
  \includegraphics[width=1\linewidth]{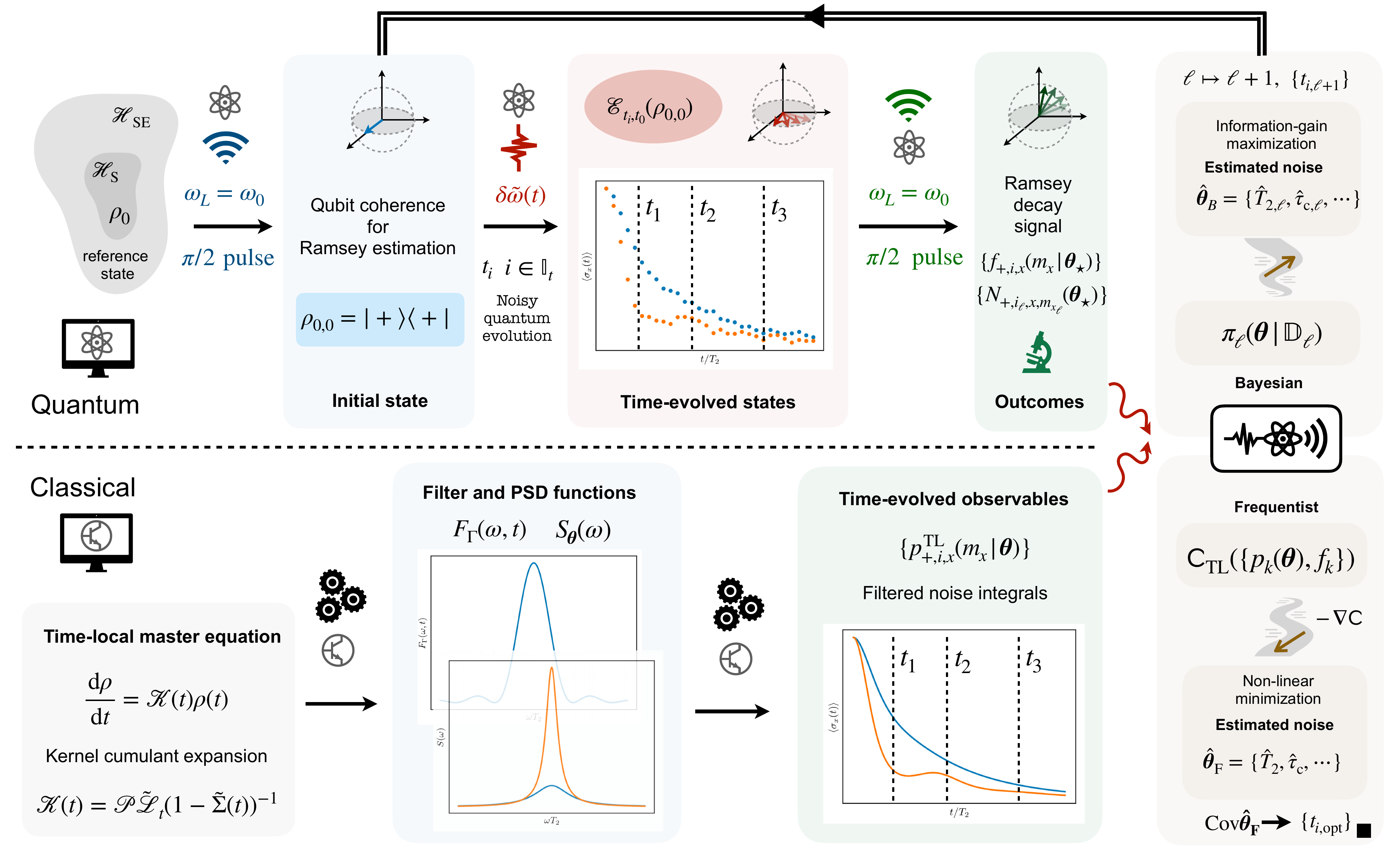}
  \caption{{\bf Scheme for  L$\ell$QT protocol of a dynamical dephasing map:} In the case of pure dephasing, the L$\ell$QT scheme reduces to a problem of Ramsey estimation. In the QIP, the qubit is initialized in a single state in the equator of Bloch's sphere $\rho_{0,0}=\ketbra{+}$ by a resonant  driving inducing a $\pi/2$ pulse on a reference state $\rho_0=\ketbra{0}$. The qubit then evolves for different times $t_i:i\in\mathbb{I}_t$, after which it is    subjected to another resonant $\pi/2$ pulse, which effectively rotates to the Pauli measurement basis $x$. As a consequence of the device dephasing noise with actual noise $\boldsymbol{\theta}_\star$ parameters, the $m_x\in\{0,1\}$ outcomes are  collected as relative frequencies $\{f_{+,i,x}(m_x|\boldsymbol{\theta}_\star)\}$ which will decay in time. This data can be collected by distributing the measurement shots among pre-fixed evolution times, connecting to the frequentist estimation protocol. Alternatively, in a Bayesian approach, one collects the data by successive $\ell$-update steps $\delta\mathbb{D}_\ell=\{N_{+,i_\ell,x,m_{x,\ell}}(\boldsymbol{\theta}_\star)\}$ . In the lower half, we show the classical part of the estimation, which starts by solving the time-local master equation for a particular parametrized power spectral density $S_{\boldsymbol{\theta}}(\omega)$ and filter function. This leads to the theoretically predicted probabilities $\{p^{\rm TL}_{+,i}(m_x|\boldsymbol{\theta})\}$. These probabilities and the relative frequencies are either fed into a log-likelihood cost function $\mathcal{C}_{\rm TL}(\boldsymbol{\theta})$ that must be minimized in order to obtain the frequentist estimate $\boldsymbol{\theta}_{\rm F}$, or used in a Bayesian  step to update our prior knowledge of the noise parameters $\pi_\ell(\boldsymbol{\theta}|\mathbb{D}_\ell)$, converging to the final Bayesian estimate $\boldsymbol{\theta}_{\rm B}$. In the frequentist approach, by calculating the covariance matrix, we can estimate the precision and find  the optimal measurement times, which will depend on how many noise parameters we aim at estimating, and changes with their specific values. In the Bayesian approach, knowledge about the noise parameters is updated sequentially by selecting evolution times that are expected to yield the greatest information gain  at each  step, thereby implementing feedback and adaptive data acquisition from the quantum device. }
  \label{fig:LlQT_setting}
\end{figure*}

Let us formulate the L$\ell$QT for the time-local master equation of Eq.~\eqref{eq:TLC_meq} for the dephasing of a single qubit, including temporal correlations. As discussed in the Methods section, either in a semi-classical or quantum-mechanical model of pure dephasing, the qubit dynamics can be described by the time-local master equation in Eq.~\eqref{eq:TLC_meq} with a simple Hamiltonian 
$H(t)=\half\omega_0\sigma_z$ and  a single jump operator $ L(t)=\sigma_z
$, both of which are time independent
\beq
\label{eq:TCL_dephasing}
\frac{{\rm d}\rho}{{\rm d}t}=-\frac{\ii}{2}\big[\omega_0\sigma_z,\rho\big]+\gamma(t)\big(\sigma_z\rho\sigma_z-\rho\big).
\eeq
 On the contrary, the decay rate is time-dependent and contains  memory information about the noise fluctuations
\beq
\label{eq:deph_rate_autocorr}
\gamma(t)=\frac{1}{4}\int_0^t\!\!{\rm d}t' \big(C(t,t')+ C(t',t)\big).
\eeq 
In a semi-classical model, $C(t,t')=\mathbb{E}[\delta\tilde{\omega}(t)\delta\tilde{\omega}(t')]$  is the auto-correlation function  of a stochastic process $\delta\tilde{\omega} (t)$ representing frequency fluctuations. This master equation is the result of averaging over the stochastic process $\rho(t)=\mathbb{E}[\tilde{\rho}(t)]$, and is valid for a random process in a second-order cumulant expansion known as the fast-fluctuation expansion or, alternatively,   for a Gaussian random process with arbitrary correlation times $\tau_{\rm c}$, as discussed in the Methods section. Alternatively, for a fully quantum-mechanical dephasing model,   $C(t,t')=\half{\rm Tr}_B\{B_I(t)B_{I}(t')\rho^{\rm ss}_{\rm B}\}$ is the auto-correlation function on the stationary state of the environment/bath $\rho^{\rm ss}_{\rm B}$, which induces fluctuations in the qubit frequency  via the bath operators $B_I(t)$. In this case, the time-local master equation is the result of tracing over the bath degrees of freedom $\rho(t)={\rm Tr}_B\{\rho_{\rm SB}(t)\}$, and is  valid in a second-order  cumulant expansion also discussed in the Methods section. Assuming wide-sense stationarity, i.e., the mean of the stochastic process is constant and the auto-correlation function only depends on $t'-t$, we can introduce the power spectral density (PSD) of the noise
\beq
\label{eq:psd}
C(t-t')=\int_{-\infty}^{\infty}\!\!\frac{{\rm d}\omega}{2\pi}\,S(\omega)\,\ee^{\ii\omega(t-t')}.
\eeq
Therefore,   the time-dependent rate becomes
\beq
\label{eq:deph_rate_psd}
\gamma(t)=\frac{1}{2}\int_{-\infty}^{\infty}\!\!\!{{\rm d}\omega}\,S(\omega)\,f_{\gamma}(\omega,t),
\eeq
where  we have   introduced the  following modulation function
\beq
\label{eq:f_gamma}
f_{\gamma}(\omega, t)=\frac{t}{2\pi}\sinc(\omega t).
\eeq
It will be useful to define the symmetrized auto-correlation function and the symmetrized PSD as $\bar{C}(t,t')=\frac{1}{2}(C(t,t')+ C(t',t))$ and $\bar{S}(\omega)=\frac{1}{2}(S(\omega)+ S(-\omega))$, since for dephasing noise only the symmetric part of the auto-correlation function and the symmetric part of the PSD influence the time evolution, as shown in the Methods section.

As discussed  in the previous sections, we focus on an effective parametrization of the Hamiltonian and dissipation matrix that reduces the search space. Considering the expression in Eq.~\eqref{eq:deph_rate_psd}, the dephasing rate will depend on a certain number $n$ of real-valued noise parameters $\boldsymbol{\theta}_{\!\star}\in \Theta=\mathbb{R}^{n}$ via the  PSD $S_{\boldsymbol{\theta}_{\!\star}}\!(\omega)$, where we have made explicit its parametrization.   Since    the Hamiltonian of the qubit is very simple,  and only depends on the transition frequency $\omega_0$ that is known with a high precision using spectroscopic methods, it need not be included in the learning. Technically, this means that we can assume that the driving used to initialize and measure the qubit is resonant with the transition (see Fig.~\ref{fig:LlQT_setting}), and work in the rotating frame presented in the Methods section. Likewise, we have only one possible jump operator $L=\sigma_z$, such that the learning can focus directly on the estimation $\boldsymbol{\hat{\theta}}$ of  the actual noise parameters $\boldsymbol{{\theta}}_\star$, which translate into  the estimation of the dephasing rate $\hat{\gamma}(t)$ via Eq.~\eqref{eq:deph_rate_psd}. Therefore,  L$\ell$QT becomes a  non-linear minimization problem for the cost function in Eq.~\eqref{eq:NM_cost}. Before giving more details on this problem, we discuss relevant properties of the dephasing quantum dynamical map.

For a white-noise noise model with a vanishing correlation time $\tau_{\rm c}=0$, one has a  flat PSD $S(\omega)=c$ and a constant dephasing rate $\gamma(t)=c/2$. This leads to an exponential decay of the  coherences $p_{+,i,x}(m_x)=\half\big(1+(-1)^{m_x}\ee^{-t_i/T_2}\big)$, where $m_x\in\{0,1\}$ correspond to the projective measurements on $\ket{+},\ket{-}$, respectively, and we have introduced a decoherence time $T_2=2/c$. If the real system is affected by white dephasing noise, there is thus a single noise parameter to learn $\theta_{\star}=c_\star$ or, alternatively, the real decoherence time $\theta_{\star}=T_{2\star}$.   We note that this procedure is in complete agreement with the LQT based on the corresponding Lindblad  master equation in Eq.~\eqref{eq:linblad_master_diag}. On the other hand, for a time-correlated dephasing noise with  a structured PSD, the coherence decay will generally differ from the above exponential law, with the exception of the long-time regime  $t_i\gg\tau_{\rm c}$, where $p_{+,i,x}(m_x)\approx\half\big(1+(-1)^{m_x}\ee^{-t_i/T_2}\big)$ and one finds an effective decoherence time controlled by the  static part of the PSD  $T_2=2/S(0)$. As $t_i$ increases towards $\tau_{\rm c}$, the decay will no longer be a time-homogeneous exponential, which can  actually be a consequence of (but not a prerequisite for) a non-Markovian quantum evolution. In this more general  situation, we will have more noise parameters $\boldsymbol{\theta}_\star$ to learn.

Let us now connect to the formalism of filter functions~\cite{Kofman2000,PhysRevLett.87.270405,PhysRevLett.93.130406,Gordon_2007,Uhrig_2008,PhysRevB.77.174509,Biercuk_2011,Almog_2011}, which appears naturally when considering the  time evolution of the coherences at any instant of time. This follows from the exact solution of the  time-local master equation in the rotating frame, which reads
\beq
\label{eq:ps}
p^{\rm TL}_{+,i,x}(m_x)=\frac{1}{2}\left(1+(-1)^{m_x}\ee^{-\Gamma(t_i)}\right)=:p_{i}^{\rm TL}(m_x),
\eeq
where we have introduced the time integral of the decay rates
\beq
\label{eq:Gamma}
\Gamma(t)=2\int_0^t\!{\rm d}t'\gamma(t'),
\eeq
and simplified the notation by omitting the initial state and the measurement basis, as they will be unique  for the estimation of the  dephasing map.
Using the Fourier transform in Eq.~\eqref{eq:psd}, this integral can be rewritten in terms of the noise PSD as 
\beq
\label{eq:chi_filter}
\Gamma(t)=\!\int_{-\infty}^\infty\!\!\!\!{\rm d}\omega\, S(\omega)F_{\Gamma}(\omega, t), \hspace{1ex} F_{\Gamma}(\omega, t)=\!\int_{0}^t\!\!{\rm d}t'\!f_{\gamma}(\omega,t'),
\eeq
where we have introduced a   filter function that reads
\beq
F_{\Gamma}(\omega,t)=\frac{t}{2}\eta_{\frac{2
}{t}}(\omega).
\eeq
Note that, by making use of the nascent Dirac delta
\beq
\eta_{\epsilon}(x)=\frac{\epsilon}{\pi x^2}\sin^2\!\!\left(\,\frac{x}{\epsilon}\,\right),   
\eeq
where $\eta_\epsilon\!(x)\to\delta(x)$ as $\epsilon\to0^+\!\!$, and $\int_{-\infty}^{\infty}\!{\rm d}x\,\eta_\epsilon\!(x)\!=1$, one sees that  in the long-time limit  \mbox{$\epsilon=2/t\to 0^+$},  $\tilde{f}_{\Gamma}(\omega,t)\approx \frac{t}{2}\delta(\omega)$ becomes a  Dirac delta distribution, such that $\Gamma(t)\approx t S(0)/2$. This  agrees with the above coarse-grained prediction for a decoherence time $T_2=2/S(0)$. Therefore, physically, the conditions for the long-time limit to be accurate is that $t\gg\tau_{\rm c}$. 

In this article, we are not interested in this  Markovian limit, as we aim at estimating the time-local master equation that depends on the full decay rate $\gamma(t)$, including situations in which non-Markovianity becomes manifest.  To quantify this, we note that the  dephasing quantum dynamical map
\beq
\label{eq:dephasing_map}
\mathcal{E}_{t,0}^{\rm TL}(\rho_0)=(1-p(t))\rho_0+p(t)\sigma_z\,\rho_0\,\sigma_z,
\eeq
is non-Markovian when it is not CP-divisible. In Eq.~\eqref{eq:dephasing_map}, we have introduced the following time-dependent  probability for the occurrence of phase-flip errors  
\beq
\label{eq:dephaisng_rate}
p(t)=\half\big(1-\ee^{-\Gamma(t)}\big).
\eeq
Following~\cite{PhysRevLett.105.050403}, the degree of non-Markovianity of the quantum evolution can be obtained by integrating over all  times for which the rate of the time-local master equation is negative
\begin{equation}\label{eq:non_markovian_measure_cp}
\mathcal{N}_{\rm CP} = \int{\rm d}t\big(|{\gamma}(t)|-{\gamma}(t)\big).
\end{equation}

An alternatively measure of non-Markovianity 
is based on the  trace distance of two arbitrary initial states \cite{rivasQuantumNonMarkovianityCharacterization2014, RevModPhys.88.021002}, which  will decrease with time when there is a flow of information from the system into the noisy environment. When this information flows back, the trace distance increases, and the qubit can recohere for a finite lapse of time, such that one gets  a non-Markovian quantum evolution. The instantaneous variation of trace distance  is given by $\sigma(t) = \frac{\dd}{\dd{t}} D(\mathcal{E}_{t,0}^{\rm TL}(\rho_0), \mathcal{E}_{t,0}^{\rm TL}(\rho_0'))$, with  $D$ being the trace distance~\cite{nielsen_chuang_2010}, and a positive $\sigma(t)$ is thus a measure of non-Markovianity, which can be expressed in terms of the time intervals in which the phase-flip error probability decreases infinitesimally with time 
\begin{equation}
\label{eq:N_TD}
    \mathcal{N}_{\mathrm{TD}} = \max_{\rho_0, \rho_0'} \int_{\sigma(t)>0} \dd{t} \sigma(t)  = -\int_{\gamma(t)<0} \dd{t} \dot{p}(t).
\end{equation}
Here, we have rewritten this measure  in terms of  the  error probability of the phase-flip channel of Eq.~\eqref{eq:dephaisng_rate}, which changes infinitesimally with $\dot{p}(t)=\gamma(t)\ee^{-\Gamma(t)}$. Hence, the non-Markovianity condition translates into a dynamical situation in which  phase-flip errors do not increase monotonically during the whole evolution. When the dephasing rate attains negative values, the phase-flip error  probability can decrease, such that the qubit momentarily recoheres (see the orange lines in Fig.~\ref{fig:LlQT_setting}). In this simple case of pure dephasing noise, we see that both measures of non-Markovianity in Eqs.~\eqref{eq:non_markovian_measure_cp} and ~\eqref{eq:N_TD}  depend on the rate $\gamma(t)$ attaining negative values, and both are equal to 0 if $\gamma(t)$ is always positive. We define and detect non-Markovianity in this case using Eqs.\ \eqref{eq:non_markovian_measure_cp} and \eqref{eq:N_TD}. However, it is important to note that these measures provide sufficient, but not necessary, conditions for detecting all possible forms of non-Markovian behavior. For instance, certain non-Markovian dynamics may not yield an increase in trace distance.


Once these additional properties of the  dephasing quantum dynamical map have been discussed, we can move back to the estimation L$\ell$QT protocol of Eq.~\eqref{eq:LQT_ML_prob}, and how it can be simplified even further.  As noted above, in contrast to the informationally-complete set of initial states an measurements  that must be considered for the general cost  function of L$\ell$QT in Eq.~\eqref{eq:NM_cost}, we can work with a smaller number of   configurations by noting that  all  information of the dephasing map can be extracted by preparing a single  initial state $\rho_0=\ketbra{+}$, and measuring in a single basis $M_{x,\pm}$ (see Fig.~\ref{fig:LlQT_setting}). Indeed, this combination of initial state and measurement basis yields the expected value $p^{\rm TL}_{+,i,x}(m_x)=\frac{1}{2}(1+(-1)^{m_x}\ee^{-\Gamma(t_i)})$. Similarly, if we chose $\rho_0=\ketbra{+\mathrm{i}}$ and $M_{y,\pm}$, we would obtain $p^{\rm TL}_{+\mathrm{i},i,y}(m_y)=\frac{1}{2}(1+(-1)^{m_y}\ee^{-\Gamma(t_i)})$, which yields the same information. The rest of combinations of initial states and measurement bases yield constant values and do not provide any information about the decay rates.

As advanced in the introduction, we would like to know how many snapshots $|\mathbb{I}_t|$ are required, at which the system is measured  after  evolving for $\{t_i: \, i\in\mathbb{I}_t\}$  and, moreover, which are the optimal times of those snapshots in terms of the specific details of the  noise. We take here two different routes: the frequentist and the Bayesian approach. In the frequentist case, for pure dephasing we have the cost function
\begin{align}
\label{eq:ML_cost_dephasing}
    \mathsf{C}^{\rm pd}_{\mathrm{TL}}(\boldsymbol{\theta}) = -\sum_{i}  N_i\,\sum_{m_x}\, \tilde{f}_{i}(m_x\,|\boldsymbol{\theta}_\star) \log p^{\rm TL}_{i}(m_x | \boldsymbol{\theta}), 
\end{align}
which is greatly simplified with respect to the general case in Eq.~\eqref{eq:ML_cost_parameter}, as we only have a single initial state and a single measurement basis. Since we are monitoring the coherence of the qubit, this cost function corresponds to a  {\it Ramsey-type estimator}, where $\tilde{f}_{i}(0\,|\boldsymbol{\theta}_\star)=N_{i,0}/N_i$ $(\tilde{f}_{i}(1\,|\boldsymbol{\theta}_\star)=N_{i,1}/N_i)$ is the ratio of the number of outcomes observed $N_{i,0}$ ($N_{i,1}=N_i-N_{i,0}$) to the total of $N_i$  shots  collected at the instant of time $t_i$, when measuring the system with the POVM element $M_{x,0}$ ($M_{x,1}$).  Our notation remarks that these relative frequencies carry information about the real noise parameters $\boldsymbol{\theta}_{\!\star}$ we aim at estimating. In addition, the estimator depends on  $p_i(m_x | \boldsymbol{\theta})$ shown in Eqs.~\eqref{eq:ps}-\eqref{eq:Gamma}, which stand for the    probabilities obtained by solving the time-local dephasing master equation of Eq.~\eqref{eq:TCL_dephasing},   where  we make  explicit the dependence on the parametrized  noise. Minimizing $\det \Sigma_{\boldsymbol{\hat{\theta}}}$ we can determine the optimal measurement times of the estimator. For the Bayesian approach, at each  $\ell>0$ Bayesian  step, we measure the system  enlarging the data set sequentially $\mathbb{D}_{\ell-1}\mapsto\mathbb{D}_\ell=\mathbb{D}_{\ell-1}\cup\delta\mathbb{D}_{\ell}$, where $\delta\mathbb{D}_{\ell}\subset\mathbb{D}=\{N_{i,+,x,m_{x}}\}$ contains a number of measurement outcomes  $|\delta N_{\ell}|$ that is a fraction of the total $N_{\rm shot}$. These outcomes will be labelled as $\delta \mathbb{D}_\ell=\{N_{i_\ell,+,x,m_{x_\ell}}\}$. At each step, we maximize the information gain of Eq.~\eqref{eq:optimal_time_information_gain} to determine the optimal measurement time.
As shown in Fig.~\ref{fig:LlQT_setting} the scheme for L$\ell$QT in the case of pure dephasing can be divided into the following steps:

\noindent {\it a) Quantum experiment:}
    \begin{enumerate}
    \item Initialize the reference state $\rho_0 = \ketbra{0}$ (see Fig.~\ref{fig:LlQT_setting}).
    \item Prepare $\rho_{0,0} = \ketbra{+}$ by a resonant $\pi/2$ pulse.
    \item Let the qubit evolve under the pure dephasing  for different evolution times $t_i$.
    \item Apply a second resonant $\pi/2$ pulse, rotating the qubit to the Pauli-$X$ measurement basis.
    \item Collect the outcomes $m_x \in \{0, 1\}$ and  compute the frequencies $\{f_{+,i,m_x}\}$.
    \end{enumerate} 
{\it b) Classical processing and frequentist approach:}
    \begin{enumerate}
    \item Solve the time-local master equation with a parametrized power spectral density $S_{\boldsymbol{\theta}}(\omega)$ and a filter function, yielding  the predicted probabilities $\{p_{+,i,m_x}(\boldsymbol{\theta})\}$.
    \item From the quantum experiment we have a set of frequencies $\{f_{+,i,m_x}\}$ at different times $t_i$. The predicted probabilities and measured relative frequencies are used in a log-likelihood cost function, which is minimized to estimate noise parameters $\boldsymbol{\theta}$.
    \item By calculating the covariance matrix, the precision of the estimation is determined.
    \end{enumerate}
{\it c) Classical processing and Bayesian approach:}
    \begin{enumerate}
        \item Solve the time-local master equation with a parametrized power spectral density $S_{\boldsymbol{\theta}}(\omega)$ and a filter function, yielding  the predicted probabilities $\{p_{+,i,m_x}(\boldsymbol{\theta})\}$.
        \item We have some prior knowledge of the parameters $\boldsymbol{\theta}$, which is represented by a prior probability distribution.
        \item At each step, evolution time $t_i$ is chosen to maximize the information gain of the parameters. A new quantum experiment is performed with this new evolution time.
        \item After each measurement or set of measurements the relative frequencies are used to update the prior distribution of the parameters.
    \end{enumerate}

We present below a detailed comparison of the two approaches, frequentist and Bayesian, determining the regimes in which each of them is   better than the other.
For the Bayesian protocol design, we have used the Python package Qinfer \cite{Granade2017qinferstatistical}, which numerically implements the operations needed by using a sequential Monte Carlo algorithm for the updates.

Let us study some dephasing dynamics in which we can apply the two approaches we have just introduced, and make a comparative study of their performance when learning  parametrized dephasing maps with time-correlated  noise.
We study two different dephasing models: a Markovian semi-classical dephasing and a non-Markovian quantum  dephasing. These models are selected for several reasons. The semi-classical model is a well-established generalization of purely Lindbladian evolution, and in fact, contains the Lindbladian regime as a limiting case. This allows us to connect with previous results from LQT and Ramsey interferometry in the context of quantum sensing and quantum clocks. On the other hand, the non-Markovian case represents the simplest possible generalization of the Markovian scenario by introducing a shifted Lorentzian PSD, making it an ideal testbed for L$\ell$QT. Furthermore, it has experimental relevance, as we can use a laser-cooled vibrational mode in trapped ions to obtain a fully-tunable
implementation of this non-Markovian noise. These two dephasing models offer both classical and quantum noise characterization through the symmetry of the PSD, and they provide a versatile platform to demonstrate L$\ell$QT.

Regarding the choice between frequentist and Bayesian inference, the majority of quantum technologies traditionally use frequentist approaches. However, Bayesian inference is becoming increasingly relevant, particularly in situations where the number of shots is limited and  clock cycle times are large, such as in trapped-ion platforms. As we show below, Bayesian methods offer advantages in scenarios where physical priors are available and can provide better estimates in non-Markovian cases, where the dynamics are more complex.

For each one of the cases presented below, we have a PSD with some real parameters $\boldsymbol{\theta}_{\star}$ from which we can determine the time-dependent measurement probabilities $p^{\rm TL}_{i}(m_x | \boldsymbol{\theta_{\star}})$. From these probabilities, we numerically take the necessary samples to simulate the experiment by using the SciPy implementation of a binomial random variable~\cite{2020SciPy-NMeth}.

\subsection*{Markovian semi-classical dephasing}

We now apply both estimation techniques for the L$\ell$QT of a dephasing quantum dynamical map that goes beyond the Markovian Lindblad assumptions. In particular, we consider a time-correlated frequency noise $\delta\tilde{\omega}(t)$ that is described by an Ornstein-Uhlenbeck (OU) random process~\cite{gardiner2004handbook,Gillespie}. This process has  an underlying  multi-variate Gaussian joint PDF,  and incorporates a correlation time $\tau_{\rm c}>0$ above which the correlations between consecutive values of the process become very small. In fact, beyond the relaxation window $t,t'>\tau_{\rm c}$, the correlations show an exponential decay
\beq
\label{eq:OU_corr}
C(t-t')=\frac{c\tau_{\rm c}}{2}\ee^{-\frac{|t-t'|}{\tau_{\rm c}}},
\eeq
where $c>0$ is a so-called diffusion constant. Since this correlations  only depend on the time differences, the process is wide-sense stationary. Moreover, on the basis of its Gaussian joint PDF, it can be shown that the process is indeed strictly  stationary. Being Gaussian, all the information is thus contained in its two-point functions or, alternatively, in its PSD
\beq
\label{eq:lorentzain_psd}
S(\omega)=\frac{ c\tau_{\rm c}^2}{1+(\omega\tau_{\rm c})^2},
\eeq
which has a Lorentzian shape. This Gaussian process then leads to an exact time-local master equation for the dephasing of the qubit in Eq.~\eqref{eq:TCL_dephasing} with a time-dependent decay rate
\beq
\label{eq:decay_rate_OU}
\gamma(t)=\fourth S(0)\left(1-\ee^{-\frac{t}{\tau_{\rm c}}}\right).
\eeq
In the long-time limit $t\gg\tau_{\rm c}$, one recovers a constant decay rate $\gamma(t)\approx S(0)/4= c\tau_{\rm c}^2/4$, which connects to our previous discussion of the effective exponential decay of the Ramsey signal $p^{\rm TL}_i(m_x)\approx\big(1+(-1)^{m_x}\ee^{-t_i/T_2}\big)/2$ and the decoherence time $T_2=2/c\tau_{\rm c}^2$. On the other hand, for  shorter time scales, we see the effects of the noise memory through a  time-inhomogeneous evolution of the coherences that goes beyond a Lindbladian description. 
The time-dependent decay rate is always positive, such that the two measures of non-Markovianity in Eqs.~\eqref{eq:non_markovian_measure_cp}-\eqref{eq:N_TD} vanish exactly. The pure dephasing quantum dynamical map of a qubit subjected to OU frequency noise is thus Markovian albeit not Lindbladian.

From the perspective of L$\ell$QT, we have two  parameters to learn $\boldsymbol{\theta}=(c,\tau_{\rm c})\in\Theta=\mathbb{R}_+^2$, which fully parametrize the  PSD, the decay rate or, alternatively, the Ramsey  attenuation factor
\beq
\label{eq:OU_decay}
\Gamma(t) = \half S(0) \left( t - \tau_{\rm c} \big(1-\ee^{-\frac{t}{\tau_{\rm c}}}\big) \right).
\eeq

\begin{figure}
    \centering
    \includegraphics[width=.9\linewidth]{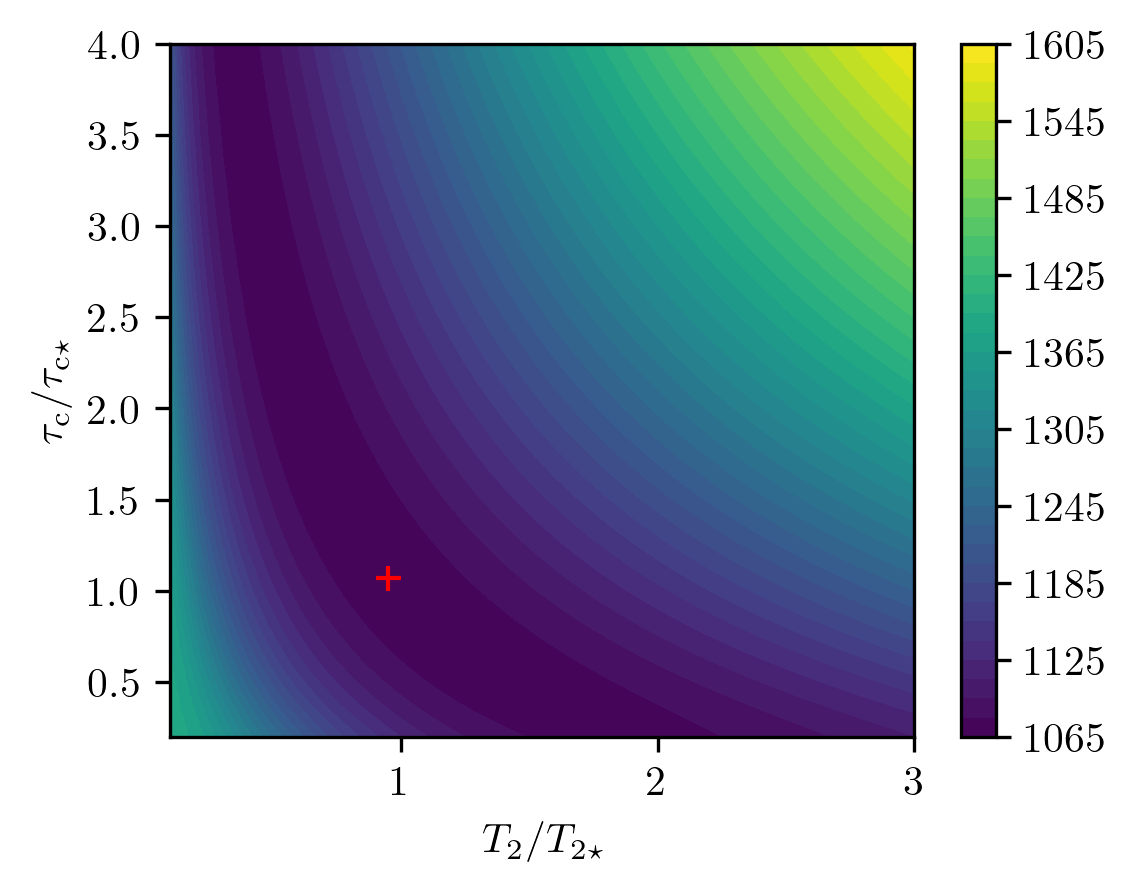}
    \caption{ {\bf Gradient descent for  OU dephasing L$\ell$QT:} Contour plot of the Ramsey cost function  $\mathsf{C}^{\rm pd}_{\mathrm{TL}}(\boldsymbol{\theta})$ of Eq.~\eqref{eq:ML_cost_dephasing} as a function of the noise parameters $(T_2,\tau_{\rm c})$, where we recall that $T_2=2/c\tau_{\rm c}^2$ plays the role of an effective decoherence time in the long-time limit.   We choose  two evolution times  and, for illustration purposes, fix them at $t_1=\tau_{\rm c\star}$, $t_2=2\tau_{\rm c\star}$, fixing the OU diffusion constant such that  $T_{2\star}/\tau_{\rm c\star}=1$. We distribute $N_{\rm shot}=2\times 10^3$ shots equally per time step. The gradient descent of this convex problem converges towards the global   minimum, and  is marked with a red cross at $T_2/T_{2\star} \approx 0.95$ and $\tau_{\rm c}/\tau_{\rm c\star}\approx 1.07$, lying close to the real noise parameters $\boldsymbol{\theta}\mapsto\boldsymbol{\theta}_\star$.}
    \label{fig:cost_function_ou_noise}
\end{figure}

\textit{a. Frequentist Ramsey estimators.} We can now evaluate the L$\ell$QT cost function $ \mathsf{C}^{\rm pd}_{\mathrm{TL}}(\boldsymbol{\theta})$ in Eq.~\eqref{eq:ML_cost_dephasing} by substituting the attenuation factor in Eq.~\eqref{eq:OU_decay} in the likelihood function $p^{\rm TL}_{i}(m_x | \boldsymbol{\theta})=\big(1+(-1)^{m_x}\ee^{-\Gamma(t_i)}\big)/2$ after a certain set of evolution times $\{t_i,i\in\mathbb{I}_t\}$, and the relative frequencies for the measurement outcomes $\tilde{f}_{i}(m_x\,|\boldsymbol{\theta}_\star)$. For a Lindbladian dynamics, a single measuring time $t_1$ would suffice for the estimation~\cite{dobrynin2024compressedsensing}, which can actually be solved for analytically in the present pure dephasing context, as discussed in Sec.~\ref{app:Lindblad_case} of the Supplementary Material. For the OU dephasing, this is no longer the case, and we actually need at least two times $t_1,t_2$. In order to assess the performance of the frequentist minimization problem in Eq.~\eqref{eq:min_freq} under shot noise, we numerically generate the   relative frequencies $\tilde{f}_{1}(m_x\,|\boldsymbol{\theta}_\star),\tilde{f}_{2}(m_x\,|\boldsymbol{\theta}_\star)$ at two instants of time by sampling the probability distribution with the actual OU parameters $\boldsymbol{\theta}_{\star}=(c_{\star},\tau_{\rm c{\star}})$ a number of times $N_{\rm shot}=N_1+N_2$. In the following, rather than learning $(c_\star,\tau_{\rm c\star})$, we will focus on two noise parameters with units of time $(T_{2\star},\tau_{\rm c\star})$, where we recall that $T_{2\star}=2/c^{\phantom{2}}_\star\tau_{{\rm c}\star}^2$ is an effective decoherence time in the long-time limit. In Fig.~\ref{fig:cost_function_ou_noise}, we present a contour plot of this two-time cost function $ \mathsf{C}^{\rm pd}_{\mathrm{TL}}(\boldsymbol{\theta})$, which is actually convex and allows for a neat visualization of its global minimum. We also depict with a red cross  the result of a gradient-descent minimization, where one can see that the estimates $\boldsymbol{\hat{\theta}}_{\rm F}=(\hat{T}_{2}, \hat{\tau}_{\rm c})$ are close to the real noise parameters. The imprecision of the estimate is a result of the shot noise, which we now quantify.

\begin{figure}
        \centering
        \includegraphics[width=.85\linewidth]{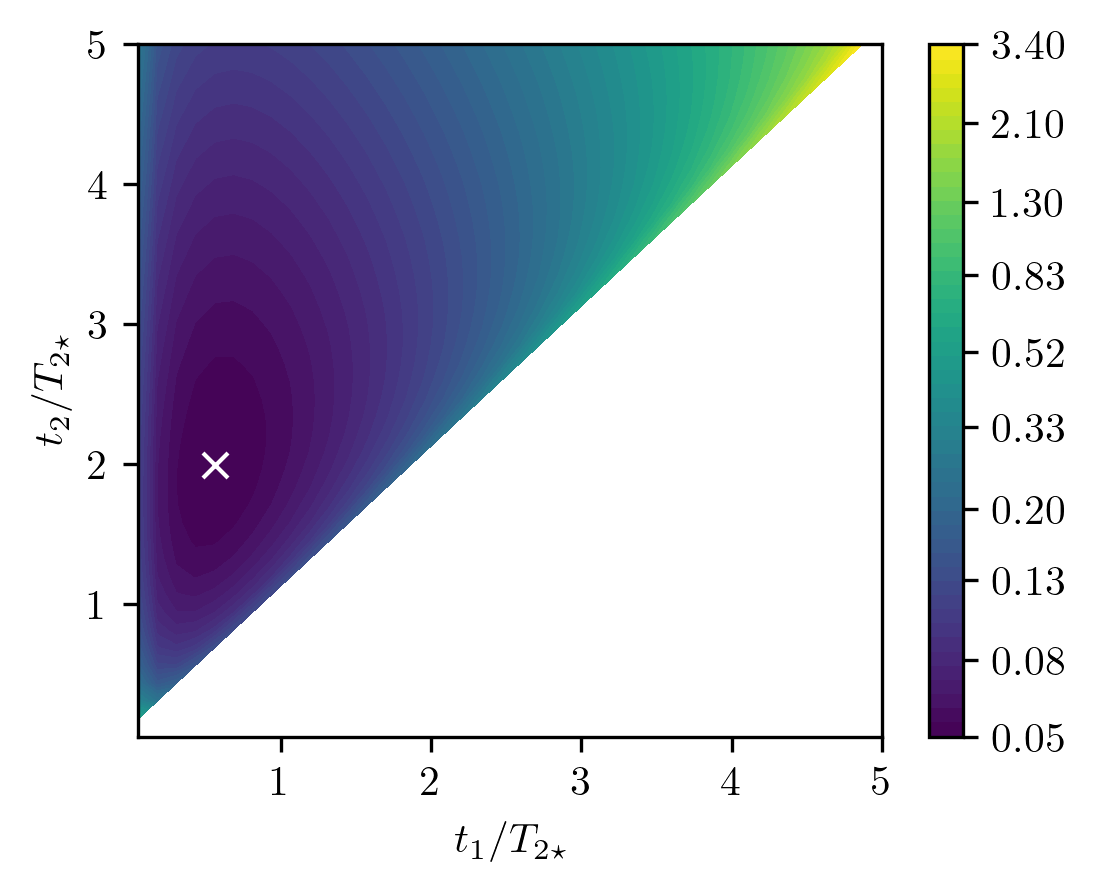}
    \caption{ {\bf Asymptotic covariance matrix for OU dephasing:} We represent $\det(\Sigma_{\boldsymbol{\hat{\theta}}}(t_{1},t_{2}))^{1/2}$ as a function of the $t_1$ and $t_2$ measurement times selected. For $t_1=t_2$ the determinant diverges, since it is not possible to determine the parameters by just measuring at a single time. The true parameters are $\tau_{\rm c\star}=T_{2\star}/2$ and the number of shots is $N_1=N_2=5\cdot10^3$. The optimal times that minimize the determinant are indicated with a white cross, $t_1\approx0.56 T_{2\star}$, $t_2\approx1.99 T_{2\star}$. The determinant increases rapidly when moving away from this minimum. Note the color bar scale is logarithmic. The plot is symmetric with respect to the line $t_1=t_2$, therefore we just represent the part $t_2>t_1$.}
            \label{fig:det_times_OU}
\end{figure}
In order to find the optimal evolution times $t_1,t_2$ that maximize the precision of our estimates,  we can minimize the covariance in Eq.~\eqref{eq:asymptotic_covariance} which, in turn, requires maximizing the Fisher information matrix in Eq.~\eqref{eq:fisher}. By Taylor expanding the  cost function, we can actually find a linear relation between the estimate difference $\delta{\boldsymbol{\hat{\theta}}}=\boldsymbol{\hat{\theta}}_{{\rm F}}-\boldsymbol{\theta}_{\star}$, and the differences between the parametrized probabilities and the relative frequencies $\delta\tilde{f}_i=p_i^{\rm TL}(m_x | \boldsymbol{\hat{\theta}})-\tilde{f}_i(m_x\,|\boldsymbol{\theta}_\star)$, namely
\begin{equation}\label{eq:delta_t2_tau_delta_f}
    \begin{pmatrix}
        \delta \hat{T}_2 \\
          \delta \hat{\tau}_{\rm c}
    \end{pmatrix}
   \! =\!
    \frac{1}{\mathcal{N}}\!\!
    \begin{pmatrix}
        -\frac{\Gamma'_{\tau_{\rm c}}\!(t_2) (\ee^{2\Gamma(t_1)}-1)}{\sinh \Gamma(t_1)} & \frac{\Gamma'_{\tau_{\rm c}}\!(t_1) (\ee^{2\Gamma(t_2)}-1)}{\sinh \Gamma(t_2)} \\
        \frac{\Gamma'_{T_2}\!\!(t_2) (\ee^{2\Gamma(t_1)}-1)}{\sinh \Gamma(t_1)}   & -\frac{\Gamma'_{T_2}\!\!(t_1) (\ee^{2\Gamma(t_2)}-1)}{\sinh \Gamma(t_2)}
    \end{pmatrix}\!\!\!  \begin{pmatrix}
        \delta \tilde{f}_1 \\
        \delta \tilde{f}_{2}
    \end{pmatrix}\!,
\end{equation}
where $\mathcal{N}=\Gamma'_{T_2}\!(t_1)\Gamma'_{\tau_{\rm c}}\!(t_2) - \Gamma'_{\tau_{\rm c}}\!(t_1)\Gamma'_{T_2}\!(t_2)$, and we have introduced a shorthand notation for the partial derivatives  $\Gamma'_{\tau}\!(t_i)=\partial\Gamma(t_i)/\partial\tau_{\rm c}$,  $\Gamma'_{T_2}\!(t_i)=\partial\Gamma(t_i)/\partial T_2$.  The first thing one notices is that, fixing $t_2=t_1$, the factor $\mathcal{N}=0$, and  the difference between the estimation and the true value of the parameters diverges,  signaling  the fact that one cannot learn two noise parameters using a Ramsey estimator with a single instant of time. The second result one finds is that, in the asymptotic limit $N_i\rightarrow \infty$, the estimate differences  will follow a bi-variate normal distribution. This follows from the fact that Eq.~\eqref{eq:delta_t2_tau_delta_f}  is a linear combination of the differences between the finite frequencies and the binomial probabilities, which are known to follow a  normal distribution $N(\boldsymbol{0},{\rm diag}\big(\sigma^2_{f_1},\sigma^2_{f_2})\big)$ with  binomial variances $\sigma^2_{ f_i}$  defined in Eq.~\eqref{eq:bin_variance}.
Therefore, the frequentist estimates will also be normally distributed $\delta{\boldsymbol{\hat{\theta}}}\sim N(\boldsymbol{0},\Sigma_{\boldsymbol{\hat{\theta}}}\big)$ according to  
\beq
\label{eq:cov_matrix}
\Sigma_{\boldsymbol{\hat{\theta}}}(t_1,t_2)=M_{\boldsymbol{\hat{\theta}}}(t_1,t_2) \begin{pmatrix}
        \sigma^2_{f_1}& 0 \\
         0& \sigma^2_{f_2}
    \end{pmatrix}M_{\boldsymbol{\hat{\theta}}}^{\rm T}(t_1,t_2),
\eeq
where $M_{\boldsymbol{\hat{\theta}}}(t_1,t_2)$ is the matrix in Eq.~\eqref{eq:delta_t2_tau_delta_f}. A more detailed derivation of the relationship between shot noise and the uncertainty in  estimation, as well as the asymptotic covariance matrix, is provided in the Methods section. From this perspective, the aforementioned divergence for $t_1=t_2$  is a consequence of the singular nature of this matrix, which cannot be thus  inverted. 

\begin{figure}
        \centering
        \includegraphics[width=.85\linewidth]{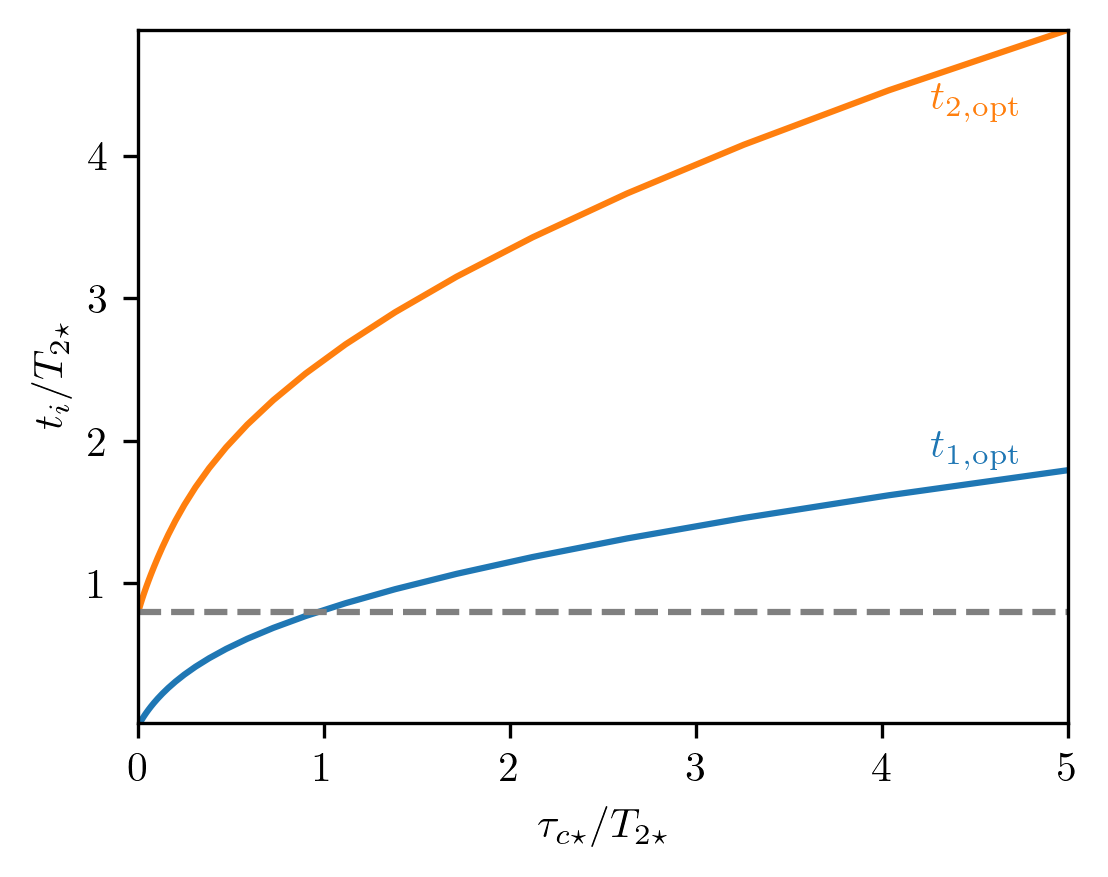}
    \caption{ {\bf Optimal  times for  OU dephasing L$\ell$QT:} We represent $(t_{1, \rm opt},t_{2, \rm opt})=\texttt{argmin}\{\det(\Sigma(t_1,t_2))\} $ as a function of the ratio of the  real noise parameters $\tau_{\rm c\star}/T_{2\star}$.
    In the regime where $\tau_{\rm c\star} \ll T_{2\star}$, we find that $t_2 \approx 0.8 T_{2\star}$ and $t_1 \approx 0$. The  optimal time of this measurement lies very close to the Lindbladian  result $t=0.797 T_{2\star}$ (gray dashed line) found when there is no time correlation and we have a single parameter $T_c$, which is an exact analytical result discussed in Sec.~\ref{app:Lindblad_case} of the Supplementary Material.}
            \label{fig:optimal_time_OU}
\end{figure}

A measure  of the imprecision of the estimation is then obtained from $\det(\Sigma_{\boldsymbol{\hat{\theta}}}(t_1,t_2))=(\det M_{\boldsymbol{\hat{\theta}}})^2\sigma^2_{f_1}\sigma^2_{f_2}\propto 1/N_1N_2=1/N_1(N_{\rm shot}-N_1)$ which, in this bi-variate case,  can be related to the area enclosed by a covariance ellipse. We thus clearly see that the maximum precision will be obtained when $N_1=N_2=N_{\rm shot}/2$. Turning to the optimal measurement times, we can now numerically minimize the  determinant of the asymptotic covariance matrix
\beq
\label{eq:opt_times}
\{t_{i,{\rm opt}}\}=\texttt{argmin}\left\{\det\Sigma_{\boldsymbol{\hat{\theta}}}\big(\{t_i\}\big)\right\},
\eeq
finding the two  optimal values at which the signal shows the highest sensitivity to changes in the OU noise (see Fig.~\ref{fig:det_times_OU}). The optimal times obtained in this way are depicted  in Fig.~\ref{fig:optimal_time_OU} as a function of the noise correlation time. In the limit where this correlation time is much smaller than the effective decoherence time $\tau_{\rm c\star}\ll T_{2\star}$, the signal only carries  important information about the noise for times that are much larger than the correlation time. Hence, we are in the long-time limit where the decay rate is constant $\gamma(t)\approx 1/2T_2$ and one expects to find agreement with a purely Lindbladian dephasing noise. As discussed in Sec.~\ref{app:Lindblad_case} of the Supplementary Material, the LQT for pure dephasing requires a single measuring time, and can be analytically found by minimizing the standard deviation of the estimated noise parameter. This solution  yields an optimal  time $t_{\rm opt}=0.797T_{2\star}$, which is actually very close to the intercept of the curve of $t_{2,{\rm opt}}$ shown in Fig.~\ref{fig:optimal_time_OU}. In this long-time regime, we  find $t_{1,{\rm opt}}\approx 0$ indicating that measurements at time $t_{2, \rm opt}$ will be mainly used to determine parameter $T_2$, while those  at $t_{1,{\rm opt}}\approx 0$  contribute to determine the much smaller $\tau_{\rm c}$. 

In the more general case in which the time correlation of the OU noise yields important memory effects, the measurement times have to be adapted to specific optimal values, which are in general larger than the purely Lindbladian limit as shown in Fig.~\ref{fig:optimal_time_OU}. Since these optimal times depend on the parameters we aim at learning, it is not straightforward to devise a practical strategy to minimise the imprecision of the frequentist estimates. In the pure Lindbladian case, one may foresee that the experimentalist will have an accurate prior knowledge of the $T_2$ time, such that the measurements can all be implemented close to the predicted optimal time. On the other hand, for the OU noise, one has the additional noise correlation time $\tau_{\rm c}$, which is related to deviations from the time-homogeneous exponential decay of the coherences and is not typically characterised experimentally. The frequentist procedure to operate at the optimal regime of estimation would then need to distribute the total $N_{\rm shot}$ in smaller groups that are applied in sequence, each time shifting the measurement times to try to get to the optimal point. One can foresee that this procedure will  not  be optimal, as one will loose many measurements along the way and, moreover, not scalable to other situations in which one aims at learning more noise parameters also optimally.

\textit{b. Bayesian Ramsey estimators.} Let us now describe how a Bayesian inference for OU dephasing L$\ell$QT would proceed, which will provide an experimental procedure to operate at the optimal estimation times. We start by commenting on the fully-uncorrelated Lindbladian limit discussed in Sec.~\ref{app:Lindblad_case} of the Supplementary Material, where the optimization of the measurement time for each Bayesian step in Eq.~\eqref{eq:optimal_time_information_gain} can also be solved analytically. Considering that the prior probability distribution $\pi_{\ell}(\gamma)$ for our knowledge about the decay rate at the $\ell$-th step is Gaussian, we can focus on how its mean and variance change as one takes the next Bayesian step. In Sec.~\ref{app:Lindblad_case} of the Supplementary Material, we show that, minimizing the  Bayesian variance of the next step, one finds optimal measurement times  that agree with the above frequentist prediction, albeit for the knowledge of the decay rate that we actually have at each particular step $t_\ell=0.797/2\hat{\gamma}_{\ell}$ or, alternatively, of the decoherence time $T_{2,\ell}$. This result is very encouraging, as the experimentalist may only have a crude guess of this value, but it gets automatically updated towards the optimal regime.  This motivates  an extension to time-correlated  dephasing such as the OU noise. 

\begin{figure}
    \centering
    \includegraphics[scale=.8]{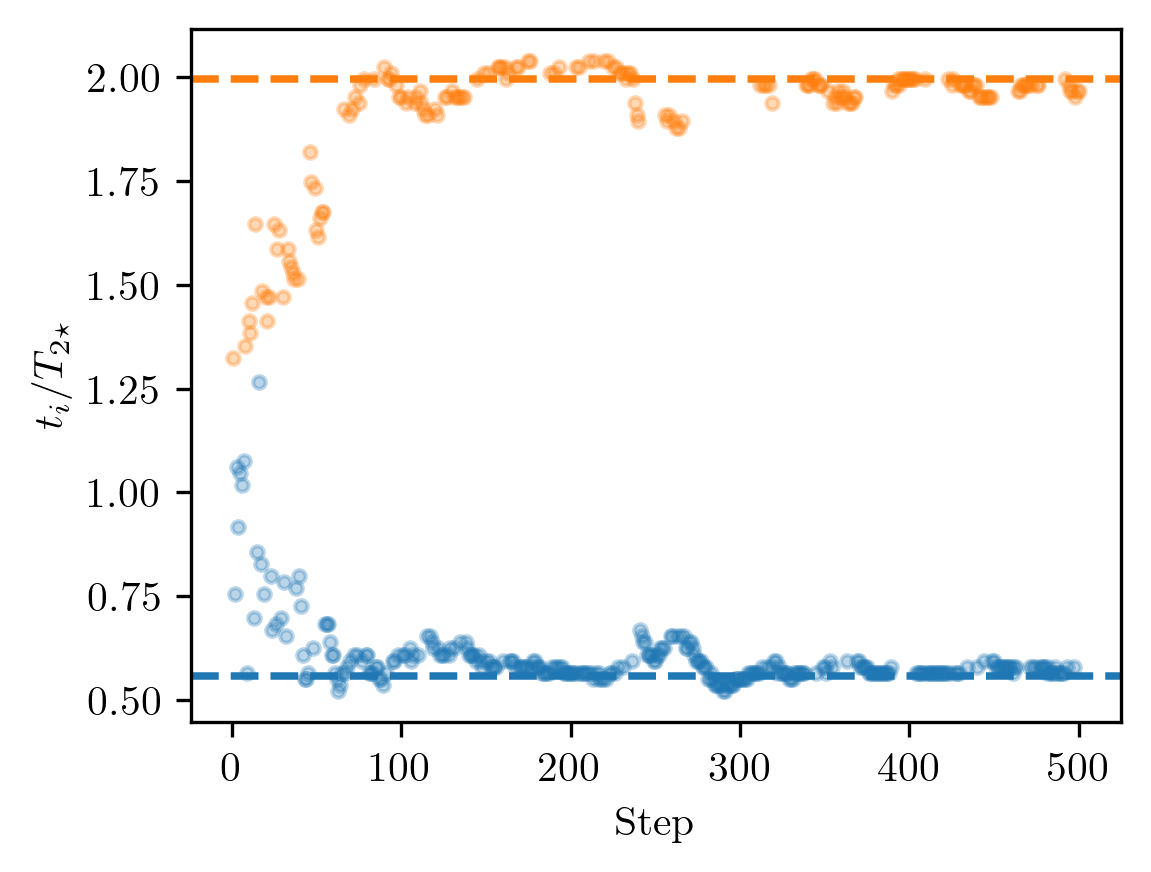}
    \caption{{\bf Bayesian optimal measurement time at each step:} We consider OU noise and set  $\tau_{\rm c\star} = T_{2\star}/2$. The prior of the parameters is a continuous uniform distribution with $T_2 \in [T_{2\star}/3, 3T_{2\star}], \tau\in [\tau_{\rm c\star}/3, 3\tau_{\rm c\star}]$. At each Bayesian step, we increase the data set with $|\delta\mathbb{D}_\ell|=50$ new outcomes, which are  used to compute the Bayesian update. The total number of measurements after 500 steps is $500\times50=25000$. In the Bayesian steps at the beginning, we see that the evolution times lie around $t\sim T_{2\star}$. Later on, the algorithm tends to alternate between two times, which correspond to the optimal times $t_{1, \rm opt} \approx 0.56 T_{2\star}$, $t_{2, \rm opt} \approx1.99 T_{2\star}$ in the case of $\tau_{\rm c\star} = T_{2\star}/2$. As the number of steps increases and and the posterior gets closer to the true values of $\tau_{\rm c\star}$ and $T_{2\star}$, the Bayesian algorithm tends to select measurement times which are closer to the optimal times.} \label{fig:bayesian_experiment_optimal_times}
\end{figure}

For the OU dephasing, we have two parameters to learn, and we can maximize the Kullback-Leibler divergence of Eq.~\eqref{eq:optimal_time_information_gain}  to obtain the subsequent optimal  time $t_\ell$ for the next Ramsey measurement(s), and the corresponding extension of the data set $\mathbb{D}_{\ell-1}\mapsto \delta\mathbb{D}_{\ell}=\mathbb{D}_{\ell-1}\cup\delta\mathbb{D}_{\ell-1}$. We then  proceed by measuring at this time,   updating the prior,  and starting the optimization step  all over again to finally find the estimates in Eq.~\eqref{eq:exp_value}  $
\boldsymbol{\hat{\theta}}_{{\rm B}}=(\hat{\tau}_{{\rm c},\ell},\hat{T}_{2,\ell})$. 
As shown in Fig.~\ref{fig:bayesian_experiment_optimal_times}, as one collects more and more data, the Bayesian measurement times cluster at two single times, and tend to alternate between them. Remarkably, these times are  the optimal $t_{1,{\rm opt}}$ and $t_{2,{\rm opt}}$ predictions of the frequentist approach shown in Fig.~\ref{fig:optimal_time_OU}. We can see how the Bayesian approach automatically finds the optimal measurement setting to learn a time-correlated dephasing noise. 

\begin{figure}
    \includegraphics[width=0.85\columnwidth]{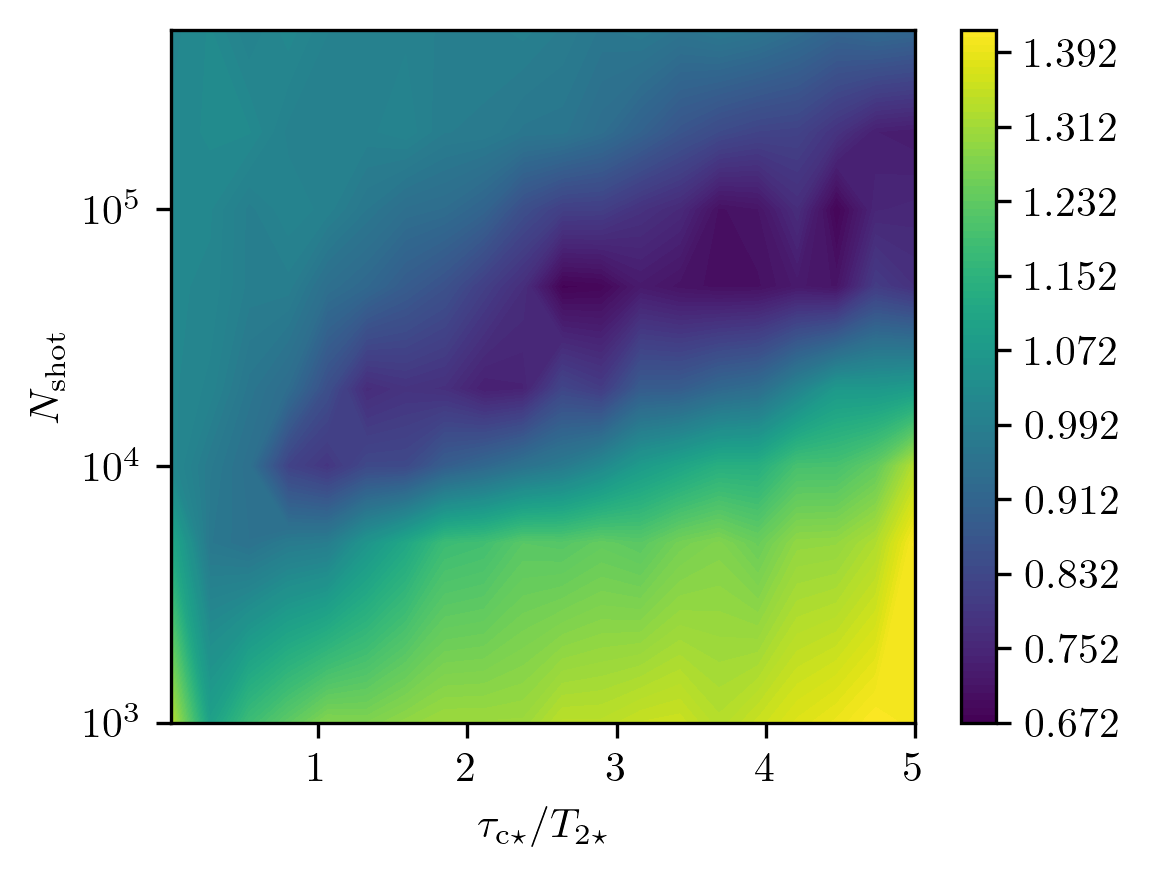}
    \caption{{\bf Comparison of frequentist and Bayesian approaches for OU noise:} The ratio of Eq.~\eqref{eq:det_ratio_OU} computed for different parameter values $\tau_{\mathrm{c}\star}$ and total number of shots $N_{\rm shot}$. 5000 frequentist runs were done to estimate $\det \mathrm{Cov}(\hat{\boldsymbol{\theta}}_{\mathrm{F,opt}})$, while 200 Bayesian runs were done to estimate $\det \Sigma_{\ell_{\rm max}}$ for each point $(\tau_{\rm c,\star}, N_{\rm shot})$. In the Bayesian approach 100 shots were taken at each step and the prior is a continuous uniform distribution with $T_2 \in [T_{2\star}/3, 3T_{2\star}], \tau\in [\tau_{\rm c\star}/3, 3\tau_{\rm c\star}]$. Analogously, for the least-squares minimization algorithm used in the frequentist approach (trust-region reflective algorithm), we set the same parameter bounds as the ones of the uniform distribution.}
    \label{fig:det_optimal_bayesian_OU}
\end{figure}

Let us now present a detailed comparison of the  precision of the frequentist and Bayesian approaches. For the frequentist approach, we can obtain the expected covariance  of the estimator ${\rm Cov}(\boldsymbol{\hat{\theta}}_{\rm F})$ by performing several runs, and computing the covariance matrix of the results. We emphasise that this is not the asymptotic $\Sigma_{\boldsymbol{\hat{\theta}}}$ discussed previously, and does not require a very large number of measurement shots. For the Bayesian approach we obtain a posterior probability distribution after $\ell$ steps, $\pi_\ell(\boldsymbol{\theta})$, and we can directly compute the covariance matrix $\Sigma_{\ell}$ of this posterior distribution. A good measure of the uncertainty of each one of the approaches can be obtained by taking the determinant of the corresponding covariance matrix. For a Gaussian distribution, this quantity $\det\Sigma$ gives us the elliptical area associated to the bi-variate Gaussian covariance, and  we can define an average radius $\bar{R}=\sqrt{\det\Sigma}/\pi$. In order for this to scale as $1/\sqrt{N_{\rm shot}}$ and that it has the same units as the standard deviation, we will use the square root of this radius $\det \Sigma^{1/4}$. Therefore, we will compare $\det \Sigma^{1/4}$ for both frequentist and Bayesian approaches by taking the ratio of the determinants,
\beq\label{eq:det_ratio_OU}
r_{\rm OU}=\sqrt[4]{\frac{\det ( \mathrm{Cov}(\hat{\boldsymbol{\theta}}_{F,\mathrm{opt}}))}{\det \Sigma_{\ell_{\rm max}}}},
\eeq
with $\hat{\boldsymbol{\theta}}_{F,\mathrm{opt}}$ the frequentist estimator taking measurements at optimal times and considering for both approaches the same number of total measurements $N_{\mathrm{shot}}$. This ratio is represented in Fig.~\ref{fig:det_optimal_bayesian_OU} as a function of $\tau_{\rm c,\star}/T_{2\star}$. As we can see, the Bayesian approach is better for small number of measurements, since it has some prior knowledge of the parameters. As we make more measurements and the frequentist estimator keeps measuring at optimal times we get to the opposite situation. Finally, in the limit of big $N_{\mathrm{shot}}$ the Bayesian approach takes also most measurements at these optimal times and the ratio saturates to $r_{\mathrm{OU}}\approx 1$, indicating that both approaches offer a  similar precision. Let us emphasize, however, that the frequentist approach will not operate in practice at the optimal times, as these depend on the noise parameters one aims at estimating. It is also interesting to note that, as the correlation time of the noise increases, the region where the Bayesian strategy overcomes the frequentist one grows also in terms of the required number of shots. In regimes in which the time correlations are much larger than the effective  decoherence time, the Bayesian approach will always be preferred unless one can perform a prohibitively-large number of measurements.

\subsection*{Non-Markovian quantum dephasing} \label{sec:non_mark_quantum_dephasing}

Let us now move on to the discussion of L$\ell$QT for a non-Markovian dephasing dynamics. In the previous section, we have shown that a semi-classical dephasing with  OU noise, an archetype for time-correlated Gaussian random processes, yields a dephasing map that, although departing from the time-homogeneous Lindbladian case,  does not fall under the class of non-Markovian quantum dynamical maps. We have shown how both the frequentist and Bayesian approaches can learn the time-local master equation, which is parametrized in terms of an effective decoherence time $T_{2}$ and a correlation time $\tau_{\rm c}$. In this section, we focus on a quantum-mechanical dephasing noise that can actually lead to  non-Markovianity in the qubit evolution, and see how the degree of non-Markovianity affects the precision of both the frequentist and  L$\ell$QT. 

We consider an apparently mild modification of the noise PSD with respect to the OU case in Eq.~\eqref{eq:lorentzain_psd}. In particular, we use 
\beq
\label{eq:displayed_lorentzain_psd}
   S(\omega) = 4g^2_{\bar{n}}\frac{\kappa}{(\omega + \Delta_{\rm c})^2 + (\kappa/2)^2},
\eeq
which is a  Lorentzian of width $\kappa $ centered around $-\Delta_{\rm c}$, and reaching a maximum of $16g_{\bar{n}}^2/\kappa$. We note that for $\Delta_{\rm c}=0$, we recover the previous OU case~\eqref{eq:lorentzain_psd} with $\tau_{\rm c}=2/\kappa$ and $c=4g^2_{\bar{n}}\kappa$. On the other hand, for $\Delta_{\rm c}\neq 0$,  this PSD is not an even function $S(\omega)\neq S(-\omega)$, and the associated frequency noise cannot arise from a semi-classical stochastic model~\cite{RevModPhys.82.1155}. Instead, this particular PSD can be deduced from a quantum-mechanical dephasing model as discussed in the Methods section, and applied to a qubit coupled to a  dissipative bosonic mode.

In the context of superconducting circuits~\cite{PRXQuantum.1.010305}, $\Delta_{\rm c}$ is the detuning of a bosonic microwave resonator with respect to the frequency of an external driving, which is considered to be resonant with the qubit, such that $\Delta_{\rm c}=\omega_{\rm c}-\omega_0$. In addition, $g$ is a qubit-resonator cross-Kerr coupling that leads to a bosonic enhancement $g^2_{\bar{n}}= g^2\bar{n}$, where $\bar{n}$ is the average bosonic occupation of the driven resonator,  and $\kappa$ is the rate of spontaneous emission/loss of photons into the electromagnetic environment. We note that a similar dynamics can be engineered in a two-ion crystal, in analogy to~\cite{PhysRevLett.110.110502}, such that one of the ions encodes the qubit in a pair of ground state/metastable levels, while the other one is continuously Doppler cooled via a laser that is red-detuned with respect to a dipole-allowed transition. This laser then drives the carrier and motional sidebands, and effectively laser cools the common vibrational modes, one of which will play the role of the above dissipative bosonic mode, such that  the above $\omega_{\rm c}$ will now be its  vibrational frequency. The role of the above $\kappa$ is then played by the laser cooling rate, and the phonon population in the steady state $\bar{n}$ depends on  the difference of laser cooling and heating processes~\cite{PhysRevA.46.2668}, which can be controlled by the Rabi frequency and detuning of the laser that drives the dipole-allowed transition. The dissipative phonons will then act  as an effective Lorentzian bath for the qubits~\cite{PhysRevLett.110.110502,Cormick_2013,Bermudez_2016}. We consider the qubit to be subjected to a far-detuned sideband coupling, which induces a second-order cross-Kerr coupling of strength $g$ describing a phonon-dependent ac-Stark shift on the qubit levels. 

In any of the two architectures  discussed, when the coupling between the bosonic mode and the qubit is weaker than the dissipative rate $g_{\bar{n}}\ll\kappa$, one can truncate the cumulant expansion of a time-convolutionless master equation of the qubit at second order such that, after tracing over the driven-dissipative mode in its stationary state, one  arrives at a time-local dephasing master equation of the form given in Eq.~\eqref{eq:TCL_dephasing}. This  master equation will be controlled by an auto-correlation function for
the  bath operator $B(t) = 2g(a^\dagger a - \bar{n})$, following the notation used below Eq.~\eqref{eq:deph_rate_autocorr} and  in the Methods section. In particular, making use of the quantum regression theorem~\cite{Carmichael1993Open}, this auto-correlation can be expressed as
\begin{equation}\label{eq:autocorrelation_g_k_d}
    C(t-t') = 4 g^2_{\bar{n}} \ee^{-\frac{\kappa}{2} |t-t'|} \ee^{-\mathrm{i}\Delta_{\rm c} (t-t')},
\end{equation}
which coincides with the OU auto-correlation function in Eq.~\eqref{eq:OU_corr} when $\Delta_{\rm c}=0$. Being wide-sense stationary, one can  Fourier transform this function as shown in Eq.~\eqref{eq:psd},   leading  to the  displaced Lorentzian PSD in Eq.~\eqref{eq:displayed_lorentzain_psd}.  Following Eq.~\eqref{eq:deph_rate_psd}, one can obtain the following time-dependent decay rate
\beq
\label{eq:gamma_displaced}
\gamma(t) = \frac{1}{4} S(0) \left( 1 -  \ee^{-\frac{\kappa}{2} t} \Bigr(\cos \Delta_{\rm c} t 
     + \frac{2\Delta_{\rm c}}{\kappa} \sin \Delta_{\rm c} t \Bigl)\right).
\eeq
In comparison to Eq.~\eqref{eq:decay_rate_OU},  this decay rate presents additional  oscillatory terms for $\Delta_{\rm c}\neq 0$ that will play an important role for  the non-Markovianity of the quantum dynamical map.
The attenuation factor  that controls the decay of the coherence is
\begin{align}\label{eq:chi_g_k_d}
    \Gamma(t) = & \frac{S(0)}{2} \bigg( t+ \frac{2_{\phantom{_c}\!\!}}{\kappa_{\phantom{_c}\!\!}}\frac{(2\Delta_{\rm c}/\kappa)^2-1}{(2\Delta_{\rm c}/\kappa)^2+1}     +\frac{2}{\kappa} \ee^{-\frac{\kappa}{2} t} \cos\varphi(t)  \bigg),
\end{align}
where we have introduced $\varphi(t)=\Delta_{\rm c} t-2\arctan \frac{2\Delta_{\rm c}}{\kappa}$.

\begin{figure}
    \includegraphics[width=0.8\columnwidth]{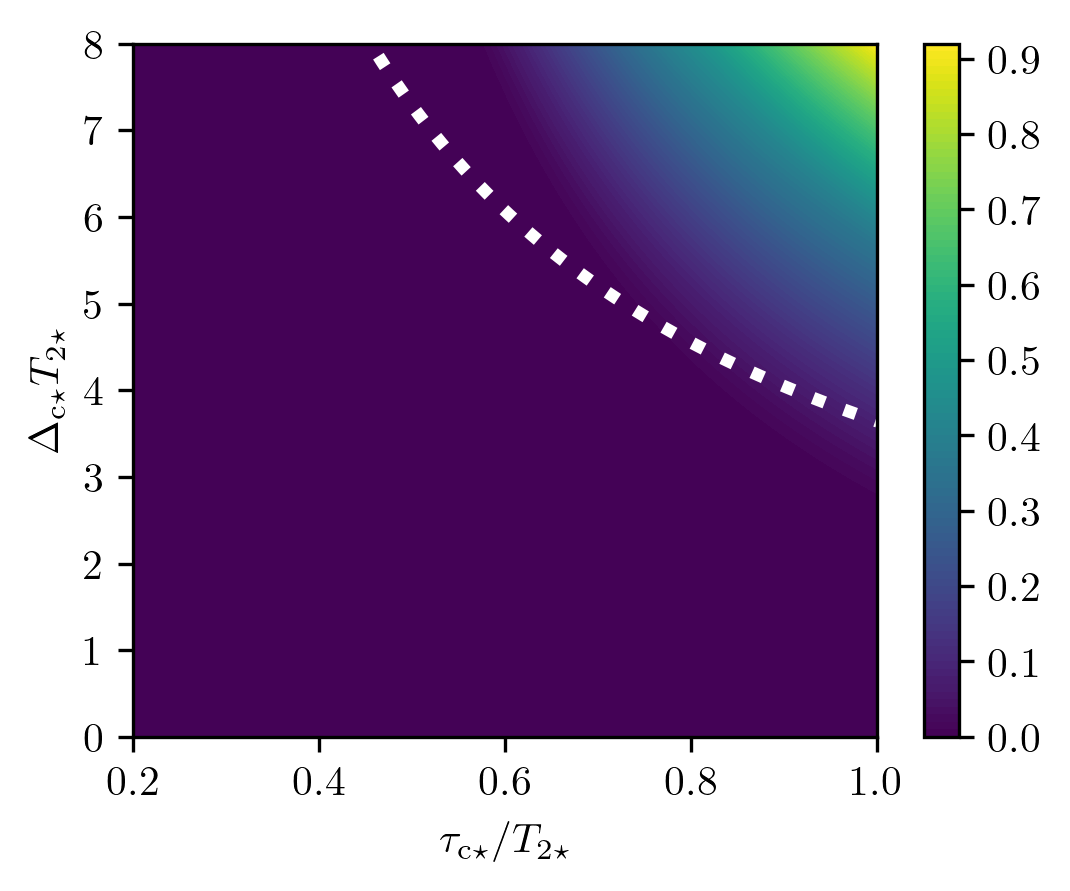}
    \caption{{\bf Non-Markovianity measure for the quantum dephasing map:} non-Markovianity measure of Eq.~\eqref{eq:non_markovian_measure_cp} for different values of $\tau_{c}=2/\kappa$ and $\Delta_{\rm c}$. The dashed line signals the limit between the Markovian and the non-Markovian regions and corresponds to the equation $\Delta_{\rm c} \approx 1.82 \kappa$. The non-Markovian region is located in the upper right corner.}
    \label{fig:non_mark_measure}
\end{figure}

Recalling that we can assign a correlation time to this noise by $\tau_{\rm c}=2/\kappa$, one would expect to recover a Markovian Lindbladian description for  $t\gg\tau_{\rm c}$. In this limit,  the term linear in $t$ in Eq.\ \eqref{eq:chi_g_k_d} is the dominant one, which leads to a time-homogeneous exponential decay of the Ramsey probabilities $p^{\rm TL}_{i}(m_x | \boldsymbol{\theta})\approx\big(1+(-1)^{m_x}\ee^{-t_i/T_2}\big)/2 $ with an associated decoherence time $T_2 =2/S(0)= ({(\kappa/2)^2+\Delta_{\rm c}^2})/{2g^2_{\bar{n}}\kappa}$. As in the OU case, for shorter times, the memory effects will start playing a bigger role in the qubit dynamics, such that the Ramsey decay is no longer a time-homogeneous exponential. Moreover, in this particular case, these memory effects can give rise to a non-Markovianity that can be understood as a backflow of information from the environment into the system. 
According to Eq.~\eqref{eq:non_markovian_measure_cp} or~\eqref{eq:N_TD}, non-Markovianity  occurs when the decay rate takes negative values $\gamma(t)<0$. This can only happen if the second contribution in Eq.~\eqref{eq:gamma_displaced} dominates over the first one, which cannot happen if $\Delta_{\rm c}=0$. Since this contribution is suppressed by $\ee^{-\kappa t/2}$, we will need the frequency of the oscillations $\Delta_{\rm c}$ to be sufficiently large  in comparison to $\kappa$, such that one can get a non-vanishing  degree of non-Markovianity. In Fig.~\ref{fig:non_mark_measure}, we depict the measure of non-Markovianity of Eq.~\eqref{eq:non_markovian_measure_cp} as a function of $\Delta_{\rm c}$ and $\tau_{\rm c}=2/\kappa$. We see that the  parameter regime  $\Delta_{\rm c} \gtrsim 1.82 \kappa$ (white dashed line) is where $\gamma(t)$ can become negative  at some time during the  evolution, leading to larger  non-Markovianity as both $\Delta_{\rm c}$ and $\tau_{\rm c}$ are further increased. 

Let us now discuss the statistical inference for the L$\ell$QT of this non-Markovian dephasing map, and
compare the frequentist and Bayesian approaches to the statistical estimation.

\textit{a. Frequentist Ramsey estimators.} As in the  OU case,  we need to minimize $ \mathsf{C}^{\rm pd}_{\mathrm{TL}}(\boldsymbol{\theta})$ in Eq.~\eqref{eq:ML_cost_dephasing}, where the likelihood function $p^{\rm TL}_{i}(m_x | \boldsymbol{\theta})=\big(1+(-1)^{m_x}\ee^{-\Gamma(t_i)}\big)/2$  now depends on  the  new attenuation factor in Eq.~\eqref{eq:chi_g_k_d}.  Since  there are three noise parameters $\boldsymbol{\theta}_{\star} = (g^2_{\bar{n}\star}, \kappa_\star, \Delta_{{\rm c}\star})\in\Theta=\mathbb{R}^3_+$,  we shall at least  need to measure  at three different times. To assess the performance of the frequentist approach  under shot noise, we numerically generate the   relative frequencies $\tilde{f}_{1}(m_x\,|\boldsymbol{\theta}_\star),\tilde{f}_{2}(m_x\,|\boldsymbol{\theta}_\star),\tilde{f}_{3}(m_x\,|\boldsymbol{\theta}_\star)$ at these times by sampling the probability distribution with the real noise parameters $\boldsymbol{\theta}_{\star}$ a number of times $N_{\rm shot}=N_1+N_2+N_3$. In order to find the three optimal times, we minimize the  determinant of the covariance matrix in Eq.~\eqref{eq:opt_times}, which will have an similar expression as Eq.~\eqref{eq:cov_matrix}, but now expressed in terms of $3\times3$ matrices for the underlying trivariate normal distribution.  
Since we are optimizing $t_1$, $t_2$ and $t_3$, one may wonder if we also could improve the estimation by redistributing the total number of measurements differently at each of these times. However, according to our prescription in which the imprecision is quantified by the long-run Gaussian PDF, which defines an elliptical volume in this case, one gets $\det\Sigma_{\boldsymbol{\hat{\theta}}}(t_{1, \rm opt},t_{2, \rm opt},t_{3, \rm opt})\propto {1}/{N_{1}N_{2}N_{3}}$. Therefore,  the number of measurements must be equally distributed  between the three different times $N_{1}=N_{2}=N_{3}=N_{\rm shot}/3$.

\begin{figure}
    \includegraphics[width=0.85\columnwidth]{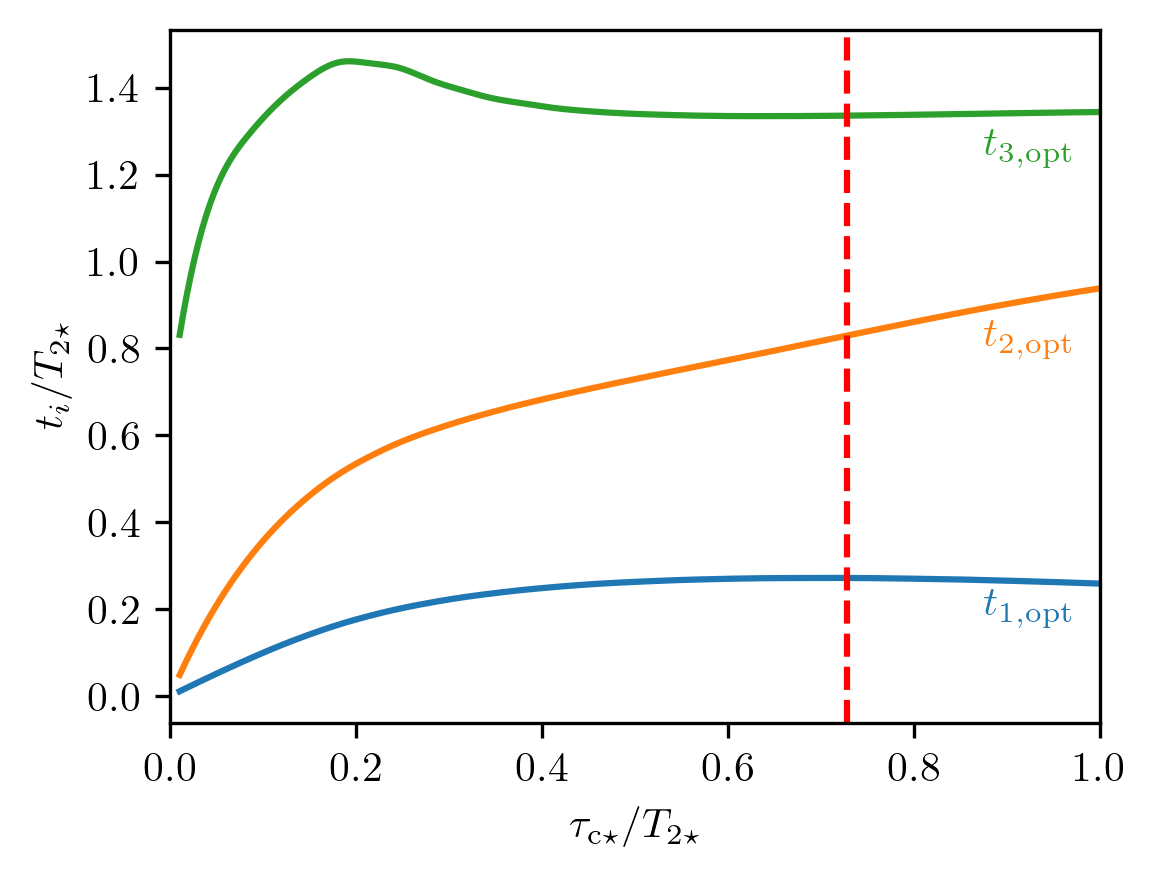}
    \caption{{\bf Optimal times for quantum dephasing L$\ell$QT:} We represent the three optimal times $(t_{1, \rm opt},t_{2, \rm opt},t_{3, \rm opt})=\texttt{argmin}\{\det(\Sigma_{\boldsymbol{\hat{\theta}}}(t_1,t_2,t_3))\} $ as a function of the ratio of the  real noise parameters $\tau_{\rm c\star}/T_{2\star}$, setting $\Delta_{{\rm c}\star}=5/T_{2\star}$. The vertical red dashed line separates the Markovian regime (left) from the non-Markovian one (right). On the Markovian regime, as $\tau_{{\rm c}\star}\ll T_{2\star}$, we see that the $t_1,t_2$ in blue and orange tend to zero, while $t_3$ in green tends to the optimal value of the pure Lindbladian case $t_3\approx 0.797 T_{2\star}$. On the other hand, as $\tau_{{\rm c}\star}$ increases and memory effects become more relevant, the three optimal times start to depart from each other. In the  non-Markovian regime, the optimal time $t_{2, \rm opt}$ is closer to $t_{3, \rm opt}$ than it is to $t_{1, \rm opt}$. Moreover, this optimal $t_{1, \rm opt}$ time starts to decrease as $\tau_{{\rm c}\star}$ goes deep into the non-Markovian regime.}
    \label{fig:optimal_times_g_k_d}
\end{figure}

The minimization in Eq.~\eqref{eq:opt_times}  then  yields the optimal times $\{t_{1,{\rm opt}},t_{2,{\rm opt}},t_{3,{\rm opt}}\}$ shown in Fig.\ \ref{fig:optimal_times_g_k_d}, which have been represented as a function of $\tau_{{\rm c}\star}/T_{2\star}$ for a fixed value of $\Delta_{\rm c\star}$. The non-zero value of the later is responsible for the fact that, for $\tau_{{\rm c}\star}\gtrsim 0.73 T_{2\star}$, we cross the red-dashed line and enter a non-Markovian regime in which the effective dephasing dynamics is no longer CP-divisible.  In this non-Markovian regime, the optimal times tend to be close to the local maxima of $\tilde{f}_{i}(m_x|\boldsymbol{\theta}_{\star})$,  providing a large amount of information about the dephasing noise. Conversely, deep in the Markovian regime  $\tau_{\rm c\star}\ll T_{2\star}$, it suffices to measure at $t_{3,{\rm opt}}\approx 0.797T_{2\star}$ in order to determine $T_{2\star}$, in agreement with the analytical Lindblad result, while $t_{1,{\rm opt}}$ and $t_{2,{\rm opt}}$ tend to zero and would be used to determine the two other noise parameters.

\begin{figure}
    \centering
    \includegraphics[scale=.8]{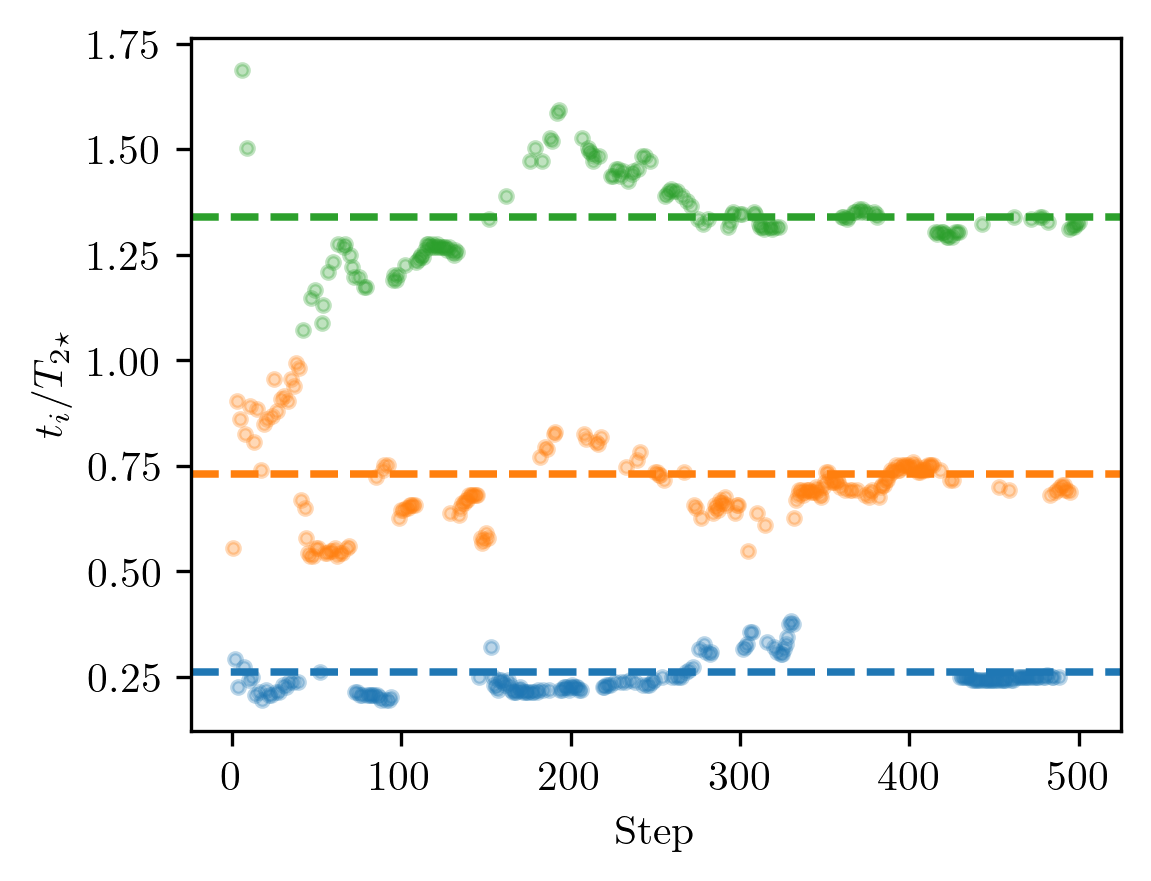}
    \caption{{\bf Bayesian optimal measurement time at each step:} We consider the non-Markovian quantum dephasing  with noise parameters  $\tau_{\rm c,\star} = T_{2\star}/2$ and $\Delta_{\rm c\star}=5/T_{2\star}$. We consider as initial prior a continuous uniform distribution with $g_{\bar n}^2 \in [g_{\bar{n} \star}^2/3, 3g_{\bar{n}\star}^2], \kappa\in [3\kappa_{\star}, \kappa_{\star}/3], \Delta_{\rm c}\in [3\Delta_{\rm{c}\star}, \Delta_{\rm{c}\star}/3]$. At each Bayesian step, we increase the data set with $|\delta\mathbb{D}_\ell|=50$ new outcomes. As the Bayesian protocol proceeds, we see that the Bayesian update times cluster in three groups of data, which lie around the optimal times $t_{1,{\rm opt}},t_{2,{\rm opt}},t_{3,{\rm opt}}$ of the frequentist approach, which are represented by dashed lines.} \label{fig:bayesian_experiment_optimal_times_nm}
\end{figure}

\textit{b. Bayesian Ramsey estimators.} Let us now move to the Bayesian approach, where we have some prior knowledge as shown in Eq.~\eqref{eq:posterior} of the parameters $\boldsymbol{\theta}=(g^2_{\bar{n}},\kappa,\Delta_{\rm c})$ that gets updated at each Bayesian step $t_{\ell}$ by enlarging the data set with  $\delta\mathbb{D}_\ell$. Minimizing the relative entropy between the prior and the posterior in Eq.~\eqref{eq:optimal_time_information_gain}, we obtain the subsequent evolution time $t_\ell$, and update our knowledge about the noise parameters in the best possible way. As shown in Fig.~\ref{fig:bayesian_experiment_optimal_times_nm}, the Bayesian procedure starts by using evolution update times that are scattered in a broad range of values. However, as the number of iterations increases and we gain more knowledge, they tend to cluster around three well-defined times. In fact, as shown by the corresponding dashed lines, these times coincide with the  optimal measurement times of the frequentist approach  in Fig.~\ref{fig:optimal_times_g_k_d}. 
Therefore, if we start a Bayesian experiment with some prior, and we let the experiment run for sufficiently long number of steps,  we will learn the optimal  times automatically. 

\begin{figure}
    \includegraphics[width=0.85\columnwidth]{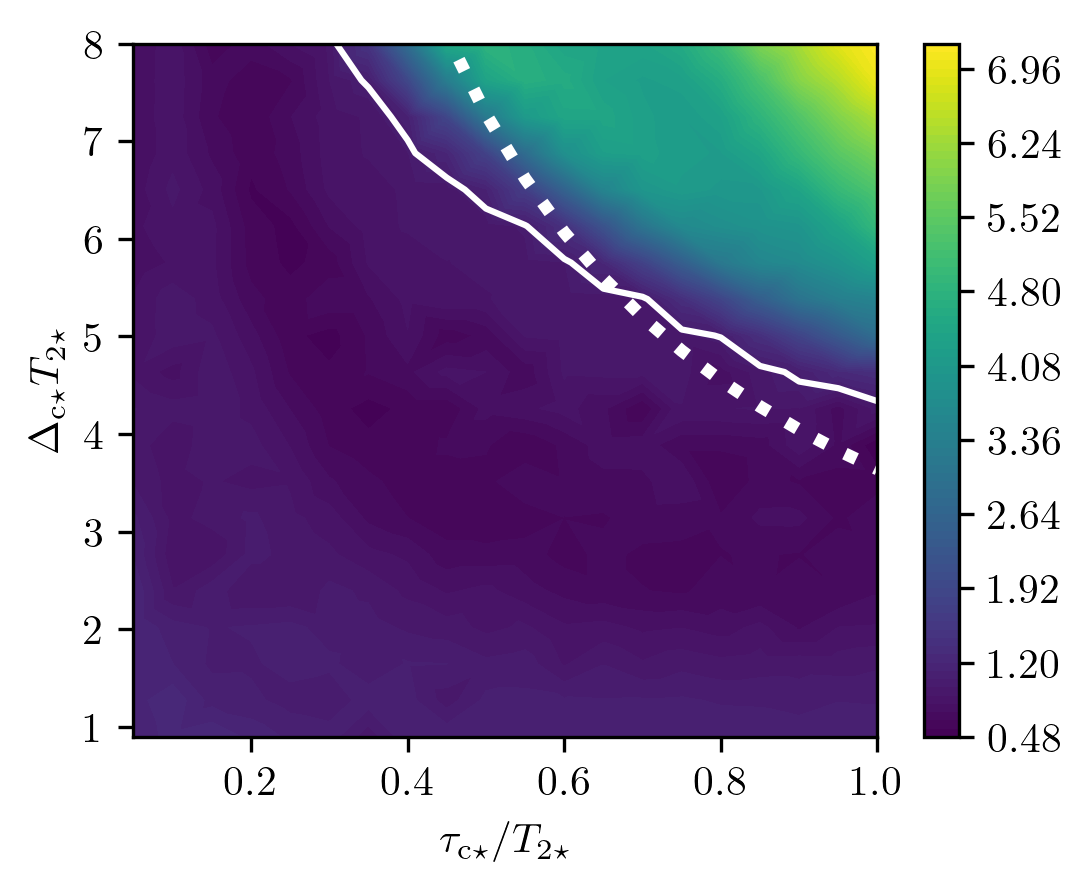}
    \caption{{\bf Ratio between determinants of covariance matrices of frequentist and Bayesian approaches:} The ratio of Eq.~\eqref{eq:det_ratio_non_markovian} computed for different parameter values $\tau_{\mathrm{c}\star}$ and $\Delta_{\mathrm{c}\star}$, and a fixed number of total measurements $N_{\rm shot}=2\cdot10^4$. The dashed white line separates the Markovian from the non-Markovian regime, with the non-Markovian region located in the upper right corner. The continuous white line indicates the level where the ratio equals 1. 5000 frequentist runs were done to estimate $\det \mathrm{Cov(\hat{\boldsymbol{\theta}}_{\mathrm{F, opt}})}$, while 30 Bayesian runs were done to estimate $\det \Sigma_{\ell_{\rm max}}$ for each point $(\tau_{\rm c\star}, \Delta_{\rm c\star})$. In the Bayesian approach 100 shots were taken at each step and the prior is a continuous uniform distribution with $\boldsymbol{\theta} \in [\boldsymbol{\theta}_{\star}/3, 3\boldsymbol{\theta}_{\star}]$. Analogously, for the least-squares minimization  used in the frequentist approach (trust-region reflective algorithm), we set the same parameter bounds as the ones of the uniform distribution.}
    \label{fig:det_optimal_bayesian}
\end{figure}

In order to compare  the  precision of the frequentist and Bayesian approaches, we proceed in analogy to the  OU noise by looking for a parameter that captures the relative precision of the two approaches in Eq.~\eqref{eq:det_ratio_OU}. We now have to consider that the determinant of the trivariate Gaussian covariance $\det\Sigma$ is a volume, and  we can define an average radius as $\bar{R}=\sqrt[3]{\det\Sigma}/(4\pi/3)$. We can quantify the precision  by taking the square root of this radius, which scales like a standard deviation. Altogether, the relative precision of  the frequentist and Bayesian  approaches is defined by the ratio 
\beq \label{eq:det_ratio_non_markovian}
r_{\rm NM}=\sqrt[6]{\frac{\det (\mathrm{Cov(\hat{\boldsymbol{\theta}}_{\mathrm{F, opt}})})}{\det \Sigma_{\ell_{\rm max}}}}.
\eeq
Here, $\hat{\boldsymbol{\theta}}_{F,\mathrm{opt}}$ is the frequentist estimator taking measurements at the optimal times, and we consider the same number of total measurements $N_{\mathrm{shot}}$ for both approaches. This ratio is represented in Fig.~\ref{fig:det_optimal_bayesian} for $N_{\rm shot}=2\times 10^4$, which is still far from the asymptotic regime of large $N_{\rm shot}$ where one would obtain   $r_{\rm NM}\approx 1$ in similarity to the results found for the OU noise in Fig.~\ref{fig:det_optimal_bayesian_OU}. Instead of exploring how this ratio changes with the number of measurement shots, we are here interested in understanding how the degree of non-Markovianity can affect the performance of the two estimation strategies. We thus set $N_{\rm shot}=2\times 10^4$, since the variance is already good enough but the cost in terms of number of measurements is still not too big, and plot the precision ratio as a function of the real noise parameters. As we can see in Fig.~\ref{fig:det_optimal_bayesian}, the blue region represents a regime in which the frequentist approach is slightly better than the Bayesian one, and coincides with the regime of Markovian dephasing that is delimited by the dashed white line. The continuous white line marks the ratio contour line with $r_{\rm NM}=1$, and thus delimits the part of the blue region  in which the frequentist approach with optimal times is preferable. In the green and yellow areas, which coincide with the  non-Markovian regime, the  Bayesian approach becomes preferable and the advantage can actually be  quite significant. As we go deeper in the non-Markovian regime, the oscillating term in Eq.~\eqref{eq:chi_g_k_d} becomes bigger and the decay of the coherence exhibits an increasing number of local maxima. These local maxima represent times that provide a significant amount of information in terms of the Kullback-Leibler divergence of Eq.~\eqref{eq:optimal_time_information_gain}. Therefore, the presence of more local maxima in the non-Markovian regime makes it easier for the Bayesian method to, even if the prior information is minimal, select a time as useful as the asymptotically optimal times.

 It is also useful to quantify the advantage of one estimator with respect to the other in terms of number of measurements that is required to reach a target precision, which amounts to reaching the same value of the above covariance determinant. The values shown for the ratio of the determinants in Fig.\ \ref{fig:det_optimal_bayesian} can be converted into the ratio of number of measurements by assuming that both covariance determinants in Eq.~\eqref{eq:det_ratio_non_markovian}  scale as $1/\sqrt{N_{\rm shot}}$ even in the non-asymptotic regime. Although there can be corrections to this scaling, this can give us in most cases an idea of the proportion of measurement shots one can save by using the best estimator. With this assumption, we get $N_{\rm shot, F}/N_{\rm shot, B} \approx r_{\rm NM}^2$, with $r_{\rm NM}$ the ratio of the determinants of two different approaches. Thus, if for instance $r_\mathrm{\rm NM} = 1.5$, we obtain that $N_{\rm shot, F}/N_{\rm shot, B} \approx 2.25$. Therefore, we get more than a $2\times$ reduction in the number of  measurements with the Bayesian approach in comparison to the frequentist one to reach the same precision. Going back to the values of Fig.~\ref{fig:det_optimal_bayesian}, we see that the Bayesian approach can result in a considerable improvement as one goes deep in the non-Markovian limit. Before closing this subsection, it is worth recalling once more that, in practice, the frequentist estimation will never be performed at the three optimal times, and the advantage of the Bayesian approach can be even bigger. In Sec.~\ref{app:uniform_estimator} of the Supplementary Material, we present a detailed comparison of these estimators with another one in which the shots are evenly distributed between the measurements after evolution times that cover uniformly the whole time interval $T$. As discussed in Sec.~\ref{app:uniform_estimator} of the Supplementary Material, the Bayesian approach is preferable for most of the parameter values, and can again show a big advantage as one enters into the non-Markovian regime.

\section*{\bf Discussion}\label{sec:conclusion}

We have presented L$\ell$QT, a new tool designed to characterize non-Markovian dephasing noise in QIPs, building upon the established framework of Lindblad quantum tomography. L$\ell$QT extends the applicability of Lindblad learning to scenarios where temporal correlations and non-Markovian dynamics play a significant role. In particular, it allows us to extend the characterization of the  generators of  quantum dynamical maps that go beyond the time-homogeneous Lindblad limit, which connect to a time-local master equation that can display  negative decay rates in certain time intervals  and, thus, strictly non-Markovian quantum evolutions. Through a detailed comparative study, both frequentist and Bayesian approaches to L$\ell$QT are presented, offering insights into the accuracy and precision of noise  estimation under different  conditions.

By focusing on the time-correlated dephasing quantum dynamical map of a single qubit, we show that L$\ell$QT can be formally expressed as a parameter estimation process, which simplifies the most general learning scheme to a single initial state and a single measurement basis. In particular, the problem reduces to a time-correlated Ramsey estimator for a parametrized decay rate, which depends on the noise parameters via a filtered power spectral density  of the noise.   In the frequentist approach, the focus lies on optimizing measurement times to reduce the number of necessary measurements while minimizing error in parameter estimation. By leveraging statistical inference techniques, the frequentist approach provides valuable insights into the efficiency and effectiveness of L$\ell$QT, particularly in scenarios with varying degrees of temporal correlations and non-Markovianity. Conversely, the Bayesian approach offers a more dynamic and adaptive framework, allowing for the incorporation of prior knowledge and iterative updates to refine noise estimates over time.

We have compared the performance of both approaches for two different dephasing quantum dynamical maps, either for a semi-classical or for a quantum-mechanical noise model. In both cases, the microscopically-motivated parametrization allows one to interpolate between a fully Markovian Lindblad limit, for which we derive analytical solutions for the optimal estimation, and a time-correlated and even non-Markovian regime which require a different distribution of the optimal and Bayesian measuring times. Interestingly, in the quantum-mechanical dephasing model, which can be obtained from a microscopic model of a qubit that is coupled to a dissipative bosonic mode in both superconducting-circuit and trapped-ion architectures,  the best of the two approaches  depends on whether we are in the Markovian or the non-Markovian regime. The Bayesian approach yields much better results in the non-Markovian regime, showing that it is able to automatically adapt to the particularities of the non-Markovian evolution to make much better estimations with a limited number of shots. Moreover, we also compare to more standard schemes considered in the context of Lindbladian quantum tomography, in which the measurements are distributed uniformly  (see Sec.~\ref{app:uniform_estimator} of the Supplementary Material). In this case we show an advantage of our schemes that again becomes more appreciable in the non-Markovian regime for the Bayesian approach. However, we also find that for the frequentist approach, the optimal distribution of measurements does not always outperform the uniform distribution when the number of measurments is small and the asymptotic regime has not yet been reached. These results may suggest that the observed patterns  extend to other similar cases, but further exploration would be required to confirm this.

 The semi-classical dephasing model we have discussed has also been considered in the context of central spin models, where the central spin/qubit is dipolarly-coupled to an ensemble of environmental spins that play the role of a mesoscopic bath. For the case of NV centers in diamond \cite{doi:10.1126/science.1192739}, the bath corresponds to   nearby substitutional nitrogen atoms, also known as P1 centers, which have a much smaller splitting than the qubit. Due to the mismatch, energy exchange processes can be neglected, and only longitudinal ones survive leading to a time-correlated dephasing. For these systems, assuming incoherent dynamics of different bath spins and negligible back action is a good approximation, such that the above semi-classical model with an Ornstein-Uhlenbeck process turns out to be a reasonably-good approximation. More generally, if the bath consists of weakly-interacting nuclear spins that get dipolarly-coupled to the qubit, bath correlations might be relevant. Although this is not generally valid, when the leading bath correlations can be reduced to pairs of spins, a similar Gaussian semi-classical model can still be used \cite{PhysRevB.79.245314}, albeit considering other PSDs. In more general situations, one should consider larger groups of spins, and explore other spin-bath dephasing models. Although this goes beyond the scope of this work, we believe that incorporating a detailed microscopically-motivated parametrization of this type of non-Markovian dephasing should allow for a more efficient parametrization of the noise that could be incorporated in a L$\ell$QT  that is similar in spirit to the one considered in this work.

Future research shall explore the extension of these non-Markovian  characterization techniques to larger quantum systems, combining the effect of spatial and temporal correlations. More importantly, our work sets the stage to generalize to  more complex  situations beyond pure dephasing, specially focusing on scalability and robustness, and eventually targeting the noise in full universal gate sets of  QIPs.

\section*{\bf Methods}

\subsection*{Time-local master equation for pure dephasing}
\label{app:pure_dephasing}

For the sake of completeness, we present here a derivation  of the time-local master equation in Eq.~\eqref{eq:TCL_dephasing} for a qubit subjected to time-correlated dephasing noise, both in a semi-classical and a fully quantum-mechanical model. This serves to introduce well-known concepts and set our notation following~\cite{velazquez2024dynamical}. 

\textit{a. Semi-classical time-correlated dephasing.-} The qubit evolves under a stochastic rotating-frame Hamiltonian
\beq
\label{eq:stoch_H}
	\tilde{H}(t) =  \frac{1}{2}\delta\tilde{\omega}(t) \sigma_z, 
\end{equation}
where $\delta\tilde{\omega}(t)$ is the detuning of  the qubit with respect to the frequency of a driving used in the initialization/measurement stages with respect to, and we have set $\hbar=1$. We use a tilde to highlight the  
random nature of  $\delta\tilde{\omega}(t)$, 
which is modeled as a stochastic process with zero mean $\mathbb{E}[\delta\tilde{\omega}(t)]=0$, thus  assuming that the driving frequency is resonant with the qubit transition  on average. We recall that the averages  are taken with respect to the underlying joint PDF of the process for any finite set of times $p_{\delta\tilde{\omega}}(\delta\boldsymbol{\omega})= p_{t_1,t_2,\cdots, t_n}(\delta\omega_1,\delta\omega_2,\cdots, \delta\omega_n)$, $\forall n\in\mathbb{Z}^+: \delta\omega_n=\delta\omega(t_n)\in\mathbb{R}, \, \{t_i\}_{i=1}^n\in T$, which fulfills the conditions  $p_{\delta\tilde{\omega}}(\delta\boldsymbol{\omega})\geq 0$ and $\int\prod_n{\rm d}\delta\omega_np_{\delta\tilde{\omega}}(\delta\boldsymbol{\omega})=1$
~\cite{vankampen2007spp,gardiner2004handbook}. Physically, these stochastic fluctuations can either stem from frequency/phase noise of the drive, or from additional external fields that shift the energy of the qubit. 
 For each individual trajectory of the noise $\delta\tilde{\omega} (t)$, the  evolution of an initial qubit state $\rho_0=\ket{\psi_0}\!\bra{\psi_0}$ in the rotating frame is purely unitary but random, giving rise to $\tilde{\rho}(t)$, and expectation values will thus depend on  stochastic averages $\mathbb{E}$,  leading to a completely-positive trace-preserving (CPTP)  map after averaging $\rho(t)=\mathbb{E}[\tilde{\rho}(t)]=\mathcal{E}_{t,t_0}({\rho}_0)$~\cite{nielsen_chuang_2010,KRAUS1971311,CHOI1975285,watrous_2018}.
The corresponding stochastic differential equations  are 
\beq
\label{eq:LVN}
\frac{\rm d\tilde{\rho}}{{\rm d}t}=\tilde{\mathcal{L}}_t(\tilde{\rho}(t)), \hspace{1ex}\tilde{\mathcal{L}}_t(\bullet)=-\ii[\tilde{H}(t),\bullet\,],
\eeq
where one sees that the noise $\delta\tilde{\omega}(t)$ thus enters multiplicatively.
Using the Nakajima-Zwanzig~\cite{Breuer2002}   projection operators  $\mathcal{P}=\mathbb{E}$, and $\mathcal{Q}=1-\mathcal{P}$, we can find differential equations for the averaged density matrix using
\beq
\label{eq:proj}
\rho(t)=\mathcal{P}(\tilde{\rho}(t)):\hspace{2ex}\mathcal{Q}(\rho_0)=0=\mathcal{P}(\tilde{\mathcal{L}}_t).
\eeq
In fact, this averaged evolution can be written as a time-local master equation~\cite{Chaturvedi1979,TERWIEL1974248,zoller_course}, namely
\beq
\label{eq:TCNZ_eq}
\frac{\rm d{\rho}}{{\rm d}t}=\mathcal{K}\!(t)\rho(t),
\eeq
where  $\mathcal{K}\!(t)$ is the so-called  time-convolutionless kernel that encapsulates the effects that the finite memory of  the time-correlated noise has on the qubit. In particular, this kernel can be expressed as follows
\beq
\label{eq:TCL_kernel}
\mathcal{K}\!(t)=\mathcal{P}\tilde{\mathcal{L}}_t(1-\tilde{\Sigma}(t))^{-1},
\eeq
where we have 
used a super-operator playing the role of a `self-energy', which can be expanded as
\beq
\label{eq:power_expansion}
\tilde{\Sigma}(t)=\sum_m\alpha^m\tilde{\Sigma}_m(t),\hspace{1ex}\mathcal{K}\!(t)\!=\sum_{n}\mathcal{P}\tilde{\mathcal{L}}_t\!\left(\!\sum_m\alpha^m\tilde{\Sigma}_m(t)\!\!\right)^{\!\!n}\!\!\!.
\eeq
In this way, the kernel is organised in 
a power series of  a  microscopic coupling  $\alpha$ that characterizes the order of magnitude of the coupling of the system to the external  noise, and Eq.~\eqref{eq:proj} can be used to show that only even terms contribute
\beq
\label{eq:power_series}
\mathcal{K}\!(t)=\sum_{n}\mathcal{K}_{2n}(t).
\eeq

This series agrees with the Kubo and Van Kampen cumulant expansion~\cite{doi:10.1143/JPSJ.17.1100,VANKAMPEN1974215}, and one finds that 
the $n$-th order    term can be expressed in terms of $n-1$ nested time-ordered integrals~\cite{Breuer2002}, being the lowest-order contribution $\mathcal{K}_2(t)=\int_{0}^t\!{\rm d}t'\mathcal{P}(\tilde{\mathcal{L}}_t\tilde{\mathcal{L}}_{t'})$. This term is  controlled by the  auto-correlation  of the stochastic process, leading to
\beq
\label{eq:dephasing_tcl}
  \frac{{\rm d}{ \rho}}{{\rm d}t}  =
    \frac{1}{4} \int_0^t\!{\rm d}t' \left(C(t,t') + C(t', t) \right) \big( \sigma^z  \rho(t)  \sigma^z -  \rho(t)  \big), 
\eeq 
which, for wide-sense stationary processes, can be expressed in terms of the   PSD of the stochastic process
\beq
\label{eq:wide_sense}
C(t-t')=\mathbb{E}[\delta\tilde{\omega}(t)\delta\tilde{\omega}(t')]=\int_{-\infty}^{\infty}\!\!\frac{{\rm d}\omega}{2\pi}\,S(\omega)\,\ee^{\ii\omega(t-t')}.
\eeq

We note that for any wide-sense stationary classical noise, the PSD is even $S(\omega)=S(-\omega)$~\cite{RevModPhys.82.1155}, and $C(t,t')=C(|t-t'|)=C(t',t)$ such that the symmetrized autocorrelation function and the symmetrized PSD introduced below Eq.~\eqref{eq:f_gamma} already contain all of the required information for a second-order approximation. The truncation at this order is justified by first noting that
the autocorrelation is typically concentrated within  $|t-t'|\leq {\tau}_{c}$, where ${\tau}_{\rm c}$ is a characteristic correlation time. Due to the cluster property~\cite{TERWIEL1974248,10.1063/1.523041}, one finds that $\mathcal{K}_2(t)\sim\alpha\zeta$ with $\alpha=\sqrt{C(0)}$ and    a small  parameter 
 \beq
 \label{eq:small_param_stand}
 \zeta=\alpha{\tau}_{\rm c}=\sqrt{\frac{\tau_{\rm c}}{T_2}},
 \eeq
 where we have defined a characteristic  time as $T_2=2/S(0)$.
The cluster property for the higher $n$-th order contributions, which have $(n-1)$ nested integrals, states that the corresponding kernels scale with  $\mathcal{K}_n(t)\sim \alpha\zeta^{n-1}$, justifying a low-order truncation whenever the condition $\zeta\ll 1$ is met. This is known as a fast-fluctuation expansion and,   back from the rotating frame, yields the time-local master equation in Eq.~\eqref{eq:TCL_dephasing}.

Let us note that the above  truncation rests on the importance of the memory effects $\tau_{\rm c}$ within the $T_2$ time. As  discussed in more detail in the ``Results'', this $T_2$ time  controls the time scale for the decay of coherences $\langle\sigma_x(t)\rangle\approx\ee^{-t/T_2}$ in a  long-time Lindbladian limit $t\gg\tau_{\rm c}$. However,  for shorter times, the structure of the noise can actually lead to deviations from this  limit, leading to a coherence decay that is not exponential. As emphasized in the main text, this is not an univocal signal of non-Markovianity for the qubit evolution. We note that there is an exception to the $\tau_{\rm c}\ll T_2$ requirement for Gaussian random processes, which are defined by a joint PDF that  is a multivariate normal distribution for any set of times. In this case, the time-local master equation in Eq.~\eqref{eq:TCL_dephasing}  is actually an exact result, independently of the value of $\zeta$. In fact, the higher-order contributions  to the kernel~\cite{Breuer2002}  vanish identically $\mathcal{K}_n(t)=0, \forall n>2$, due to Isserlis' theorem, most commonly referred to as Wick's theorem in the context of physics $\mathbb{E}[\delta\tilde{\omega}(t_1)\delta\tilde{\omega}(t_2) \cdots \delta\tilde{\omega}(t_{n})]=\sum_{\sigma\in S_n} \mathbb{E}[\delta\tilde{\omega}(t_{\sigma(1)})\delta\tilde{\omega}(t_{\sigma(2)})]\cdots\mathbb{E}[\delta\tilde{\omega}(t_{\sigma(n-1)}) \delta\tilde{\omega}(t_{\sigma(n)})]$, where ${S}_n$  is the group of all possible permutations of $n$ elements, e.g. $\sigma(1,2,\cdots,n-1,n)=(n,1,2,\cdots, n-1)$. 

 \vspace{1ex}
\textit{b. Quantum-mechanical time-correlated dephasing.-}
We consider a single qubit coupled to an environment, and evolving ${\rm d }\rho_{\rm SB}/{\rm d}t=\mathcal{L}_t(\rho_{\rm SB})$ under the following Liouvillian 
\begin{equation}\label{eq:basic_dephasing_hamiltonian}
    \mathcal{L}_t =    \mathcal{L}_{\rm SB}+\mathcal{L}_{\rm B}, \hspace{1ex}\mathcal{L}_{\rm SB}(\bullet)=-\frac{\ii}{2}\big[\big(\omega_0 + B(t)\big)\sigma^z,\bullet\big],
\end{equation}
where $B(t)$ is an environment/bath  operator that introduces fluctuations on the qubit frequency, and $\mathcal{L}_{\rm B}(\bullet)$ is the Liouvillian of the bath. In the standard description of quantum master equations, the environment is macroscopically large and subject to a purely-unitary evolution $\mathcal{L}_{\rm B}(\bullet)=-\ii[H_{\rm B},\bullet]$.   When the system-environment  coupling is weak, one can assume that the environment remains unaltered, such  that  the evolution takes place on the qubit but there is no back action $\rho_{\rm SB}=\rho(t)\otimes\rho_B$. A Born-Markov approximation then yields a non-unitary master equation for the qubit~\cite{Carmichael1993Open}. This can be expressed as  a time convolutionless master equation as the one discussed in the previous subsection in Eq.~\eqref{eq:TCNZ_eq}, also truncated at second order, where $\mathcal{P}$ is now a super-operator tracing over the bath degrees of freedom $\mathcal{P}(\rho_{\rm SB}) = \mathrm{Tr}_B(\rho_{\rm SB})=\rho$ ~\cite{Breuer2002}. The non-unitary evolution of the qubit results from the large number of degrees of freedom in the environment, such that the purity of the state can only decrease with no recurrences. 

Let us note, however, that the conditions under which these assumptions are made can be more general, and the degrees of freedom playing the role of an environment need not be macroscopically large. The crucial requirement is that the time with which the effective environment reaches its steady state $\mathcal{L}_{\rm B}(\rho^{\rm ss}_{\rm B})=0$ must be much shorter than the timescale of interest in which the system evolves $\rho(t)$. In the present context, this is the case of a single bosonic mode that  exchanges energy with  a larger electromagnetic bath with a certain rate $\kappa$. The bath Liouvillian reads
\beq
\mathcal{L}_{\rm B}(\bullet)=-\ii[\,H_{\rm B}(t),\bullet\,]+\kappa \left(a\rho a^{{\dagger}} -\half\{a^{{\dagger}}a, \bullet\}\right),
\eeq
where $H_{\rm B}(t)$ is the bosonic mode Hamiltonian, which can include external drivings,  and $a^\dagger,a$ are the bosonic creation and annihilation operators, respectively.  The condition for this single driven-dissipative mode to act as an environment is that $\kappa$ must be  much larger than the coupling strength inside $B(t)$. In the context of the superconducting circuits discussed in the main text, $\kappa$ is the rate of photon loss in a resonator, and $H_{\rm B}$ must contain a linear resonant microwave driving of the resonator that controls the non-zero number of photons in the steady state~\cite{PRXQuantum.1.010305}. For trapped ions, $\kappa$ will be the rate of sympathetic cooling of a vibrational model in a two-ion crystal, which will also be supplemented with a smaller heating rate ~\cite{PhysRevA.46.2668}. The difference of these two rates controls the population of phonons in the steady state, and can be controlled by an external laser.

We now move to the interaction picture with respect to the bare system Liouvillian $\rho_I(t)=\ee^{t\mathcal{L}_{\rm S}}(\rho(t))$ with  $\mathcal{L}_{\rm S}(\bullet)=-\ii[\half\omega_0\sigma_z,\bullet]$, and the bare bath Liouvillian, i.e., $ B_I(t)= \ee^{t\mathcal{L}_{\rm B}}(B(t))$. The key step is that, due to the fast decay of the bath, for the  timescales of interest $t\gg1/\kappa$,  one can assume that $\rho_{\rm SB}(t)=\rho(t)\otimes \rho_B^{\rm ss}$, the second-order time-convolutionless master equation can be expressed as in Eq.~\eqref{eq:dephasing_tcl} with 
\beq
\label{eq:quanum_corr_function}
    C(t,t')=\mathbb{E}[  B_I (t)  B_I (t')]= \mathrm{Tr}_B\{ B_I (t)  B_I (t')  \rho_{\rm B}^{\rm ss}\}.
\eeq 
Here, we have assumed that $\mathcal{P}({B}_I(t))=0$, and  we note that  $ B (t)$ need not commute with itself at different times. Once more, if these quantum-mechanical auto-correlation functions are wide-sense stationary, $C(t,t')=C(t-t')$. We note that, in contrast to the semi-classical case where $S(\omega)=S(-\omega)$, this is not necessarily the case in the quantum-mechanical case $S(\omega)\neq S(-\omega)$ \cite{RevModPhys.82.1155}. However, in the case of pure dephasing, the time evolution in Eq.~\eqref{eq:dephasing_tcl} only depends on the symmetrized auto-correlation function $\bar{C}(t, t')=\frac{1}{2}(C(t,t') + C(t', t))$ and therefore only the symmetric part of the auto-correlation function  influences the time evolution
\begin{equation}
    \frac{{\rm d}{ \rho}}{{\rm d}t}  =
    \frac{1}{2} \int_0^t\!{\rm d}t' \bar{C}(t,t')\big( \sigma^z  \rho(t)  \sigma^z -  \rho(t)  \big).
\end{equation}
Moving back to the Schr\"odinger picture, we obtain the  master equation in Eq.~\eqref{eq:TCL_dephasing}, which will only depend on the symmetrized noise PSD defined below Eq.~\eqref{eq:f_gamma}.

\subsection*{Error analysis and asymptotic statistics}\label{sec:error_analysis}

We show here the asymptotic normality of the maximum-likelihood estimator in Eq.~\eqref{eq:min_freq} when $N_{\rm shot}$ tends to infinity and how this is related to the Fisher information in Eq.~\eqref{eq:asymptotic_covariance}. The asymptotic covariance matrix is used in the main text to obtain the optimal times of the frequentist estimator and can be used to obtain approximate confidence intervals of estimations. To simplify the notation and reduce complexity, we consider the pure dephasing case here, where the cost function is simplified. However, the conclusions remain the same, and the general case can be recovered by substituting $i \rightarrow i,s,b$ and $m_x \rightarrow m_b$. In this case, instead of considering only measurement points in time, we account for measurement configurations, including evolution times, initial states, measurement bases, and measurement outcomes.

Our estimation $\boldsymbol{\hat \theta}_F$ is affected by the limited number of measurement shots $N_{i}$ at each point in time $t_i$, which will cause $\tilde{f}_{i}$ to behave as a normal random variable and will produce a random error in the final estimation of the parameters. The maximum-likelihood cost function reads
\begin{align}
    \mathsf{C}^{\rm pd}_{\mathrm{TL}}(\boldsymbol{\theta}) = -\sum_{i, m_x}  N_i \tilde{f}_{i,m_x} \log p_{i}(m_x | \boldsymbol{\theta}),
\end{align}
where $\tilde{f}_{i,m_x}$ denotes $\tilde{f}_{i}(m_x\,|\boldsymbol{\theta}_\star)$ and $p_{i}(m_x | \boldsymbol{\theta})$ denotes $p^{\rm TL}_{i}(m_x | \boldsymbol{\theta})$.
The minimum of the cost function satisfies
\beq \label{eq:cost_function_minimum}
\partial_{\boldsymbol{\theta}} C_{\rm TL}^{\rm pd}= -\sum_{i, m_x} N_{i} \tilde{f}_{i, m_x} \frac{\partial_{\boldsymbol{\theta}} p_{i}(m_x | \boldsymbol{\theta})}{p_{i}(m_x | \boldsymbol{\theta})} = 0.
\eeq
Since $\tilde{f}_{i, m_x} = p_{i}(m_x | \boldsymbol{\theta}_{\star}) + \Delta \tilde{f}_{i,m_x}$, with $\Delta \tilde{f}_{i,m_x}$ small, the minimum of the cost function is slightly displaced from the true minimum $\boldsymbol{\theta}_{\star}$ to $\boldsymbol{\theta}_{\star} + \Delta\boldsymbol{\theta}$. Taylor expanding Eq.\ \eqref{eq:cost_function_minimum} around $\boldsymbol{\theta}_{\star}$ to first order we have
\begin{align}
    \partial_{\theta_j} C_{\rm TL}^{\rm pd} \approx & -\sum_{i, m_x} N_{i} \left[ p_{i}(m_x | \boldsymbol{\theta}) + \Delta \tilde{f}_{i,m_x}\right] \left[ \left.\frac{\partial_{\theta_j} p_{i}(m_x | \boldsymbol{\theta})}{p_{i}(m_x | \boldsymbol{\theta})}\right|_{\boldsymbol{\theta}=\boldsymbol{\theta}_{\star}} \right. \nonumber\\ 
    &\left. + \sum_k \left.\partial_{\theta_k}\frac{\partial_{\theta_j} p_{i}(m_x | \boldsymbol{\theta})}{ p_{i}(m_x | \boldsymbol{\theta})}\right|_{\boldsymbol{\theta}=\boldsymbol{\theta}_{\star}}\Delta \theta_k \right] = 0.
\end{align}
Keeping only first-order terms in $\Delta \boldsymbol{\theta}$ and $\Delta \tilde{f}_{i, m_x}$ and simplifying we obtain
\begin{equation}
    \sum_{i,j} \Delta \boldsymbol\theta_j N_{i} [I_{i} (\boldsymbol{\theta}_{\star} ) ]_{jk} = \sum_{i, m_x} N_{i} \Delta \tilde{f}_{i,m_x} \left. \frac{\partial_{\theta_k} p_{i}(m_x | \boldsymbol{\theta})}{ p_{i}(m_x | \boldsymbol{\theta})}\right|_{\boldsymbol{\theta}=\boldsymbol{\theta}_{\star}},
\end{equation}
where we have defined the matrix 
\begin{align}
    [I_{i}& (\boldsymbol{\theta}_{\star} ) ]_{jk} =\\ 
    & \sum_{m_x} p_{i}(m_x | \boldsymbol{\theta}_{\star}) \partial_{\theta_j}\log p_{i}(m_x | \boldsymbol{\theta}) \partial_{\theta_k}\log p_{i}(m_x | \boldsymbol{\theta}) |_{\boldsymbol{\theta}=\boldsymbol{\theta}_{\star}}. \nonumber
\end{align}
Note that this is the Fisher information given in Eq.~\eqref{eq:fisher}. Taking into account that $\Delta \tilde{f}_{i,1} = - \Delta \tilde{f}_{i,0}$ and $\tilde{f}_{i,1} = 1-\tilde{f}_{i,0}$, and defining the matrices 
\begin{equation}
    J_{jk} = \sum_{i} N_{i} [I_{i}(\boldsymbol{\theta}_{\star})]_{jk},\quad\ F_{ji} = N_{i} \frac{\left.\partial_{\theta_j} p_{i}(0 | \boldsymbol{\theta}) \right|_{\boldsymbol{\theta}=\boldsymbol{\theta}_{\star}}}{ p_{i}(0 | \boldsymbol{\theta}_{\star})(1-p_{i}(0 | \boldsymbol{\theta}_{\star}))},
\end{equation}
we arrive at the expression
\beq \label{eq:delta_c_delta_f_relation}
\Delta \theta_k = \sum_{i,j} J^{-1}_{jk} F_{ji} \Delta \tilde{f}_{i,0},
\eeq
which relates differences between the expected and measured values $\Delta \tilde{f}_{i,m_x}=\tilde{f}_{i, m_x} - p_{i}(m_x | \boldsymbol{\theta}_{\star})$ to differences between the estimated and the true parameters $\Delta \theta_k$. When $N_{i}$ is sufficiently large, $\tilde{f}_{i,0}$ behaves as the normal distribution $N\left[\mu=p_i(0|\boldsymbol{\theta}_{\star}), \sigma_f^2=p_{i}(0|\boldsymbol{\theta}_{\star}) p_{i}(1|\boldsymbol{\theta}_{\star})/N_{i}\right]$. Thus, we have
\begin{align}
    \Delta \tilde{f}_{i,0} \sim  N \left[\mu=0 , \sigma^2_{f_{i}}=p_{i}(0|\boldsymbol{\theta}_{\star}) p_{i}(1|\boldsymbol{\theta}_{\star})/N_{i}\right].
\end{align}
After the linear transformation of Eq.~\eqref{eq:delta_c_delta_f_relation}, we obtain that $\Delta \boldsymbol{\theta}$ behaves as
\beq
\Delta \boldsymbol{\theta} \sim N\left[\mu=0,\  \Sigma_{\hat{\boldsymbol{\theta}}} = J^{-1}\right],
\eeq
where we have used that $J^{-1} F \mathrm{diag}(\sigma^2_{f})F^T (J^{-1})^T=J^{-1}$. Thus, the shot noise produces a normal random error with covariance matrix $\Sigma_{\boldsymbol{\theta}}=J^{-1}$ which scales as $1/N_{\rm shot}$, with $N_{\rm shot}=\sum_i N_{i}$, which is the result presented in Eq.~\eqref{eq:asymptotic_covariance}.


\section*{Code availability}
The code used to conduct the analyses and generate the results presented in this work is openly accessible at \url{https://github.com/varona/llqt}.

\begin{acknowledgements}
The project leading to this publication has received funding from the US Army Research Office through Grant No. W911NF-21-1-0007. A.B acknowledges support from PID2021-127726NB- I00 (MCIU/AEI/FEDER, UE), from the Grant IFT Centro de Excelencia Severo Ochoa CEX2020-001007-S, funded by MCIN/AEI/10.13039/501100011033, from the CSIC Research Platform on Quantum Technologies PTI-001, and from the European Union’s Horizon Europe research and innovation programme under grant agreement No 101114305 (“MILLENION-SGA1” EU Project). M.M. furthermore acknowledges support by the European Union’s Horizon Europe research and innovation program under Grant Agreement No. 101046968 (BRISQ), the ERC Starting Grant QNets through Grant No. 804247, by the Germany ministry of science and education (BMBF) via the VDI within the project IQuAn, and by the Deutsche Forschungsgemeinschaft (DFG, German Research Foundation) under Germany’s Excellence Strategy “Cluster of Excellence Matter and Light for Quantum Computing (ML4Q) EXC 2004/1” 390534769. This research is also part of the Munich Quantum Valley (K-8), which is supported by the Bavarian state government with funds from the Hightech Agenda Bayern Plus.

\end{acknowledgements}




\section{\bf Supplementary Material: Analytical results for the frequentist and Bayesian estimation of  Markovian Lindblad dephasing}\label{app:Lindblad_case}

In this section, we provide an analytical solution of both the frequentist and Bayesian estimation for the LQT of single-qubit Lindblad dephasing. This  serves to benchmark  some of the limiting results of subsections ``Markovian semi-classical dephasing'' and ``Non-Markovian quantum dephasing'' of the main text. We start here from the equations and derivations shown in subsection ``Lindblad-like quantum tomography for non-Markovian dephasing'' of the main text.
In the Lindblad approximation to pure dephasing, we can substitute $\gamma(t)=\gamma$ in the time-local master equation,
\beq
\frac{{\rm d}\rho}{{\rm d}t}=-\frac{\ii}{2}\big[\omega_0\sigma_z,\rho\big]+\gamma(t)\big(\sigma_z\rho\sigma_z-\rho\big).
\eeq
obtaining a constant dephasing rate. This leads to the decay rate  $\Gamma(t) = t/T_2$, where the dephasing time is $T_2=1/2\gamma$,  such that the likelihood function  $p^{\rm L}_i(m_x|T_2) = \frac{1}{2}(1+(-1)^{m_x}\ee^{-t_i/T_2})$ shows  a purely-exponential coherence decay. Assuming that the actual qubit dephasing   follows this pdf with a real value of $T_{2\star}$, or equivalently $\gamma_{\star}$,  we   have a single noise parameter to learn, and  could thus estimate it by measuring  at a single time $t_i=t$.  Note that the relative frequencies  in this case would follow $\tilde{f}_i(m_x|T_{2\star}) = \frac{1}{2}(1+(-1)^{m_x}\ee^{-t_i/T_{2\star}})$ for $N_{\rm shot}\rightarrow \infty$.

For the frequentist approach, we can actually solve  the minimization of the cost function in Eq.~(34) of the main text analytically, since
$\dd{\mathsf{C}_{\rm L}}/\dd{T_2} = 0$ yields $\hat{T}_2 = -t_i/\log(2 f_i(0|T_{2\star})-1)$. We can also estimate the precision of $\hat T_2$ using Fisher's information and Eq.\ (13) of the main text, which shows the   standard quantum limit~\cite{PhysRevA.47.3554} scaling of the standard deviation with the number of measurement shots
\begin{equation}\label{eq:variance_markovian_t2}
    \mathrm{Var}\left(\hat{T}_2\right) = \frac{T_{2\star}^4 }{ N_{\rm shot} t_i^2}\left(\ee^{\frac{2 t_i}{T_{2\star}}}-1\right)\!\!.
\end{equation}
We can then obtain the optimal evolution time at which one should perform the measurements by minimizing the variance, which yields $t_i \approx0.797T_{2\star}$, which is where the signal has the highest sensitivity to changes in the dephasing rate. For very short evolution times, the signal is still not highly influenced by the dephasing, whereas  the decoherence is too large to extract any useful information for much larger times. In the frequentist approach, if we have some prior knowledge of $T_2$, we should then measure close to this optimal time, instead of distributing the measurements uniformly in time and making a fit to a purely exponential decay.

We note that, in the context of Ramsey estimators  for frequency standards, an atomic qubit will acquire a phase that depends on the detuning of its frequency with respect to that  of the field  that drives the initial and final $\pi/2$ rotations in the initialization and measurement steps~\cite{PhysRevA.47.3554,HaaseSmirneHuelgaKołodynskiDemkowiczDobrzanski+2016+13+39,RevModPhys.89.035002,10.1116/1.5119961}. In the absence of dephasing, both the signal and the error have the same sinusoidal dependence on the elapsed time, and the signal to noise decays with  $1/t_i\sqrt{N_{\rm shot}}$ regardless of the time at which one measures, leading to the aforementioned standard quantum limit. If there are other sources of noise beyond shot noise, this is no longer the case, and the error is minimized by maximizing the slope of the signal, i.e., maximizing the Fisher information in Eq.~(12) of the main text.

Let us now switch to the Bayesian approach in which, at each step, we select the best time for the next measurement considering our current knowledge of the parameters. We here show that these optimal times can be obtained analytically for the Lindbladian evolution. We will assume that we are trying to determine $\gamma_{\star} = 1/2T_{2\star}$ and the prior probability distribution for our knowledge about $\gamma_{\star}$ is  normal $\pi_{0}(\gamma) \sim {N}(\hat{\gamma}_{0}, \sigma_0^2)$ with  a small dispersion around the mean $\sigma_0 \ll \hat{\gamma}_0$.  Therefore, the parameter space is $\theta=\gamma\in\Theta=\mathbb{R}^+$, and  the likelihood function for obtaining a single new outcome $\delta\mathbb{D}_\ell=\{m_x\}$ conditioned on our statistical knowledge of the noise is  $p_{\rm TL}(\delta\mathbb{D}_\ell|\gamma_\ell) = \frac{1}{2}(1+(-1)^{m_x} \ee^{-2\gamma_\ell t})$. We can thus calculate the posterior $\pi_{\ell}(\gamma|\mathbb{D}_{\ell-1}\cup\delta\mathbb{D}_\ell)$ at each $\ell>0$ step using the Bayesian update in Eq.~(16) of the main text for each of the two possible outcomes $m_x\in\{0,1\}$. Accordingly, the expected variance of the updated distribution can be expressed as follows 
\begin{equation}
    \mathbb{E}[ \tilde{\sigma}_\ell^2]  = p^{\rm TL}_{\ell-1}(\{0\})\sigma^2_{\ell,0} + p^{\rm TL}_{\ell-1}(\{1\})\sigma^2_{\ell,1},
\end{equation}
where we have introduced the variances for the posteriors for each of the two possible outcomes 
\begin{widetext}
\begin{equation}
    \sigma^2_{\ell,m_x} = \int_0^\infty\!\!\! \dd{\gamma}\, \gamma^2\, \pi_{\ell}\big(\gamma|\mathbb{D}_{\ell-1}\cup\{m_x\}\big) - \left(\int_0^\infty\!\! \dd{\gamma}\, \gamma\,\pi_{\ell}\big(\gamma|\mathbb{D}_{\ell-1}\cup\{m_x\}\big)\right)^{\!\!2}.
\end{equation}
\end{widetext}
Minimizing the expected variance of the posterior  produced similar results to the maximization of the information gain of Eq.~(19) of the main text in the  case of a normal prior. Moreover, assuming that the priors also follow a Gaussian distribution $\pi_{\ell}\sim N(\hat{\gamma}_\ell,\sigma^2_\ell)$, which is justified in the asymptotic limit of a large number of shots, one can solve the integrals analytically \cite{martinez_garcia_2019}, obtaining a simple update rule for the expected variances  
\begin{equation}\label{eq:expected_variance_diff}
     \mathbb{E}[ \tilde{\sigma}_\ell^2]  -  \sigma^2_{\ell-1} = \frac{-4t_\ell^2 \sigma_{\ell-1}^4}{\ee^{4 \hat{\gamma}_{\ell-1} t_{\ell} - 4\sigma^2_{\ell-1} t^2_\ell} - 1}.
\end{equation}
Minimizing this variance with respect to $t_\ell$ and using the approximation $\sigma_{\ell-1} \ll \hat{\gamma}_\ell$, we obtain again the results $t_\ell\approx 0.797/2\hat{\gamma}_\ell$, which agrees with the optimal timing in the frequentist approach that was discussed right below Eq.~\eqref{eq:variance_markovian_t2}. This agreement is reassuring of the validity of the analytical derivations for the Bayesian approach. We note that our assumption $\sigma_{\ell-1} \ll \hat{\gamma}_\ell$ implicitly requires that we are in the asymptotic regime in which the variance is small and the posterior is also a Gaussian, and therefore Eq.~\eqref{eq:variance_markovian_t2} should apply. However, in contrast to the frequentist approach for the Linbladian dephasing, where the we would optimally set a unique time in which all the $N_{\rm shot}$ measurements are performed, here the evolution time changes as we update our knowledge of $\{\hat{\gamma}_\ell\}$ in the consecutive steps. 

\begin{figure}
    \centering
    \includegraphics[width=0.9\linewidth]{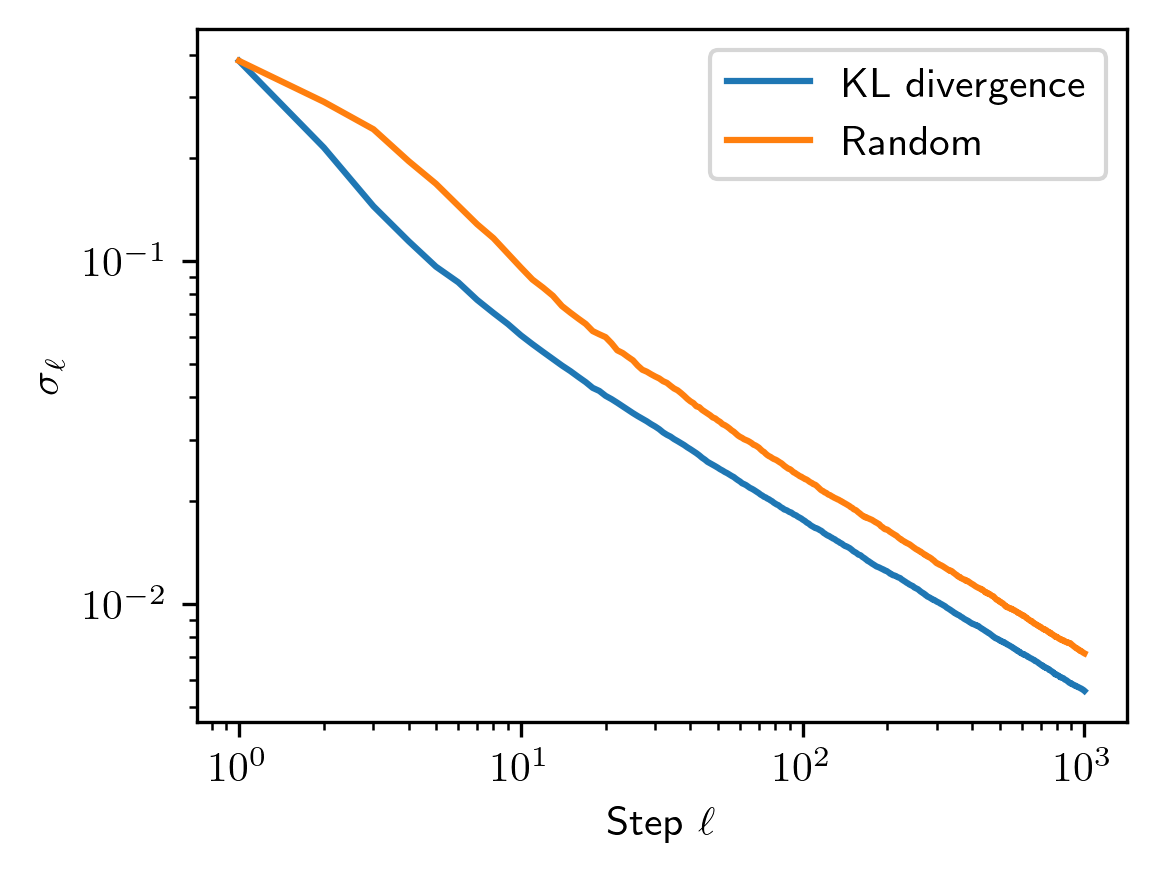}
    \caption{Standard deviation of the posterior $\sigma_\ell$ at each Bayesian step for the dephasing rate $\gamma$ in Markovian dephasing. The blue line represents the standard Bayesian approach in which at each step we find the optimal time using Eq.\ (19) of the main text. The orange line simply chooses a random time $t_\ell\in \left[0.1, 3\right]$. In both cases we start from the same uniformly distributed prior $U(\gamma_\star/3, 3\gamma_\star)$. At the beginning, the standard approach obtains an advantage, later on, once the number of measurements is sufficiently large, both instances become straight lines with slope $-\frac{1}{2}$ in this log-log plot, reflecting the asymptotic behavior $\sigma_\ell\sim \ell^{-1/2}$. A total of 30 Bayesian runs were done in each case to estimate the expected $\sigma_\ell$. 50 measurement shots were taken at each step.}
    \label{fig:bayesian_random_vs_kl}
\end{figure}

Let us close this section by discussing how $\sigma_\ell^2\approx \mathbb{E}[ \sigma_\ell^2]$ scales with the Bayesian step according to Eq.~\eqref{eq:expected_variance_diff}. We also assume that 
$2 \hat{\gamma}_{\ell-1} t_\ell \gg \sigma^2_{\ell-1} t^2_\ell$, and that we have applied a sufficiently large number of Bayesian steps such that the changes in our knowledge are small and $\hat{\gamma}_{\ell} \simeq \hat{\gamma}_{\ell-1}$, all of which is reasonable in the asymptotic regime when we have done many measurements and $\sigma^2_{\ell-1}$ is small. We note that there might be  small deviations of $\sigma_{\ell}^2$ from $\mathbb{E}[ \sigma_\ell^2]$ but, after some additional measurements, they should eventually become vanishingly small. Under these assumptions, we have $\sigma^2_{\ell} = \sigma^2_{\ell-1} - k(t_\ell, \hat{\gamma}_{\ell-1}) \sigma_{\ell-1}^4$,
with $k(t_\ell, \hat{\gamma}_{\ell-1}) = t_\ell^2/(\ee^{2\hat{\gamma}_{\ell-1} t_\ell} - 1)$. The recursive sequence  $a_{\ell} = a_{\ell-1} - c a_{\ell-1}^2$ scales as $O(1/\ell)$ for $c$ a positive constant, $ca_0 \ll 1$ and $0<a_0 \ll 1$. Therefore $\sigma_\ell^2$ will scale as $O(1/\ell)$ independently of the time $t$ we choose to measure (i.e. we assume here that one chooses the same time  $t$ for every step $\ell$). Therefore, differences between the Bayesian optimal approach and any other inference that chooses non-optimal times can only be noticed when the variance of the prior is still large. Once we have taken sufficient steps, and the standard deviation has decreased to a  sufficiently small value, both the optimal measurement and the non-optimal measurement should look like two parallel lines in a log-log plot of standard deviation against $\ell$, see Fig.~\ref{fig:bayesian_random_vs_kl}.

\section{\bf Supplementary Material: Comparison to uniform-time Ramsey estimation } \label{app:uniform_estimator}

\begin{figure}
    \includegraphics[width=0.9\columnwidth]{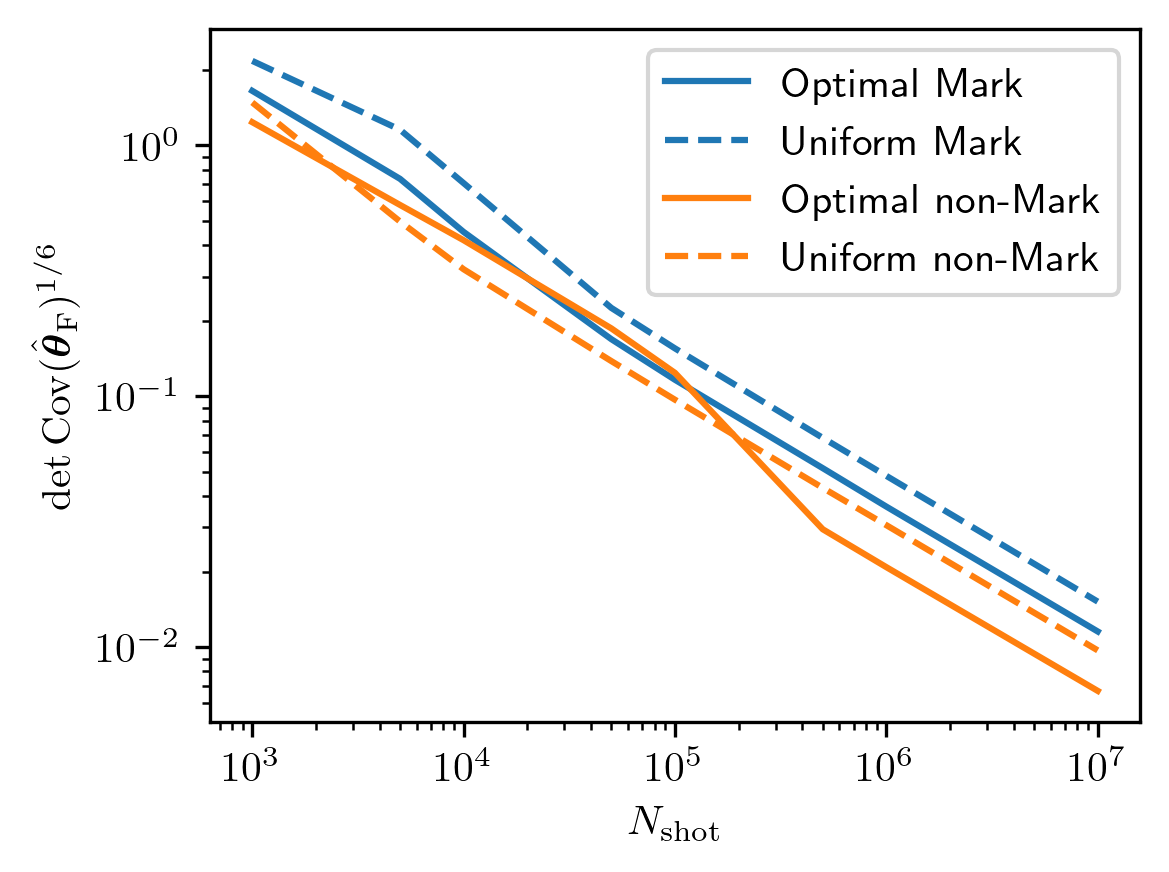}
    \caption{The efficiency of the optimal (solid) and uniform (dashed) estimators displayed in terms of $\det \mathrm{Cov}(\boldsymbol{\hat{\theta}}_{\rm F})^{1/6}$, with parameters $\boldsymbol{\theta} = (g^2_{\bar{n}}, \kappa, \Delta_{\rm c})$ for different number of total measurements $N_{\rm shot}$. We consider two cases: (\textit{i}) the Markovian case (blue) in which $\tau_{\rm c\star}=2/\kappa_{\star}=0.3 T_{2\star}$, $\Delta_{\rm c\star}=5/T_{2\star}$, (\textit{ii}) the non-Markovian case (orange) where $\tau_{\rm c\star}=2/\kappa_{\star}=0.9T_{2\star}$, $\Delta_{\rm c\star}=5/T_{2\star}$. $T_{2\rm \star}$ is set to 1, which makes $g^2_{\bar{n}\star}\approx 2.7$ for the first and $g^2_{\bar{n}\star}\approx 5.9$ for the second case. The uniform estimator measures time in 20 different equidistant points in the interval $t\in[0.02, 3]$. Asymptotically $\det \mathrm{Cov}(\boldsymbol{\hat{\theta}}_{\rm F})^{1/6}$ scales as $1/\sqrt{N_{\rm shot}}$ for both estimators. For each value $N_{\rm shot}$, data is sampled and fitted to obtain an estimation. This is done a total of 5000 times to obtain afterwards the expected value of the covariance at that point. Bounds for the least squares algorithm were set to $(\boldsymbol{\theta}_{\!\star}/3,3\boldsymbol{\theta}_{\!\star})$. The initial guess was randomly chosen in that range.}
    \label{fig:frequentist_comparison}
\end{figure} 
\begin{figure}
    \centering
    \includegraphics[width=0.9\linewidth]{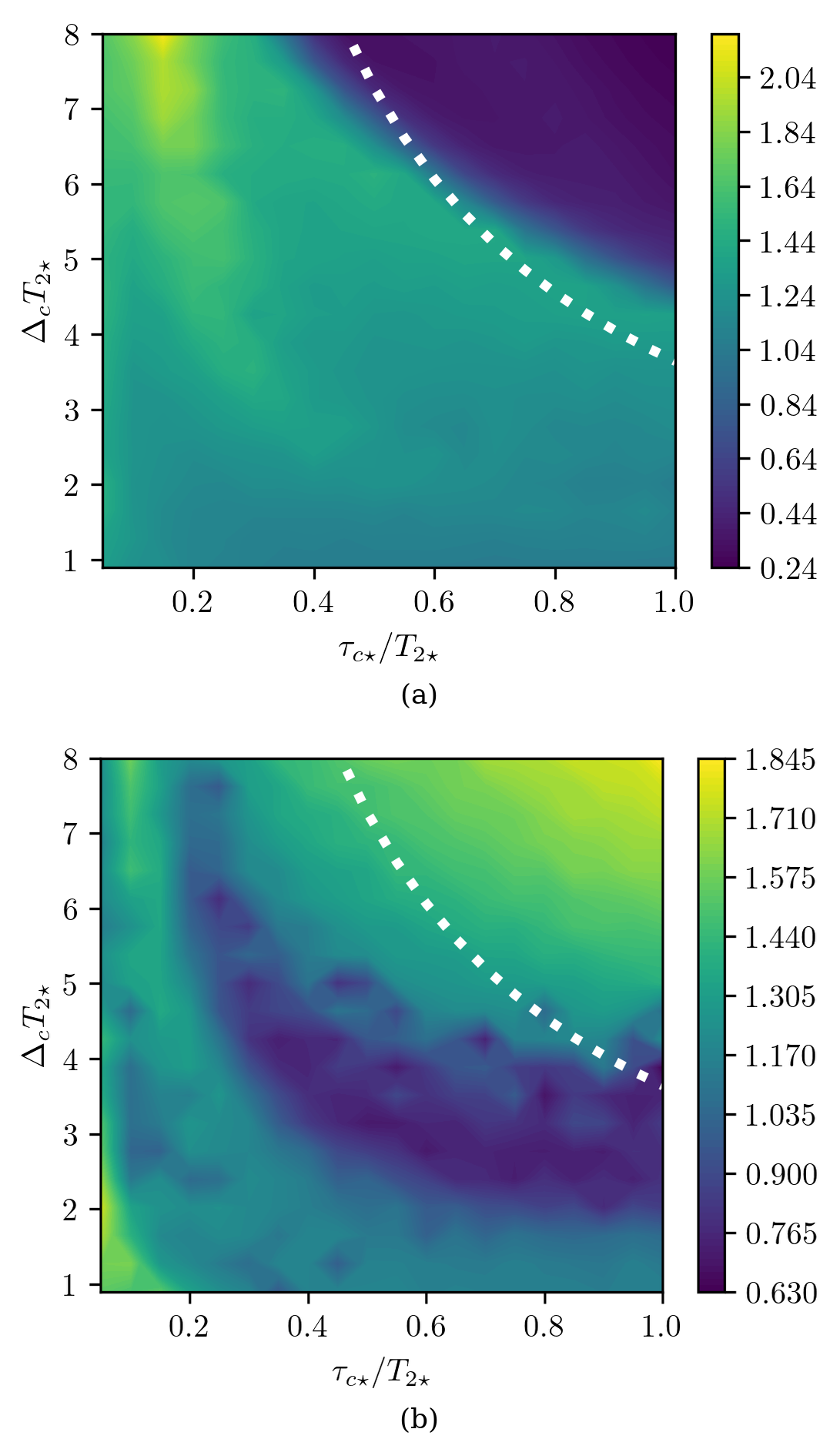}
    \caption{Ratios between determinants of covariance matrices of different estimators for different parameter values $\boldsymbol{\theta}_{\!\star}=(g^2_{\bar{n}\star}, \kappa_{\star}, \Delta_{\rm c\star})$ and a fixed number of total measurements $N=2\cdot10^4$. $T_2=1$ for all combinations of $\tau$ and $\Delta_{\rm c}$. (a) The ratio between the uniform and the optimal estimator $r_\mathrm{uni,opt}=(\det \mathrm{Cov}(\boldsymbol{\hat{\theta}}_{\rm F,uni}) / \det \mathrm{Cov}(\boldsymbol{\hat{\theta}}_{\rm F,opt}))^{1/6}$. (b) Ratio between the uniform estimator and the Bayesian approach, $r_\mathrm{uni,Bay}=(\det \mathrm{Cov}(\boldsymbol{\hat{\theta}}_{\rm F,uni}) / \det \Sigma_{\ell_{\max}})^{1/6}$. For the frequentist estimators the same settings as in Fig.\ \ref{fig:frequentist_comparison} were used to determine the expected value of the determinant. For the Bayesian approach a total of 30 experiments were run, with starting prior given by the continuous uniform distribution $U(\boldsymbol{\theta}_{\!\star}/3, 3\boldsymbol{\theta}_{\!\star})$, i.e., the same bounds that were given to the least squares algorithm. The dashed white line separates the Markovian from the non-Markovian regime. The non-Markovian region is located in the upper right corner. \label{fig:det_ratios}}
\end{figure}

As discussed at the end of subsection ``Non-Markovian quantum dephasing'', one can also compare the efficiency of the  estimators introduced in the main text with a different one in which the measurement times are uniformly distributed over the whole time interval $T$. Such time distributions have been considered in previous works for Lindblad quantum tomography~\cite{PhysRevApplied.18.064056} and spin-locking  spectroscopy of correlated noise~\cite{PRXQuantum.1.010305}. In this section, we explore them in the context of the non-Markovian quantum dephasing.
 
 This comparison can be done by calculating the determinants of the covariance matrix of the estimators, as is shown in Fig.~\ref{fig:frequentist_comparison} for two examples with different microscopic parameters. One parameter set shows a Markovian evolution, while the other leads to a non-Markovian one. The frequentist  estimator with three optimal times is depicted by solid lines, and is always the best one in the asymptotic limit of an infinitely-large number of measurements $N_{\rm shot}\rightarrow \infty$. However,  before reaching this regime, the uniform estimator displayed by dashed lines can become better. We are specially interested in the region with $10^4<N_{\rm shot}<10^5$, where the variance is already good enough to stop making more measurements, but the cost in terms of number of measurements is still not too big. In this region and for the Markovian case (blue), the uniform estimator is better than the frequentist one, which only considers three evolution times, set at the optimal values of the asymptotic limit. For the non-Markovian case (orange),  the comparison is the opposite.

In order to extend this comparison between the uniform and the optimal-time estimators for a broader set of parameters, we compute the ratios between the determinant of the covariance matrix of the two estimators, $r_{\mathrm{uni,opt}}=\det \mathrm{Cov}(\boldsymbol{\hat{\theta}}_{\rm F,uni})^{1/6} / \det \mathrm{Cov}(\boldsymbol{\hat{\theta}}_{\rm F,opt})^{1/6}$, for different values of parameters $\tau_{\rm c \star}$ and $\Delta_{\rm c\star}$, as shown in Fig.~\ref{fig:det_ratios}a and analogously to what we do in Fig.~10 of subsection ``Non-Markovian quantum dephasing'' in the main text. This determinant ratios are computed for $N_{\rm shot}=2\cdot 10^4$, before reaching the region with asymptotic behavior, where the optimal estimator is always better than the uniform one. We can clearly see that the uniform estimator outperforms the optimal one in the non-Markovian regime (upper right region), while the opposite happens in the non-Markovian regime. While the optimal estimator is not realistic, since we need to know beforehand the true value of the parameters so that we can determine the optimal times, we may use it as a baseline to compare against other estimators or try to approximate it if we have some prior knowledge of the parameters. This comparison shows that the uniform estimator might be a good choice in the non-Markovian regime when doing a small number of shots.

In Fig.~\ref{fig:det_ratios}b we compare the uniform estimator with the Bayesian estimation. We can see that the Bayesian approach is better than the uniform estimator except for a small region in the Markovian regime.

\bibliography{citations}
\end{document}